\documentstyle[12pt,eqsecnum,aps,epsf,amsfonts]{revtex}
\tightenlines

\def\gcm3{{\rm g}/{\rm cm}^3}
\def\smgt{{\scriptscriptstyle > }}
\def\smlt{{\scriptscriptstyle < }}
\def\smD{{\scriptscriptstyle D }}
\def\smS{{\scriptscriptstyle {\rm S} }}
\def\smR{{\scriptscriptstyle {\rm R} }}
\def\smC{{\scriptscriptstyle {\rm C} }}
\def\smQ{{\scriptscriptstyle {\rm Q} }}
\def\smG{{\scriptscriptstyle {\rm G} }}
\def\smT{{\scriptscriptstyle {\rm T} }}
\def\smH{{\scriptscriptstyle {\rm H} }}
\def\smB{{\scriptscriptstyle {\rm B} }}
\def\smI{{\scriptscriptstyle {\rm I} }}
\def\smLP{{\scriptscriptstyle {\rm LP} }}
\def\smLP{{\scriptscriptstyle {\rm LP} }}
\def\smBPS{{\scriptscriptstyle {\rm BPS} }}

\def\smD{{\scriptscriptstyle D }}
\def\smC{{\scriptscriptstyle {\rm C} }}
\def\smQ{{\scriptscriptstyle {\rm Q} }}

\def\mmbf#1{\mbox{\boldmath${#1}$}} 

\makeatletter
\def\footnotesize{\@setsize\footnotesize{15pt}\xpt\@xpt
\abovedisplayskip 10pt plus2pt minus 5pt
\belowdisplayskip \abovedisplayskip
\abovedisplayshortskip \z@ plus 3pt
\belowdisplayshortskip 6pt plus 2pt minus 2pt
\def\@listi{\topsep 6pt plus 2pt minus 2pt
\parsep 3pt plus 2pt minus 1pt \itemsep \parsep}}
\makeatother
\def\footnoterule{\kern-3pt \hrule width \hsize \kern6.2pt}
\begin{document}
\def\floatpagefraction{.1}

\title
{
\baselineskip 18pt plus 1pt minus 1pt
Charged Particle Motion in a Highly Ionized Plasma
}

\author
{
Lowell S. Brown$^{a,b,c}$, Dean L. Preston$^c$,
and Robert L. Singleton Jr.$^c$
}

\address
{
\baselineskip 15pt plus 1pt minus 1pt
$^a$ Department of Physics, Box 351560,\\
     University of Washington, Seattle, WA 98195
\\[0.2cm]
$^b$ Institute of Theoretical Physics, University 
     of California\\
     Santa Barbara, CA 93106
\\[0.2cm]
$^c$ Los Alamos National Laboratory,\\
     Los Alamos, New Mexico 87545
}

\date{January 6, 2005}

\maketitle
\thispagestyle{empty}
\setcounter{page}{1}

\begin{abstract}
\baselineskip 15pt plus 1pt minus 1pt
  {A recently introduced method utilizing dimensional
  continuation is employed to compute the energy loss
  rate for a non-relativistic particle moving through
  a highly ionized plasma. No restriction is made on
  the charge, mass, or speed of this particle. It is,
  however, assumed that the plasma is not strongly coupled
  in the sense that the dimensionless plasma coupling
  parameter $g= e^2 \kappa_\smD / 4 \pi T $ is small,
  where $\kappa_\smD$ is the Debye wave number of the
  plasma.  To leading and next-to-leading order in this 
  coupling, $dE/dx$
  is of the generic form $g^2 \, \ln[C g^2] $.  The
  precise numerical coefficient out in front of the
  logarithm is well known.  We compute the constant
  $C$ under the logarithm exactly for arbitrary particle
  speeds.  Our exact results differ from approximations 
  given in the literature.  The differences are in the 
  range of 20\% for cases relevant to inertial 
  confinement fusion experiments. The same method is 
  also employed to compute the rate of momentum loss 
  for a projectile moving in a plasma, and the rate at
  which two plasmas at different temperatures come into
  thermal equilibrium. Again these calculations are done 
  precisely to the order given above.  The loss rates of 
  energy and momentum uniquely define a
  Fokker-Planck equation that describes particle motion
  in the plasma.  The coefficients determined in this way
  are thus well-defined, contain no arbitrary parameters
  or cutoffs, and are accurate to the order described. 
  This Fokker-Planck equation describes the 
  straggling --- the spreading in the longitudinal 
  position of a group of particles with a common initial 
  velocity and position --- and the transverse diffusion 
  of a beam of particles.  It should be emphasized that
  our work does not involve a model, but rather it is a
  precisely defined evaluation of the leading terms in a
  well-defined perturbation theory.}
  
\end{abstract}

\vfill
\noindent LA-UR-042713 \hfill Phys. Rep. 410/4 (2005) 237

\newpage
\tableofcontents

\newpage
\baselineskip 18pt plus 1pt minus 1pt

\section{Introduction} 

The methods of quantum field theory (QFT), originally developed to
describe the interactions of elementary particles, have since been
successfully applied in several other fields of physics, including plasma
theory.  Condensed matter physics in particular has been a proving ground
for the utility of QFT methodology. The Kondo problem, phase
transformations in magnetic metals, and a host of other problems in
condensed matter theory have been tackled using QFT 
methods\cite{kondo,mj,wk}.
Classical non-equilibrium reaction-diffusion systems have been studied 
using QFT techniques since the pioneering work of Doi in 1976\cite{doi}. 
More recently the methods of effective field theory were applied
to multicomponent, fully ionized plasmas by Brown and Yaffe\cite{BY}.  
These
methods significantly simplify high-order perturbative calculations and
clarify the structure of the theory.  Here we use the method of
dimensional continuation, which was originally developed as a means of
regularizing divergent integrals that arise in perturbative calculations
in gauge-invariant QFT, to calculate the stopping power and temperature
equilibration in highly ionized plasmas.  The well-known Lenard-Balescu
kinetic equation\cite{len,bal}  describes the long-distance, collective 
excitations
of the plasma, whereas the Boltzmann equation for pure Coulomb scattering
describes the short-distance, hard collisions of the plasma particles. A
complete description of the plasma includes both the long- and
short-distance physics encoded in the Lenard-Balescu and Boltzmann
equations, but each contribution is divergent if integrated over all of
three-dimensional space.  A finite sum, the large Coulomb logarithm and
its coefficient, can be obtained by introducing cutoffs, but this 
approach
does not determine the constants that accompany the logarithm. These
additional constants are given by the convergent kinetic equation 
method\cite{gould},
but spurious higher-order terms are introduced.  In contrast, our
method, which is based on a rigorous expansion in a small parameter and
dimensional continuation of the Lenard-Balescu and Boltzmann equations,
gives the constants accompanying the Coulomb logarithm but no spurious
higher-order terms.

In the course of presenting our calculations of the stopping power we 
make frequent contact with the existing literature and unify many previous
results; thus this work serves as a review.  On the other hand, we do
present some original results: we introduce new methods, obtain an
analytic expression for the stopping power in a fully ionized plasma that
is more accurate than all previous expressions, and we provide a more
precise definition of the Fokker-Planck equation for dilute plasmas.  In
order to make this Report accessible to readers who are not experts in
plasma theory or have no familiarity with dimensional continuation or
both, it is largely self-contained and is written in a pedagogical style.

\section{Method}

Many physical problems involve both large and small length scales
and are governed by a small parameter $g$, with $ 0 < g \ll 1 $.  
Two different physical mechanisms dominate at short and large
distances.  An infrared (IR) mechanism dominates at large
distances or low energies; an ultraviolet (UV) mechanism
dominates at short distances or high energies. In plasma
physics, the long-distance, collective effects (described
to leading order by a dielectric function) are the
dominant infrared effects that set the long-distance
scale.  Hard Coulomb scattering, cut off by either
the classical minimum approach distance or by the quantum
maximum momentum transfer, is the ultraviolet mechanism
that sets the short-distance scale.  A novel application
of dimensional continuation has recently been 
introduced\cite{brown1} to treat such problems when they
can be formulated in spaces of arbitrary dimensionality
$\nu$. If the spatial dimensionality $\nu$ is analytically
continued
below the physical $\nu = 3$, then the infrared mechanism
dominates for all scales and, since it is dominant, its
contribution is thus easy to compute in the lower
spatial dimensions.  On the other hand, if $\nu$ is
continued above $\nu = 3$, then the ultraviolet mechanism
dominates for all scales, and this different contribution
is easy to compute in the higher spatial dimensions.
As a simple example, let us consider a case in which
the dominate infrared mechanism for $\nu<3$ gives the
leading contribution of order $g^{ 2 - (3-\nu)}$, while
the dominate ultraviolet  mechanism for $\nu > 3$ gives 
the leading contribution of order $ g^{ 2 - ( \nu -3)} $:
\begin{eqnarray}
\nonumber
  g^{2-(\nu-3)} && ~{\rm for}~ \nu > 3 ~{\rm (UV)}\ ,
\\ 
  g^{2-(3-\nu)} && ~{\rm for}~ \nu < 3 ~{\rm (IR)}\ .
\label{simpleg}
\end{eqnarray}
The actual $\nu$ dependence is slightly more complicated
for the problem we will study; therefore, we shall look
at this simple example first since it illustrates the
point more concisely. From Eq.~(\ref{simpleg}), we see that 
when the infrared contribution is analytically
continued from $\nu < 3$ to $\nu > 3$ it becomes subleading
compared to the ultraviolet contribution since $ g^{- (3 
-\nu)} < g^{- ( \nu - 3)} $ for $ \nu > 3$.  Conversely,
when continued to $\nu < 3$, the ultraviolet mechanism 
becomes subleading compared to the infrared mechanism.  
One concludes that the {\it sum} of the two processes 
contains both the leading and the (first) subleading 
terms for all spatial dimensionality $\nu$ near the 
physical value $\nu = 3$, and hence this sum provides 
the correct result in the physical limit $\nu \to 3$. 

In general, the dominant infrared mechanism will contain 
a pole $(3 - \nu)^{-1}$ reflecting an ultraviolet divergence 
that appears in this mechanism when $\nu \to 3 $ from below,  
while the dominant ultraviolet mechanism will contain a 
pole $(\nu - 3)^{-1}$ reflecting an infrared divergence
in this mechanism when $\nu \to 3$ from above. Since the
physical problem can be formulated in arbitrary
dimensionality about $\nu = 3$ with no infinities when 
$\nu = 3$, these two poles must cancel.  Residues of 
the poles bring in logarithms 
of the ratio of the relevant short and long distance 
scales (or high and low energy scales), giving a 
stopping power of the generic form $dE/dx = B\, g^2
\ln C g^2 + {\cal O}(g^3)$. Often, this 
ratio is large, giving a large logarithm. It must be 
emphasized that our method evaluates not only the
coefficient $B$ of such large logarithms, but also the 
constant term $C$ underneath the logarithm. This is so 
because it computes both the leading and first subleading 
terms. Often in a physical problem it is easy to compute 
the large logarithm, but the computation of the constant 
under the logarithm cannot be done or is very difficult 
to do. The new dimensional continuation method makes this 
easy.  Since the method just described is somewhat subtle, 
another simple but fully pedagogical example of how it 
works out is given in Appendix \ref{hank}.

One could object that we do not explicitly prove that
larger subleading terms are not present: one may ask 
if an additional term that has a power dependence between 
$g^{2-(\nu-3)}$ and $g^{2-(3-\nu)}$ can appear.  However, 
simple dimensional analysis shows that such terms of 
intermediate order cannot be formed. The point is that,
in examples such as the one we consider, 
only two physical mechanisms dominate, one at large 
and the other at small scales. Since the two mechanisms
involve different physics, it is natural that different
combinations of the basic physical parameters come into
play, and hence give quite different dependencies on the 
small parameter $g$ when the dimension $\nu$ departs from 
$\nu = 3$. 

We have illustrated the method of dimensional continuation
by a very simple model of the $\nu$-dependence of the coupling 
parameter $g$ given in (\ref{simpleg}). For the case we 
shall examine, however, the dependence on $\nu$ is somewhat 
more complex, although the same dimensional continuation arguments 
apply. For a plasma, we shall see that the leading infrared and 
ultraviolet mechanisms behave as $g^{2 -(3 - \nu)}$ and 
$g^2$, respectively. When the infrared term is analytically
continued from $\nu<3$ to $\nu>3$, it becomes subleading since 
$g^{2+(\nu-3)}< g^2$ for $\nu>3$. Conversely, when we analytically
continue from $\nu > 3$ to $\nu<3$, even though the ultraviolet 
mechanism has no $\nu$ dependence, it nonetheless becomes
subleading since $g^2 < g^{2 -(3 - \nu)}$ for $\nu<3$.
The parameter $g^2$ is a dimensionless 
constant proportional to the overall plasma number density 
$n$. The additional parameters needed to form a dimensionless 
coupling $g^2$ involve the electric charge $e$ and temperature  
$T$. At $\nu = 3$, the coupling is of the generic form $g^2 = 
e^6 \, n / T^3$, in agreement with Eq.~(\ref{dimlesss})
below. As will be seen explicitly in what follows, the
leading hard process for $\nu > 3$ involves scattering and 
is thus proportional to $n$ giving a $g^2$ contribution as 
stated here.  The leading soft process for $\nu < 3$ is 
essentially ${\bf j} \cdot {\bf E}$ heating which, for 
dimensional reasons, is proportional to $\kappa^{(\nu -1)}$, 
where $\kappa$ is the Debye wave number, giving a 
$g^{(\nu -1)}= g^{2-(3-\nu)}$ contribution as stated here.

In our case, and at $\nu =3$ dimensions, the small 
parameters are the plasma coupling parameters\footnote{
\baselineskip 15pt
In this paper we use rationalized 
cgs units so that, in three-dimensional space, the 
Coulomb potential energy has the form  $e^2 / (4\pi r)$. We
do this because then no factor of $4\pi$ appears in
Poisson's equation for the potential, a factor that 
is peculiar to three-dimensional space, and we shall 
need to work in a space with $\nu\ne 3$ dimensions.}
\begin{equation}
  g_{pb} = \beta_b \, {e_p e_b \kappa_b \over 4\pi}
  = { e_p e_b \kappa_b \over 4\pi \, T_b} \,,
\label{dimlesss}
\end{equation}
where $T_b = \beta_b^{-1}$ is the temperature of 
plasma species $b$ measured in energy units\footnote{
\baselineskip 15pt
Although we shall often graph results for a plasma
whose various components are at a common temperature,
for completeness we shall work in a general case
in which each plasma species $b$ is in thermal
equilibrium with itself at temperature $T_b$. 
},
$e_p$ is the charge of the projectile whose 
energy loss is being considered, $e_b$ is the charge 
of a plasma species labeled by $b$, and $\kappa_b$ 
is the Debye wave number of this species, which has 
density $n_b$, so that 
\begin{equation}
  \kappa_b^2 = \beta_b e_b^2 n_b \,.
\end{equation}
The total Debye wave number of the plasma is given 
by
\begin{equation}
  \kappa_\smD^2 = {\sum}_b \, \kappa_b^2 \,.
\end{equation}
The classical dimensionless parameter (\ref{dimlesss}) is the 
ratio of the electrostatic interaction energy of two 
particles of charge $e_p$ and $e_b$ a Debye length apart 
divided by the temperature $T_b$ of the plasma species $b$, 
with the 
temperature being measured in energy units (as we shall 
always do). A parameter $g$ of this form is the  correct
parameter to describe plasma effects order-by-order
because the effects come, up to logarithmic factors,
in\footnote{\label{expparam}
\baselineskip 15pt
The proper plasma expansion parameter has the generic 
form $g = \beta \, {e^2 \over 4\pi} \, \kappa $.  Thus 
it is related to the often used plasma parameter $\Gamma 
= \beta \, {e^2 \over 4\pi} \, \left( { 4\pi n \over 3}
\right)^{1/3} $ by $ \Gamma^3 = g^2/3 $. The correct
integer powers of $g$ which appear in all perturbative
expansions of plasma processes appear as fractional
powers of $\Gamma$.  That plasma perturbation theory 
involves {\em integer} powers of $g$ is discussed in 
detail, for example,  in Brown and Yaffe \cite{BY} 
in footnote 26 and in Section 3 of Appendix F.  
The plasma coupling parameter $g$ also appears 
explicitly in the BBGKY equation chain if times are 
scaled by the inverse plasma frequency and lengths 
are scaled by the Debye length.  See, for example, 
Section 12.5.1 of Clemmow and Dougherty \cite{clem} who
denote $g$ by $1/nh^3$. It is worth noting that the 
inverse of the number of particles in a sphere whose 
radius is the Debye length is given by
$[(4\pi/3) \kappa^{-3} n]^{-1} = 3 g$.} 
{\em integer} powers of $g$.  To make an 
explicit (albeit slightly arbitrary) definition of the
overall coupling of the projectile to the plasma, we define
\begin{equation}
g_p^2 = \sum_b g^2_{pb}  = {\sum}_b \beta_b^2 
\left( { e_p e_b \over 4 \pi } \right)^2 \kappa_b^2 \,.
\end{equation} 

Our calculation
gives the energy loss to the generic order $ g^2 \,
[ \ln g^2 + {\rm const} ] = g^2 \, \ln C \, g^2 $ in
the plasma parameter, including the constant $C$,
and to all orders in the parameters
\begin{equation}
  \bar\eta_{pb} = { e_p e_b \over 4\pi \hbar
  \bar v_{pb} }
\end{equation}
that measure the strength of the interaction of the
projectile of charge $e_p$ with a plasma particle of
charge $e_b$, with a typical or average relative velocity
$\bar v_{pb}$ between the projectile and plasma particle.
The presence of Planck's constant $\hbar$ in the denominator
shows that this is a quantum-mechanical, Coulomb coupling
parameter.  When $\bar v_{pb}$ becomes large, $\bar\eta_{pb}$
becomes small.  This is equivalent to the formal limit of
large $\hbar$.  Hence, when $\bar\eta_{pb}$ is small, 
quantum effects may be important. 
The plasma is taken to be composed of non-relativistic
particles that have no degeneracy so that they are
described by classical, Boltzmann statistics. We show 
in Appendix \ref{equi} that the `Convergent Kinetic 
Theory' method of Refs.~\cite{Frie}, \cite{wein}, 
and \cite{gould}, when evaluated in the leading order 
in which it was derived, produces the same results 
that are produced by our method of dimensional 
continuation.  However, that method generally produces 
spurious, higher-order corrections in the plasma parameter 
that must be discarded. They must 
be discarded because they do not include all of the 
terms of the given order in the plasma coupling parameter.
The inclusion of these terms gives, in general, misleading
results\footnote{
\baselineskip 15pt
A striking example of how the retention 
of only a part of the terms in a given order can give a 
very misleading result is provided by the calculation 
of the energy variation of the strength of the strong 
interaction in elementary particle physics. The easiest 
part of the computation is to obtain the effects of virtual 
quarks, which is akin to traditional calculations in quantum 
electrodynamics. If this is done, one concludes that the 
strength of the interaction {\em increases} with energy.  
However, these quark terms are only a part of the 
leading-order result.  In this same order, the contribution 
of virtual gluons overwhelms that of the quarks and the 
total, complete result shows that the interaction strength
{\em decreases} as the energy increases. An even more
blatant example of the error of principle entailed in 
keeping only some but not all terms of a given order is 
provided by the instruction in an elementary physics lab: 
the sum of 2.1 and 2.123456 is 4.2, {\em not 4.223456}; 
it is inconsistent to retain more decimal places than 
those of the number with least accuracy. In physics,
half a loaf is not better than none.}. Our method has 
the virtue of producing only the leading order terms 
unaccompanied by any other spurious, higher-order terms, 
terms that must be deleted in other methods.  Some authors, 
for example Refs.~\cite{gsk} and \cite{gs}, retain the 
spurious higher-order terms and thus provide inconsistent 
results.

Exactly the same considerations apply to the calculation of the
momentum loss as a projectile traverses a plasma.  This result,
taken together with the energy loss computation, uniquely
determine the coefficients in a Fokker-Planck equation
that describes the general, statistical, motion of particles in
the plasma.  Such coefficients are sometimes described as 
``Rosenbluth potentials'' which were introduced in 
Ref.~\cite{RMJ} and discussed in several places, a good reference
being Ref.~\cite{tru}. Our coefficients, however, contain no
arbitrary parameters or cutoffs and are well-defined with no
ambiguity to the order $g^2 \, \ln C g^2$ to which we work.
We also apply the same methods to calculate the rate of 
equilibration of two plasma components at different temperatures, 
again with
leading and next-to-leading accuracy in the plasma coupling
$g$. We shall postpone the derivation and description of the
Fokker-Planck equation until we have first presented our 
results for the energy loss or stopping power $dE/dx$ for 
several cases of interest. 
 
We have stated that our result is generically of 
order $g^2 [ \ln g^2 + {\rm const} ] $. This gives 
the correct order as far as the plasma density $n$ is 
concerned, namely, discarding other parameters that 
are needed to provide the right dimensions, the result
is of order $ n [ \ln n + {\rm const} ]$.  But we 
should describe the accuracy of the result in this 
paper with more care. It is of the form
\begin{equation}
  {dE \over dx} = {\sum}_b \, e_p^2 \, \kappa_b^2 \,
  \left[{\cal F}_b\left( { v_p \over \bar v_b } \right) \,
  \ln g_{pb} + {\cal G}_b\left( { v_p \over \bar v_b} ;
  \bar\eta_{pb} \right) \right] \,.
\end{equation}
Since $\kappa_b$ has the dimensions of inverse length, 
while $e_p^2\kappa_b$ has the dimensions of energy, 
the prefactor $e_p^2 \kappa_b^2 $ has the proper 
overall dimensions of energy per unit length.  Thus 
the functions ${\cal F}$ and ${\cal G}$ are dimensionless 
functions of dimensionless variables --- $v_p$ is the 
projectile velocity and the velocity
$\bar v_b$ is the average thermal velocity of the plasma
species\footnote{
\baselineskip 15pt
When $\bar\eta_{pb}$ is small, the
function ${\cal G}$ has a term of order $\ln \bar\eta_{pb}$ 
that adds to the $\ln g_{pb}$, giving $\ln( g_{pb} / 
\bar\eta_{pb} ) = \ln(\beta \hbar\kappa_b \bar v_{pb} )$ 
which converts the classical short-distance cutoff into 
a quantum cutoff.  This is in keeping with the remark above
that quantum effects may become important when 
$\bar\eta_{pb}$ becomes small.} $b$.  
The functions ${\cal F}$ and ${\cal
G}$ may also depend upon ratios of all the particle masses
that are present.  Since $e_p^2 \kappa_b^2 = { 4\pi \, 
g^2_{pb}} \,T \, (T /e_b^2) $, the overall factor is 
of the generic form $g^2$. A point to be made is that 
this leading order calculation, which has the formal
overall factor $e_p^2$ must, by simple dimensional
analysis, involve overall factors of the dimensionless 
parameters $g_{pb}^2$, which are of first order in the 
plasma density. Powers of the dimension bearing electric 
charge must appear in either a quantum-mechanical parameter 
$\eta$ or in the classical plasma coupling parameter $g$,
with $g^2$ bringing in a factor of the plasma density. 
Long ago, Barkas {\em et al.} \cite{barkas} found 
differences between the ranges of positive and negative 
pions of the same energy. This implies that there are 
corrections to the energy loss of cubic order in the
projectile charge\footnote{
\baselineskip 15pt
As we shall see, corrections
involving the quantum parameters $\eta_{pb}$ are even 
functions that are unchanged by the reflection $\eta 
\to - \eta$.}, terms of order $e_p^3$.  In a plasma, 
such `Barkas terms' must involve dimensionless parameters
$g_{pb}^3$ and are thus necessarily of order $n^{3/2}$ 
in the plasma density.  These Barkas terms, as well as 
other terms of order $g^3$, are one higher order in the 
plasma coupling $g$ to which we shall work.

The usual method for obtaining the energy loss for a
charged particle moving through matter is to divide 
the calculation into two parts: the long-distance, 
soft collisions and the short-distance, hard collisions. 
Collective effects are important in the long-distance
part, and it is evaluated from the ${\bf j} \cdot 
{\bf E}$ energy loss of a particle moving in a 
dielectric medium.  The hard collisions are described 
by Coulomb scattering. The rub is to join the disparate
pieces together. For the case of classical scattering, 
this is often done by computing the energy loss in 
Coulomb scattering out to some arbitrary long-distance, 
maximum impact parameter $B$, and 
then adding the ${\bf j} \cdot {\bf E}$ energy loss 
integrated over the space outside of a cylinder whose 
radius is this maximum impact parameter. The hard scattering
processes within the cylinder produce a logarithmic factor 
$ \ln(B / b_{\rm min}) $,
where $b_{\rm min}$ is the minimum classical distance of closest
approach in the Coulomb scattering.   
The soft, collective effects outside the cylinder produce a
factor involving the Debye radius $\kappa_\smD^{-1}$, 
$ \ln ( 1 / \kappa_\smD B) $,
which has the same overall outside factor.  Thus the 
arbitrary radius $B$ cancels when the two parts are added. 
Hence such methods must yield the correct coefficient of the 
large logarithm 
$ \ln( 1 / \kappa_\smD b_{\rm min})$,
and they do so without much difficulty of computation. 
However, the purely numerical 
constants that accompany the logarithm (which are
expected to be of order one) are harder to compute.
Here we describe an easily applied method that yields 
both the constants in front of and inside the logarithm,
with no spurious higher order terms being introduced
along the way (as is the case with other methods).
The new idea is 
to compute the energy loss from Coulomb scattering over 
all impact parameters, but for dimensions $\nu > 3$ 
where there are no infrared divergences. A separate 
calculation of the energy loss using a generalization 
of the ${\bf j} \cdot {\bf E}$ heating is done for 
$\nu < 3$, where the volume integration may be extended 
down to the particle's position without encountering 
an ultraviolet divergence. Both of these results have 
a simple pole at $\nu = 3$, but they both may be
analytically continued beyond their initial range 
of validity.  In their respective domains, $\nu > 3$
and $\nu < 3$, 
both calculations are performed to the leading 
order in the plasma density.  As will be seen, although 
the Coulomb scattering result is the leading order 
contribution for $\nu > 3$, it becomes subleading order 
when $\nu < 3$. Conversely, the ${\bf j}\cdot {\bf E}$ 
heating is subleading for $\nu > 3$ but leading for
$\nu < 3$. Hence, the sum of the two (analytically
continued) processes gives the leading and (first) 
subleading terms in the plasma density for all 
dimensions $\nu$, and thus, in the limit $\nu \to 3$, 
the pole terms of this sum must cancel with 
the remainder yielding the correct physical limit 
to leading order in the plasma density.

The highly ionized classical plasma with which we are
concerned is described exactly by a coupled set of kinetic 
equations, the well-known BBGKY hierarchy as described, 
for example, in Section~3.5 of Ref.~\cite{huang}. This 
fundamental theoretical description makes no explicit
reference to the spatial dimensionality, and hence it
is valid for a range of spatial dimensions $\nu$ about 
$\nu = 3$. Thus our 
method of dimensional continuation may be applied to a
plasma. This hierarchy holds for arbitrary plasma
densities.  We are interested, however, in the computation 
to leading order in the plasma density of the energy 
loss of a particle traversing the plasma. 
The leading low-density limit of the BBGKY hierarchy 
changes as the spatial dimensionality $\nu$ changes. 
For $\nu < 3$, the
long-distance, collective effects dominate, and the
equation derived by Lenard and Balescu 
applies\cite{len,bal}. An alternative derivation of 
their result is presented by Dupree\cite{dup}, and a 
clear pedagogical discussion appears in  
Nicholson\cite{nich}. Clemmow and Dougherty\cite{clem}
provide a derivation of the Lenard-Balescu equation 
and prove that it shares the basic features of the 
Boltzmann equation; namely that it conserves particle 
number, total momentum and energy, and that it obeys 
an H-theorem (entropy increases) like the Boltzmann 
equation with the long-time, equilibrium solution 
being a Maxwell-Boltzmann distribution. The equation 
describes the interaction of the various species that 
the plasma may contain.  In the limit in which one 
species is very dilute, as is our case in which we 
examine the motion of a single ``test particle'' or 
``projectile''  
moving through the plasma, the energy lost in the 
particle motion is described by a generalization of 
its ${\bf j} \cdot {\bf E}$ Joule heating with the 
background plasma response given by the permittivity 
of a collisionless plasma.  On the other hand, when 
the spatial dimension $\nu$ is greater than $3$, the
short-distance, hard Coulomb collisions dominate. For 
these dimensions, the leading low density limit of 
the BBGKY hierarchy is described by the familiar
Boltzmann equation.\footnote{
\baselineskip 15pt
In this case, we may go 
beyond the classical BBGKY hierarchy limit in that we 
may use the full quantum-mechanical cross section rather 
than its classical limit in the Boltzmann equation. 
The validity of this extension is, however, obvious 
on physical grounds.} The Boltzmann equation is 
derived, for example, in Section~16 of 
Ref.~\cite{Lifs} and also in Section~3.5 of 
Ref.~\cite{huang}.  We use the Boltzmann equation 
to obtain the leading order energy loss rate when
$\nu > 3$. Since we are concerned with the 
motion of a single ``projectile'', the Boltzmann 
equation reduces to the product of the energy loss 
weighted cross section times the plasma density. The
derivations that we have just described, which start 
from first principles, justify the methods outlined 
in the previous paragraph, the methods that we shall 
use.

In Ref.~\cite{brown1}, the method was illustrated by 
the simplified case in which the charged particle moved 
through a dilute plasma with a speed that is much larger 
than the 
speeds of the thermal electrons in the plasma.  Here we 
shall extend that work to the case in which the charged 
particle projectile moves with arbitrary speeds.  
As we have noted 
above, we work in the dilute limit in which the plasma
density is our small parameter.\footnote{
\baselineskip 15pt
One should work
with a dimensionless parameter so that stating that it is 
small is unambiguous. For our case, the 
dimensionless parameter is the square of the plasma coupling 
parameter, $g^2 = (e^2 \kappa_\smD / 4\pi T )^2 = (e^6 / 
16 \pi^2 T^3 ) \, n $, but to save writing we shall simply 
use the density $n$ as our parameter.} It is obvious that 
the energy loss for $\nu > 3$ as calculated for scattering 
is proportional to the first power of the plasma density, 
$dE^\smgt / dx \sim n$.  On the other hand, as we shall 
see explicitly below, the computation of the energy loss 
for $\nu < 3$ behaves as $ dE^\smlt / dx \sim n^{ 1 - 
(3 - \nu)/2 }$ Thus we see explicitly that the infrared
computation that accounts for the collective effects in 
the plasma and gives the leading term for $\nu < 3$ becomes
non-leading when it is analytically continued to $\nu > 3$.  
Conversely, the ultraviolet, hard scattering computation, 
which is leading for $\nu >3$, becomes subleading when 
analytically continued to $\nu < 3$. We conclude that the 
leading order (in density) energy loss of a projectile 
particle moving in three dimensions is given by
\begin{equation}
  { d E \over dx } = \lim_{\nu \to 3} \, \left\{
  { d E^\smgt \over dx } + { d E^\smlt \over dx } 
  \right\} \,.
\end{equation}

To the order in the coupling $g$ to which we work, a 
plasma species $b$ is described completely by a phase space 
density $f_b({\bf r}, {\bf p}, t)$.  The projectile 
$p$ may also be described by a phase space density 
$f_p({\bf r}, {\bf p}, t)$ that contains delta 
functions restricting the momenta ${\bf p}$ 
and the coordinates ${\bf r}$ to be those of the
projectile's trajectory.  The projectile energy loss 
(or ``stopping power'') that we deal with
is the rate of kinetic energy loss, the time 
derivative of (in $\nu$  dimensions)
\begin{equation}
E_p = \int d^\nu{\bf r} \int {d^\nu{\bf p} \over 
(2\pi\hbar)^\nu} \, { {\bf p}^2 \over 2 m_p } \,
f_p({\bf r}, {\bf p}, t) \,.
\end{equation}
The total kinetic energy of plasma particles of 
species $b$ is given by 
\begin{equation}
E_b = \int d^\nu{\bf r} \int {d^\nu{\bf p} \over 
(2\pi\hbar)^\nu} \, { {\bf p}^2 \over 2 m_b } \,
f_b({\bf r}, {\bf p}, t) \,.
\end{equation}
The Lenard-Balescu and Boltzmann equations that we 
use to derive the projectiles' energy loss obey an
energy conservation law that entails only these
kinetic energies,
\begin{equation}
{d \over dt } \, \left\{ E_p + {\sum}_b E_b \right\}
= 0 \,.
\end{equation}
There are, of course, additional contributions to
the total energy ---  potential energy contributions
that involve collective plasma effects.  The kinetic 
energies are of zeroth order in the plasma coupling 
$g$ and their time derivatives, which come about 
because of the Coulomb forces, are of order
$g^2 \, \ln g^2 $.  On the other hand, the potential 
energies (and possible collective effects) 
are of higher order in the coupling $g$ and their 
time derivatives are of an order that is higher than
that to which we compute. Hence these potential energy
terms do not contribute to the energy balances
accounted for by the Lenard-Balescu and Boltzmann 
equations, and only the kinetic energies are relevant.
We conclude that, to the accuracy to which we compute,
we have an unambiguous partition of the projectile's 
energy loss into energies gained by individual particle
species in the plasma.  We define the energy 
{\em loss} $dE/dx$ of the projectile to be 
{\em positive}, 
\begin{equation}
{d E \over dx }  =  - { 1 \over v_p} {d E_p \over dt} \,,
\end{equation}
and we have
\begin{equation}
{d E \over dx }  = {\sum}_b \, {d E_b \over dx} \,,
\end{equation}
where
\begin{equation}
{d E_b \over dx} = { 1 \over v_p} \, 
{d E_b \over dt} 
\end{equation}
defines the energy loss of the projectile $p$ to the 
plasma particles of species $b$ or, equivalently, the
energy {\em gain} of the plasma particles $b$ brought
about by the projectile $p$ moving through the plasma
with velocity $v_p$. 
Such an unambiguous partition into the energy gained
by the individual species in the plasma does not hold
in higher orders, and so such an accounting cannot be
done for strongly coupled plasmas which are entangled
with collective excitations.  

An important check 
of the validity of our results is that they satisfy the 
condition that the total energy loss vanishes for a 
swarm of projectile particles in thermal equilibrium with 
the plasma through which it moves.  We must have
\begin{equation}
  \left\langle { d E \over dt } \right\rangle =
  \left( { \beta m_p \over 2 \pi } \right)^{3/2} \,
  \int d^3{\bf v}_p\exp\left\{ 
  - { m_p v_p^2 \over 2 T } \right\} \, v_p \, 
  { \, d E \over dx } = 0 \,.
\label{van}
\end{equation}
Thus the energy loss $dE/dx$ must become negative 
for low projectile velocities $v_p$ so that its 
integral over all velocities can vanish.
As we shall see, this constraint is an automatic consequence 
of the method that we employ.  

We turn now to describe the results of the calculation of 
$dE/dx$.  In addition to partitioning the energy loss into 
different particle species that make up the plasma, we shall 
also increase the generality of our results by assuming that,
although each species $b$ is internally in thermal equilibrium,
it may have a private temperature $T_b$ which differs from
species to species.  We shall briefly compare our results in 
the classical limit with PIC simulation data\footnote{In the 
particle-in-cell (PIC) technique as applied to
plasmas, the plasma species are modeled as collections of
quasi-particles, each representing a large number of real
particles, moving through a numerical grid.  The particle
positions and a weighting factor are used to assign electric
charge densities to the nodes of the grid.  Poisson's equation is
then solved for the potentials at the nodes, which gives the
electric fields at the nodes, and finally an inverse weighting
factor is used to determine the electric field at the particle
positions.  The particles are then moved using Newton's
equations to start the next time step.}, illustrate our 
stopping power under solar conditions, and provide a more 
lengthy exposition relevant for inertially confined laser 
fusion experiments. These examples illustrate the calculation 
of $dE/dx$ within our method, and we can directly compare our 
results for $dE/dx$ with those that are typical of the current 
literature. The dimensional continuation method that we have 
developed and applied to the calculation of the stopping power 
can be used to calculate other physical processes. As a final 
example, we also will use this method 
to compute the rate at which Coulomb interactions in a dilute 
plasma bring two species into thermal equilibrium. The mass 
ratios and initial temperatures are arbitrary, and quantum 
corrections are included. Like the stopping power, the 
temperature equilibration rate\footnote{It should be clear from
the context in which it is used, whether the letter $\Gamma$ 
stands for a rate rather than the traditional 
plasma coupling mentioned in footnote \ref{expparam}.}
is of the form $\Gamma = B
g^2 \ln C g^2 + {\cal O}(g^3)$, and we calculate the
prefactor $B$ and the constant $C$ under the logarithm exactly.
After presenting these results, we shall
describe the general formulation that results in a 
Fokker-Planck equation.  The range of validity of the 
Fokker-Planck equation will be assessed, and then we shall
present the derivations of details of our results.

\section{Results}

For $\nu >3$, the BBGKY hierarchy reduces, in 
our dilute plasma limit, to the Boltzmann equation 
with classical Coulomb scattering. In this
higher-dimensional space, only the squared 
momentum transfer weighted Coulomb collision 
cross section enters into the rate of energy 
loss. Although quantum-mechanical corrections 
are not significant in the lower-dimensional 
$\nu < 3$ spatial regions where long-distance 
effects dominate, they may be important in the 
higher-dimensional $\nu>3$ space in which the 
short-distance collisions dominate. And although
our approach is initially based on the classical 
BBGKY hierarchy, it is physically obvious that 
quantum corrections must be incorporated in the 
higher-dimensional region when they become
important. The dimensionless parameter that 
distinguishes whether or not quantum effects 
must be taken into account is (in the $\nu\to3$
limit)
\begin{equation}
  \bar\eta_{pb} = {e_p e_b \over 4 \pi \hbar \,
  \bar v_{pb} } \,,
\label{etabar}
\end{equation}
where $\bar v_{pb}$ is a typical relative 
velocity of the projectile ($p$) of charge 
$e_p$ and a plasma particle ($b$) of charge 
$e_b$. The limit of large $\bar\eta_{pb}$ describes  
slow particles.  This limit is equivalent 
to the formal $\hbar \to 0$ limit, and thus 
the classical calculation applies here. However, 
when $\bar\eta_{pb}$ is not large, quantum effects
must be taken into account for $\nu>3$. We shall 
first examine the classical case and then later 
the quantum corrections to it.

\subsection{Classical Regime}

The classical results to the order in the
plasma coupling to which we compute are summarized
in Sec.~\ref{cclassic}. The complete energy loss
to the plasma species $b$  
in the classical case is given by
\begin{equation}
  { dE^\smC_b \over dx} = { dE^\smC_{b,\smS} \over dx} +
  { dE^\smlt_{b,\smR} \over dx} \,,
\label{doneatlast}
\end{equation}
where the two contributions are contained in
Eq's.~(\ref{wonderclassic}) and (\ref{nun}), 
and with the aid of Eq.~(\ref{aaa}) they can be 
written as:\footnote{
\baselineskip 15pt
To save writing, we use $e$
to denote the absolute value of the charge of 
a particle. Thus $e_p e_b$ is always positive 
even if projectile ($p$) and plasma ($b$) 
particles have charges of opposite sign.}
\begin{eqnarray}
  && {dE^\smC_{b,\smS} \over dx} =
  {e_p^2 \over 4\pi} \, 
	{\kappa^2_b \over m_p \, v_p} \,
  \left( { m_b \over  2\pi \beta_b  } \right)^{1/2} \,
  \int_0^1 du \, u^{1/2} \,
  \exp\left\{ - {1 \over 2} 
  \beta_b m_b v^2_p \, u \right\}
\nonumber\\[2 pt]
  && \Bigg\{
  \left[ - \ln
  \left(\beta_b  { e_p e_b \, K \over 4 \pi} { m_b \over
m_{pb} } \,
  { u \over 1-u} \right)  + 2 - 2\gamma \right]
  \left[  \beta_b \, M_{pb} \, v_p^2
  -  {1\over u} \right] + {2 \over u}  \Bigg\} \,,
\nonumber\\
  &&
\label{wonderclassicc}
\end{eqnarray}
where $\gamma \simeq 0.5772$ is Euler's constant, and
\begin{eqnarray}
  {d E^\smlt_{b,\smR} \over dx}  &=&
  {e_p^2 \over 4 \pi } { i \over 2 \pi }
  \int_{-1}^{+1} d\cos\theta \, \cos\theta \,
	{\rho_b(v_p\cos\theta) \over 
	\rho_{\rm total}(v_p\cos\theta) }
 F(v_p \cos\theta) \ln \left( { F(v_p
  \cos\theta) \over K^2 }\right)
\nonumber\\[2 pt]
  &-&  {e_p^2 \over 4 \pi } { i \over 2 \pi }
  { 1 \over \beta_b m_p v_p^2 } 
	{\rho_b(v_p) \over 
	\rho_{\rm total}(v_p) }
\Bigg[ F(v_p ) \ln
  \left( { F(v_p ) \over K^2 } \right) - F^*(v_p)
  \ln \left( { F^*(v_p) \over K^2 } \right) \Bigg] \,.
\label{nunn}
\end{eqnarray}

Here $K$ is an arbitrary wave number.  As we shall soon show,
the total result (\ref{doneatlast}) does not depend upon
$K$. However, sometimes choosing $K$ to be a suitable multiple of
the Debye wave number of the plasma simplifies the formula.

We use $v_p$ to denote the speed of the projectile 
of charge $e_p$ and mass $m_p$ whose rate of energy 
loss in the plasma we are computing. 
Rationalized units are used for the charge 
so that, for example,  the Coulomb potential energy in 
three dimensions reads $e_p^2 / (4\pi r)$.  
We write the inverse temperature of the plasma
species $b$ as 
$\beta_b = T^{-1}_b$, which we measure 
in energy units.  The charge and mass of the
plasma particle of species $b$ are written as
$e_b$ and $m_b$, with the corresponding Debye
wave number $\kappa_b$ of this species defined
by
\begin{equation}
\kappa_b^2 = \beta_b e_b^2 \, n_b \,,
\end{equation}
where $n_b$ is the number density of species
$b$. The total Debye wave number $\kappa_\smD$ is
defined by the sum over all the species 
\begin{equation}
\kappa_\smD^2 = {\sum}_b \, \kappa_b^2 \,.
\end{equation}
The relative mass of the projectile and 
plasma particles is denoted by $m_{pb}$,
with 
\begin{equation}
{1 \over m_{pb} } = {1 \over m_p }
+ {1 \over m_b } \,,
\end{equation}
while
\begin{equation}
M_{pb} = m_p + m_b 
\end{equation}
is the corresponding total mass.

The function $F(u)$ is related to the 
leading-order plasma dielectric susceptibility.
As in the discussion of Eq.~(\ref{disp}), 
it may be expressed in the dispersion form
\begin{equation}
  F(u)  =  - \int_{-\infty}^{+\infty} dv \, 
{ \rho_{\rm total}(v) \over u
  - v + i \eta } \,,
\label{dispersion}
\end{equation}
where the limit $\eta \to 0^+$ is understood.
The spectral weight is defined by
\begin{equation}
  \rho_{\rm total}(v)  = {\sum}_c \, 
	\rho_c(v) \,,
\end{equation}
where
\begin{equation}
\rho_c(v) = 
\kappa^2_c \, v\,\sqrt{ \beta_c m_c\over 2\pi } 
  \exp\left\{
  -{1 \over 2} \beta_c m_c v^2 \right\} \,.
\end{equation}
It is worthwhile noting here several properties of $F$
that will be needed
throughout for an understanding of the results.  Clearly the spectral
weight is an odd function,
\begin{equation}
\rho_c(-v) = - \rho_c(v) \,.
\end{equation}
Hence the variable change $v \to -v$ in the dispersion relation 
(\ref{dispersion}) gives the reflection property
\begin{equation}
F(-u) = F^*(u) \,.
\label{reflect}
\end{equation}
These properties imply that the total integral in the
lower-dimensional contribution (\ref{nunn}) is real as it must
be.  Since
\begin{equation}
{\rm Im} \, {1 \over x - i\eta} = \pi \, \delta(x) \,,
\end{equation}
the dispersion form (\ref{dispersion}) gives
\begin{eqnarray}
F(u) - F^*(u) &=& 2\pi i \, \rho_{\rm total}(u) 
\nonumber\\
               &=& F(u) - F(-u) \,,
\end{eqnarray}
with the second equality just a repetition of the reflection property  
(\ref{reflect}).  These results show that
\begin{equation}
K^2 \, {\partial \over \partial  K^2} 
          \left\{ { dE^{\smlt}_{b, \smR} \over dx } \right\} =
{e_p^2 \over 4\pi} \,\left\{ \int_0^1 d\cos\theta \, \cos\theta \,
\rho_b\left(v_p \, \cos\theta \right) 
   - {1 \over \beta_b \, m_p \, v_p^2} \rho\left(v_p \right) 
       \right\}\,.
\end{equation}
To show that $dE/dx$ is independent of the wavenumber $K$, 
we take the analogous derivative of Eq.(\ref{wonderclassicc}) 
and make the variable change $ u = \cos^2\theta$.  One term 
involves $ du \, u^{1/2} = 2 \, d  \cos\theta \, \cos^2\theta$,
the other $ du \, u^{1/2} \, u^{-1} = 2 \,d \cos\theta$.
For the integral involving this second term, we insert
$1 = d \cos\theta / d \cos\theta$ in the integrand, 
and integrate the result by parts. In this way we find that the total
result (\ref{doneatlast}) is indeed independent of the arbitrary 
wave number $K$.

\begin{figure}[ht]
 \begin{center}
 \leavevmode
 \epsfxsize=110mm
 \epsfbox{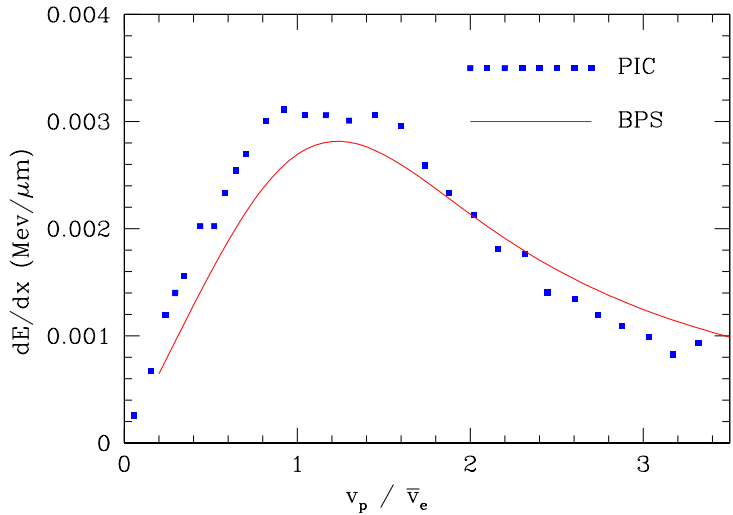} 
 \end{center}
 \vskip-1cm 
\caption{
\tenrm
Comparison of our classical result $dE^\smC/dx$ (curve) 
versus projectile velocity to PIC simulation data of Zwicknagel 
{\em et al.} cited in Ref.~\protect\cite{gs} (squares). The 
projectile is a very massive ion of charge $Z = 5$ moving 
through an electron plasma with $n= 1.1\times 10^{20}{\rm 
cm}^{-3}$ at $T = 14 \, {\rm eV} = 1.6 \times 10^5\,{\rm K}$,
giving a plasma coupling $g_p = 0.21$. The thermal 
velocity of the electron $\bar v_e=\sqrt{3T/m_e}=2.6 \times 
10^8\,{\rm cm/s}$ has been used to set the velocity scale.
\label{fig:dEdxvp.zwick1}
}
\end {figure}

\begin {figure}[h]
  \begin{center}
  \epsfxsize=110mm 
  \epsfbox{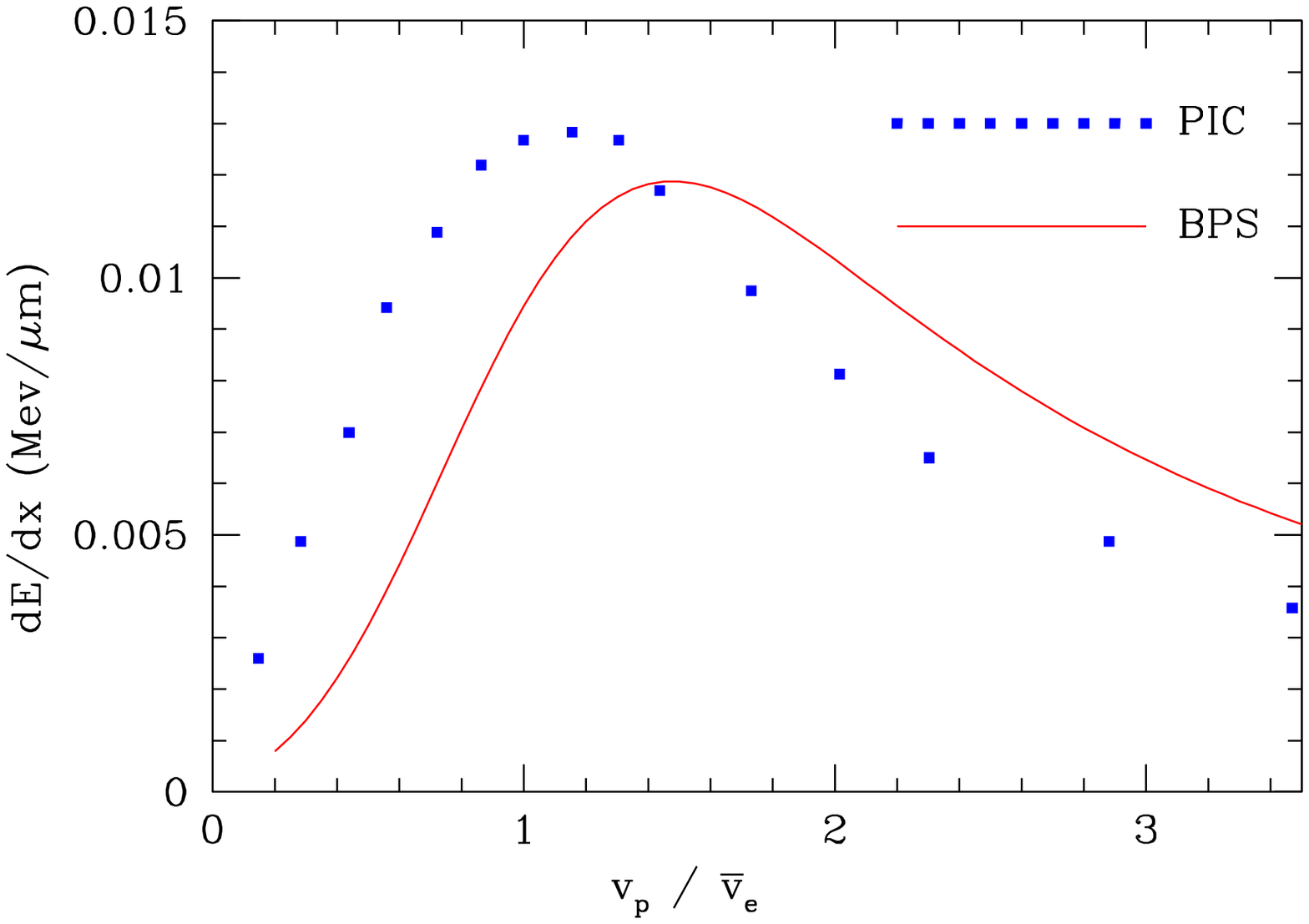} 
  \end{center}
\vskip-1cm
\protect\caption
{
\tenrm
Comparison of our classical result $dE^\smC/dx$ (curve)
versus projectile velocity to PIC simulation data of Zwicknagel 
{\em et al.} cited in Ref.~\protect\cite{gs} (squares) for a very 
massive ion of charge $Z = 10$  moving through an electron plasma 
with $n = 1.4\times 10^{20} {\rm cm}^{-3}$ at $T = 11 \, {\rm eV} 
= 1.3 \times 10^5$ K, giving a plasma coupling $g_p = 0.61$.  
The thermal  velocity of the electron is $\bar v_e=2.4 \times 
10^8\,{\rm cm/s}$. 
}
\label{fig:dEdxvp.zwick2}
\end {figure}

The classical result applies for a low velocity 
projectile moving in a relatively cool plasma.  
In addition to such physical applications, our 
results, which are rigorous and model independent
to the order $ g^2 \ln C g^2 $ to which we compute, 
may be used to check the validity of computer 
calculations such as those utilizing classical 
molecular dynamics, as illustrated in 
Figs.~\ref{fig:dEdxvp.zwick1}~and~\ref{fig:dEdxvp.zwick2}.
Such calculations must 
agree with our results in those regions where 
the plasma coupling $g$ is not large.
The following figures display our analytic results 
for the energy loss (\ref{doneatlast}) using the 
classical expressions (\ref{wonderclassicc}) and 
(\ref{nunn}). They are compared with 
the molecular dynamics calculations of Zwicknagel, 
Toeppfer, and Reinhard\cite{zwick}, as cited in 
Ref.~\cite{gs}. See also the review of Zwicknagel, 
Toeppfer, and Reinhard\cite{zwicktwo}.

\subsection{Quantum Corrections}

Thus far, our discussion has been for those 
cases in which classical physics applies.  
In these cases, the quantum parameters defined 
in (\ref{etabar}),
\begin{equation}
  \bar\eta_{pb} = {e_p e_b \over 4 \pi \hbar \,
  \bar v_{pb} } \,,
\end{equation}
are large.  In the energy loss problem, these 
are the only independent dimensionless 
parameters that 
entail the quantum unit, Planck's constant 
$\hbar$. The parameters are large when the
average relative velocity $\bar v_{pb}$ is 
small which, as far as an $\bar\eta_{pb}$ 
parameter is concerned, corresponds to the
formal limit $\hbar \to 0$. In this section, 
we treat the general case where the size of 
the quantum parameters $\bar\eta_{pb}$ has
no restriction.

According to our dimensional continuation  
method and, in particular, the discussion 
leading to (\ref{regb}), the general case 
is obtained by adding a correction to the 
classical result (\ref{classicall}). Namely, 
the energy loss to the plasma species $b$ 
in the general case appears as
\begin{equation}
  { dE_b \over dx} = { dE^\smC_b \over dx} +
  { dE^\smQ_b \over dx} \,,
\end{equation}
where, we recall, the classical contribution
$dE^\smC_b /dx $ is described by 
Eq.'s~(\ref{doneatlast}) -- (\ref{nunn}),
while Eq's.~(\ref{regb}) and (\ref{aaa}) give
\begin{eqnarray}
  { d E^\smQ_b \over dx}  &=&
  { e_p^2 \over 4 \pi } \,
  { \kappa^2_b \over 2 \beta_b m_p v_p^2 }
  \left( { \beta_b m_b \over 2\pi } \right)^{1/2}
  \int_0^\infty dv_{pb}
\nonumber\\
  && \qquad
  \Bigg\{ \left[ 1 + {M_{pb} \over m_b} { v_p \over
  v_{pb} } \left( { 1 \over \beta_b m_b v_p v_{pb} } - 1
  \right) \right] \exp\left\{ - {1 
  \over 2} \beta_b m_b \left( v_p - v_{pb} \right)^2 
  \right\}
\nonumber\\
  && \qquad
  - \left[ 1 + {M_{pb} \over m_b} { v_p \over v_{pb} }
  \left( { 1 \over \beta_b m_b v_p v_{pb} } + 1 \right)
  \right]\exp\left\{- {1 \over 2} 
  \beta_b m_b \left( v_p +
  v_{pb} \right)^2\right\} \Bigg\}
\nonumber\\
  && \qquad\qquad
  \left\{ 2\, {\rm Re} \, \psi \left( 1 + i \eta_{pb}
  \right) - \ln \eta^2_{pb}  \right\} \,.
\label{quantumm}
\end{eqnarray}
Here 
\begin{equation}
v_{pb} = | {\bf v}_p - {\bf v}_b| \,,
\end{equation}
\begin{equation}
\eta_{pb} = { e_p e_b \over 4 \pi \hbar v_{pb} } \,,
\end{equation}
and $\psi(z)$ is the logarithmic derivative 
of the gamma function. As explained in the 
derivation of (\ref{combd}), we may write
\begin{equation}
  {\rm Re} \, \psi(1+i\eta)  =
  \sum_{k=1}^\infty { 1 \over k} { \eta^2 \over k^2 +
  \eta^2 }  -\gamma \,.
\end{equation}

To illustrate our general results, we now 
present several plots of the total energy loss 
$dE/dx$ as a function of the incident projectile 
velocity $v_p$.  In order to have some comparison 
with previous work, we have chosen to plot 
the results of Li and Petrasso \cite{li} together 
with our results.  

Li and Petrasso \cite{li} have evaluated the 
energy loss $dE/dx$ to leading order in the 
plasma coupling, the order $g^2 \, \ln  g^2$ 
contribution, and they have added an 
estimation of the constant under 
the logarithm, the term of order $g^2 \, \ln C$ 
that we 
compute exactly. They do this by working with 
a Fokker-Planck approximation to the Boltzmann
equation. They then define a Coulomb logarithm 
by using a somewhat arbitrary minimal classical 
impact parameter that is then corrected in a
rather ad-hoc fashion to take account of 
quantum-mechanical corrections. A term
involving a step function is added 
to the formula to correct 
for long-distance collective effects.  
Using our notation, the final result of Li 
and Petrasso \cite{li} appears as
\begin{equation}
  {d E_\smLP \over dx} = { e_p^2 \over 4 \pi }
  { 1 \over v_p^2 } {\sum}_b \,{\kappa_b^2 \over
  \beta_b m_b } \left[G\left( {1\over2} \beta_b m_b v_p^2
  \right) \, \ln \Lambda_b + H \left({1\over2}
  \beta_b m_b v_p^2 \right) \right] \,.
\label{lii}
\end{equation}
Here
\begin{equation}
  G(y) = \left[ 1 - { m_b \over m_p }\,
  { d \over dy} \right] \mu(y) \,,
\label{G}
\end{equation}
where
\begin{equation}
  \mu(y) = { 2 \over \sqrt \pi } \int_0^y dz \,
  z^{1/2} \, e^{-z} \,,
\end{equation}
and
\begin{equation}
  H(y) = { m_b \over m_p } \left[ 1 + { d \over dy}
  \right] \mu(y) + \theta(y-1) \, \ln \left( 2
  e^{-\gamma} y^{1/2} \right) \,,
\label{H}
\end{equation}
with $\theta(x)$ the unit step function: 
$\theta(x) = 0$ for $x<0$ and $\theta(x) = 1$ 
for $x>0$. Li and Petrasso \cite{li} define a 
Coulomb logarithm in terms of the combination 
of classical and quantum cutoffs as described 
above, namely
\begin{equation}
  \ln \Lambda_b = - {1\over2} \ln \kappa_\smD^2 \,
  B_b^2 \,,
\end{equation}
where
\begin{equation}
  B_b^2 = \left( { \hbar \over 2 m_{pb} u_b }
  \right)^2 + \left( { e_p e_b \over 4\pi m_{pb} u_b^2 }
  \right)^2 \,,
\end{equation}
in which $m_{pb}$ is the reduced mass of the 
projectile ($p$) -- plasma particle ($b$) system, 
and
\begin{equation}
  u_b^2 = v_p^2 + { 2 \over \beta_b m_b }
\end{equation}
defines an average of the squared projectile 
and thermal velocities.

\begin{figure}[h] 
  \begin{center}
  \epsfxsize=110mm
  \epsfbox{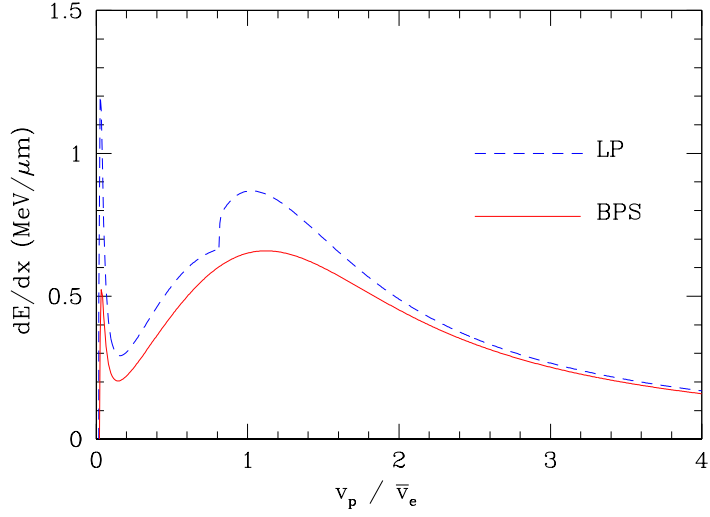} 
  \end{center}
\vskip-1cm 
\caption{
\tenrm
Proton projectile in a fully ionized hydrogen plasma: 
$dE/dx$ (in MeV/$\mu$m) {\em vs}.~$v_p/\bar v_e$. The solid 
line is the result from this work (BPS), and the dashed 
line is the result of Li and Petrasso (LP). Here the 
temperature $ T = 1$ keV and density 
$n_e=5 \times 10^{25}\,{\rm cm}^{-3}$ are chosen
to correspond approximately to values at the core
of the sun. The plasma coupling is $g_p=0.061$ and 
the thermal speed of the electron is $\bar v_e = 
\sqrt{3T/m_e} =
2.30 \times 10^9$ cm/s. The electron fugacity $z_e 
= \exp\{\beta\mu_e\}$ has the value $ z_e = 0.25 $, 
and so the relative Fermi-Dirac statistics corrections
are $z_e/2^{3/2}\sim 9\%$ for both LP and BPS. 
}
\label{fig:LPvsBPS.dEdxvp.pH010525} 
\end{figure}

\begin{figure}[h]
\centerline{
\epsfxsize=110mm
\epsfbox{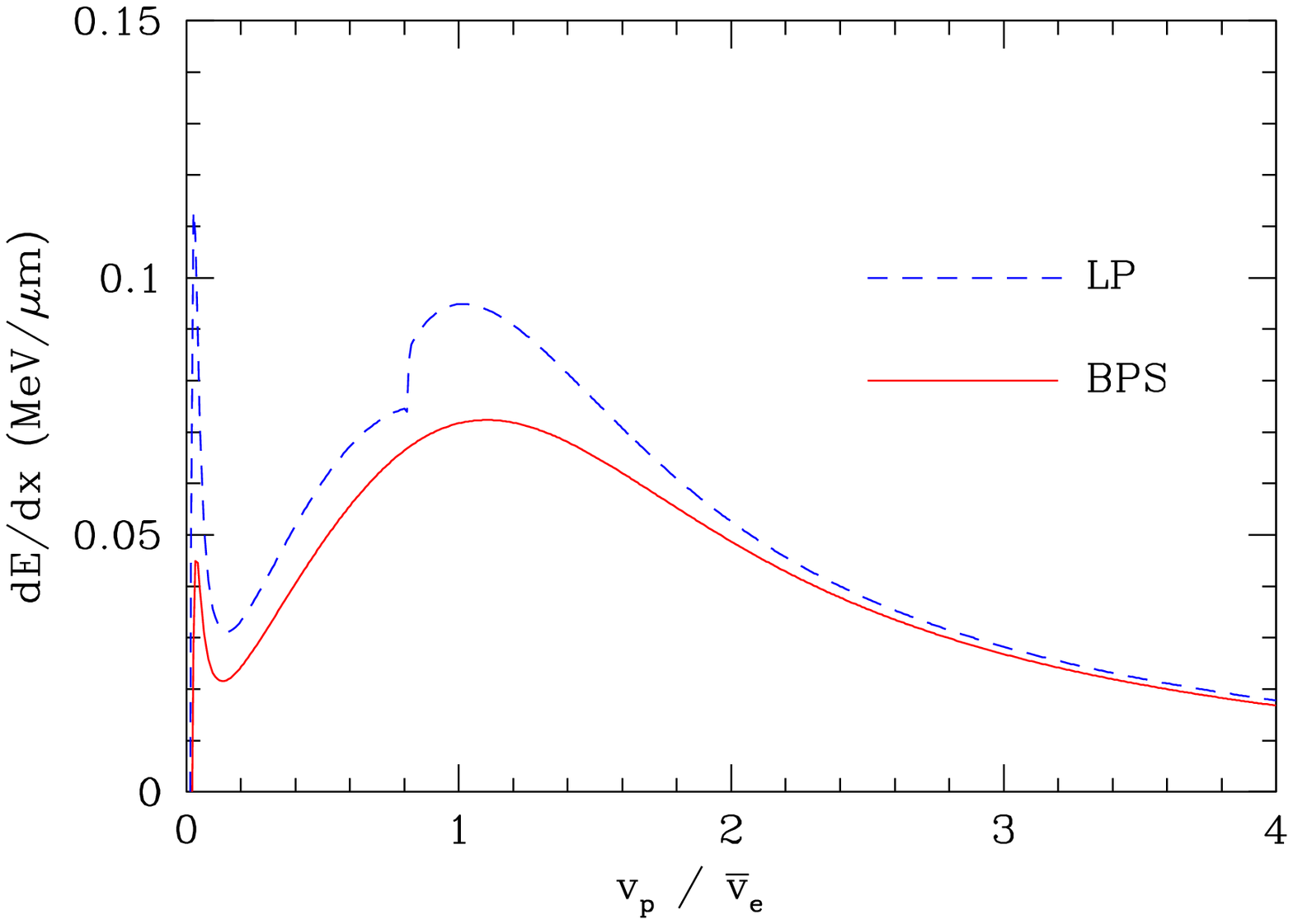} 
}
\vskip-0.3cm 
\caption{
\tenrm
Proton projectile in a fully ionized hydrogen plasma: 
$dE/dx$ (in MeV/$\mu{\rm m}$) {\em vs}.~$v_p/\bar v_e$ 
with $T=0.2\,{\rm keV}$, $n_e=10^{24}\,{\rm cm}^{-3}$, 
giving a plasma coupling $g_p=0.097$, and the thermal
speed of the electron $\bar v_e =1.03\times 10^9\, 
{\rm cm/s}$. The solid line is the result from this 
work (BPS), while the dashed line is the result of Li 
and Petrasso (LP). 
}
\label{fig:LPvsBPS.dEdxvp.pH002124}

\vskip1cm 

\centerline{
\epsfxsize=110mm
\epsfbox{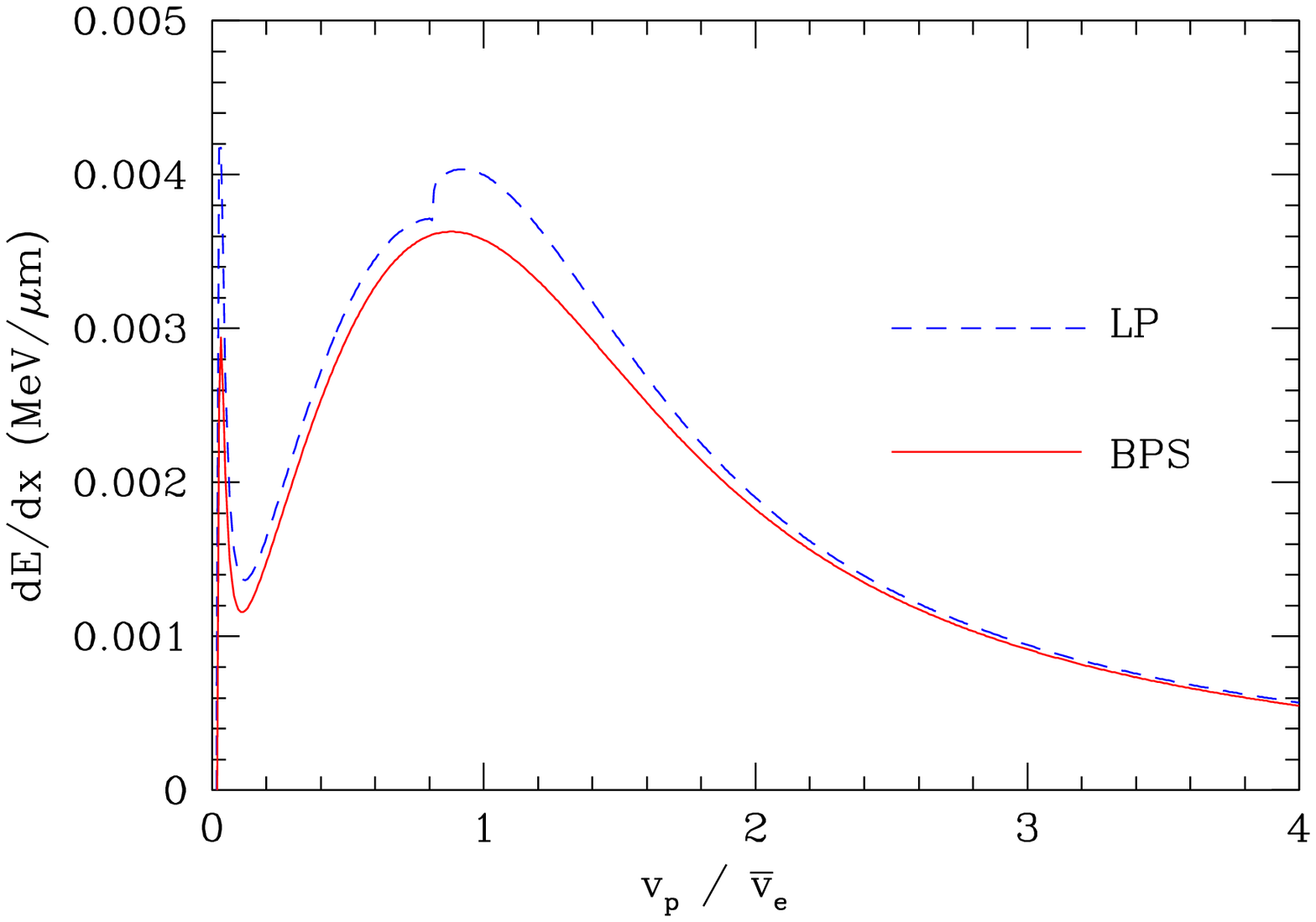} 
}
\caption{
\tenrm
Same as in Fig.~\ref{fig:LPvsBPS.dEdxvp.pH002124} except 
that the temperature is raised to $T=10\, {\rm keV}$, giving
$g_p=0.00027$ and $\bar v_e =7.26\times 10^9\,{\rm cm/s}$. 
There is less of a discrepancy between the results of Li and 
Petrasso~(LP) and our work (BPS) at higher temperatures.
}
\label{fig:LPvsBPS.dEdxvp.pH100124} 
\vskip0.2cm 
\end{figure}

Figures~\ref{fig:LPvsBPS.dEdxvp.pH010525}--\ref{fig:LPvsBPS.dEdxvp.pH100124} 
compare the results of  Li and Petrasso \cite{li} with our results 
for a proton projectile in a
fully ionized hydrogen plasma, a neutral plasma 
of electrons and protons, with the electrons
and protons at a common temperature
$T_e = T_p = T$.  These comparisons are 
made over plasma temperatures and densities 
where the plasma coupling parameter $g$ is 
reasonably small so that our approximation 
is essentially exact. 


\noindent 
The figures show fair agreement between our 
results and those of Li and Petrasso in regions
where the projectile velocities are large, but
significant discrepancies appear for small 
and intermediate projectile velocities. 
These discrepancies are
related to the fact that the rate of energy 
loss for Li and Petrasso (LP) does not keep a 
swarm of projectiles in thermal equilibrium with 
the plasma particles even though the swarm has 
the same initial temperature as that in the plasma, 
whereas we will show in Eq.~(\ref{dedtzero}) that our 
(BPS) energy loss expression does maintain thermal 
equilibrium. This is to say, the thermal average
for Li and Petrasso, given by 
\begin{equation}
\label{dedtbpsnonzero}
  \left\langle { d E_\smLP \over dt } \right\rangle_\smT
  = \left( { \beta m_p \over 2 \pi } \right)^{3/2} \,
  \int d^3 {\bf v}_p \,  e^{- {1\over2} \beta m_p
  v_p^2 }\, v_p \, { d E_\smLP \over dx } \ ,
\end{equation}
does not vanish, while the thermal average of our
energy loss does vanish
\begin{equation}
  \left\langle { d E_\smBPS \over dt } \right\rangle_\smT
  = 0 \ ,
\end{equation}
as we will show in Eq.~(\ref{dedtzero}) below.

It is easy to prove that (\ref{dedtbpsnonzero}) does not
vanish.  We write
\begin{equation}
  \left\langle { d E_\smLP \over dt } 
  \right\rangle_\smT = \left\langle {d E_\smLP 
  \over dt } \right\rangle_\smG + \left\langle 
  {d E_\smLP \over dt } \right\rangle_\smH \,,
\end{equation}
corresponding to the separate contributions 
involving the functions $G(y)$ and $H(y)$ in 
formula (\ref{lii}) of Li and Petrasso. Clearly
\begin{equation}
  \left[ 1 + {d \over dy} \right] \mu(y) \ge 0 \,.
\end{equation}
Since $2 e^{-\gamma} = 1.1229 \cdots$, the 
additional logarithmic contribution in (\ref{H}) 
(with $y \ge 1$ because of the step function 
factor) is also non-negative.  Hence $H(y) \ge 
0$, and we conclude that
\begin{equation}
  \left\langle { d E_\smLP \over dt }
  \right\rangle_\smH > 0 \,.
\end{equation}
The form (\ref{G}) of $G(y)$ gives
\begin{equation}
  \exp\left\{ - {1\over2} \beta m_p v_p^2 \right\}
  G\left( {1\over2} \beta m_b v_p^2 \right)
  = - { 1 \over \beta m_p v_p } { d \over d v_p }
  \left[\exp\left\{- {1\over2} \beta m_p v_p^2
  \right\}\, \mu\left( {1\over2} \beta m_b v_p^2
  \right) \right] \,.
\end{equation}
Thus evaluating the velocity integral in 
spherical coordinates and integrating 
by parts yields
\begin{equation}
  \left\langle { d E_\smLP \over dt } \right\rangle_\smG =
  2 {e_p^2 \over 4\pi } \hskip-0.075cm 
  \left( {\beta m_p \over 2\pi }
  \right)^{1/2} \hskip-0.1cm 
  {\sum}_b \, { \kappa^2_b \over \beta m_b }
  \int_0^\infty \hskip-0.05cm dv_p  \exp\left\{ - {1\over2} \beta m_p
  v_p^2 \right\} \, \mu\left( {1\over2} \beta m_b v_p^2
  \right) { d \over dv_p } \ln \Lambda_b \,.
\end{equation}
Here
\begin{eqnarray}
  { d \over dv_p} \ln \Lambda_b &=& - { 1 \over 2
  B^2_b } \, { d B^2_b \over dv_p }
\nonumber\\
  &=& + { 1 \over B^2_b } \, \left[ \left( { \hbar
  \over 2 m_{pb} u_b } \right)^2 + 2 \left( { e_p e_b
  \over m_{pb} u_b^2 } \right)^2\right] \, { v_p \over
  u_b^2 }
\nonumber\\
  &\ge& 0 \,,
\end{eqnarray}
and hence we have\footnote{
\baselineskip 15pt
Since $\mu(y)$ vanishes 
when $y \to 0 $ and is finite when $y \to \infty$, 
there are no end-point contributions in the partial 
integration and the resulting integral is well 
defined.}
\begin{equation}
  \left\langle { d E_\smLP \over dt }
  \right\rangle_\smG  > 0 \,.
\end{equation}
Thus the formula of Li and Petrasso violates the 
thermal equilibrium condition, a condition 
that our results always obey.

\begin{figure}[h]
  \begin{center}
  \epsfxsize=110mm
  \epsfbox{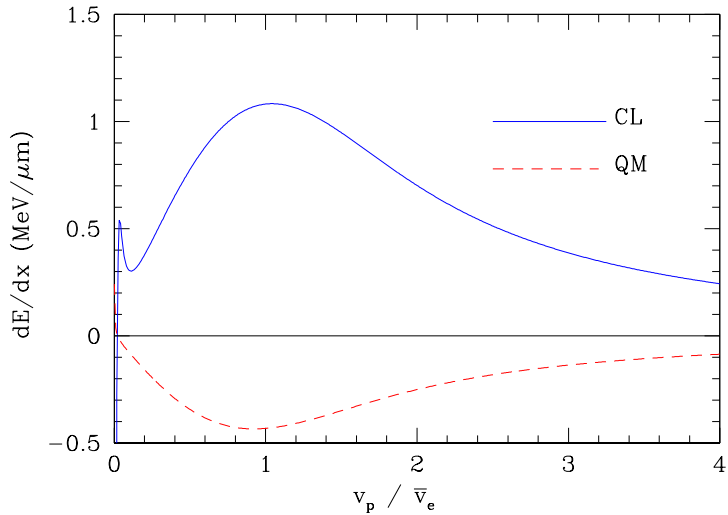} 
  \end{center}
\vskip-1cm 
\caption{
\tenrm
Proton projectile moving in a fully ionized hydrogen plasma.
The classical and quantum contributions to the BPS stopping 
power $dE/dx$ of this work (in MeV/$\mu{\rm m}$) are shown
{\em vs}.~$v_p/\bar v_e$. As in Fig.~\ref{fig:LPvsBPS.dEdxvp.pH010525},
$T=1\,{\rm keV}$ and $n_e=5 \times 10^{25}\, {\rm 
cm}^{-3}$, giving $g_p=0.061$ and 
$\bar v_e =2.30 \times 10^9\,{\rm cm/s}$. The solid line 
is the purely classical result (\ref{doneatlast}) summed 
over species, $dE^\smC /dx$, while the dashed line is the 
quantum correction (\ref{quantumm}) summed over species, 
$dE^\smQ/dx$.
}
\label{fig:LPvsBPS.dEdxvpCLQM.pH010525}
\end{figure}

\subsection{Quantum vs.~Classical Contributions}

Figure~\ref{fig:LPvsBPS.dEdxvpCLQM.pH010525} illustrates 
the size of the quantum corrections given in Eq.~(\ref{quantumm}) 
relative to the classical formula (\ref{doneatlast}). For 
the parameters listed in its
caption, the quantum correction is about 
40\% of the classical contribution for
$v_p / \bar v_e$ greater than 2.  For lesser 
values of the projectile velocity, the 
relative importance of the quantum correction
decreases.

\subsection{Quantum Limit}

We started our discussion of the stopping power 
by examining the low velocity limit.  This
limit is contained in the classical result 
(\ref{doneatlast}), which adds the regular, 
long-distance contribution (\ref{nunn}) to
the well-behaved sum of singular short- and 
long-distance contributions (\ref{wonderclassicc}).
The long-distance, collective effects are always
described by the classical dielectric properties 
of the plasma.  The short-distance effects 
presented in Eq.~(\ref{wonderclassicc}) involve
a classical description of the Coulomb scattering
which gives a classical minimum approach distance 
$b_0 = (e_p e_b / 4 \pi) ( 1 / m_{pb} v_{pb}^2)$
that provides the short-distance cutoff.  This 
classical description of the scattering is valid
in those situations where the projectile velocity
$v_p$ is sufficiently low.  The momentum transfer 
defines a quantum wave length of
order $\hbar / (m_{pb} v_{pb})$. When this length
is on the order of, or larger than, the classical 
minimum approach distance $b_0$, then the detailed
quantum-mechanical treatment that we have given is 
needed.  When the quantum wave length is much
greater than the classical minimum approach distance
$b_0$, then the full quantum description of the 
short-distance scattering simplifies, with the 
first Born approximation sufficing.  This happens
when 
$ 
\hbar / ( m_{pb} v_{pb} ) \gg
  (e_p e_b / 4 \pi) ( 1 / m_{pb} v_{pb}^2) 
$,
or when 
$
\eta_{pb} = (e_p e_b / 4 \pi \hbar v_{pb}) \ll 1
$.
The stopping power in this extreme quantum limit 
is given by 
\begin{eqnarray}
 && \eta \ll 1:
\nonumber \\[5pt]
  && {d E_b \over dx} = {d E_{b,\smQ}
  \over dx} + {d E_{b,\smR}^\smlt \over dx} \,.
\end{eqnarray}
The regular, long-distance contribution
$ d E_{b,\smR}^\smlt / dx $ 
is the classical result (\ref{nunn}) that
is not changed.  The  
$ d E_{b,\smQ} / dx $ contribution 
is obtained by adding the large velocity limit
(\ref{wonderquantum}) of the difference between 
the total and the classical hard scattering
contributions ({\it i.~e.~}the purely quantum correction) 
to the classical result (\ref{wonderclassic}).  This gives
\begin{eqnarray}
 && \eta \ll 1:
\nonumber \\[5pt]
  && {dE_{b,\smQ} \over dx} =
  {e_p^2 \over 4\pi} \, { \kappa_b^2\over m_p \, v_p}  \,
  \left( { m_b \over  2\pi \beta_b  } \right)^{1/2} \,
  \int_0^1 du \,
  \exp\left\{ - {1 \over 2} \beta_b m_b v^2_p \, u \right\}
\nonumber\\
  && \Bigg\{
  \left[ -{1\over2}  \ln
  \left(\beta_b   \hbar^2 K^2  { m_b \over 2 m_{pb}^2 } \,
  { u \over 1-u} \right)  + 1 - {1 \over 2} \gamma \right]
  \bigg[  \beta_b \, \left( m_p + m_b  \right) \, u^{1/2}
  v_p^2 -  u^{-1/2} \bigg] + u^{-1/2}  \Bigg\} \,.
\nonumber\\
&&
\label{highh}
\end{eqnarray}

This result holds when 
$ (e_p e_b / 4 \pi) / ( \hbar v_p) $
is much less than one, but there is no restriction
on the comparison of the energy of the projectile 
with the temperature.  When 
$ m_b v_p^2 \gg T_b $
for all the plasma species $b$, the results simplify
considerably.  The limit is characterized as the 
formal
$v_p \to \infty$
limit.  In this case, the contribution of the 
electrons in the plasma dominate and the limit
(\ref{semicirlim}) found below gives
\begin{eqnarray}
  v_p \to \infty \,  &:&
\nonumber\\
  {d E^\smlt_\smR \over dx} &=&
	  {d E^\smlt_{e,\smR} \over dx} 
 = {e_p^2 \over 4 \pi }
   \, { \kappa^2_e \over 2\beta_e m_e v_p^2 } \,
  \, \ln \left( {  K^2 \beta_e m_e v_p^2 
\over \kappa^2_e} \right) \,.
\end{eqnarray}
Adding this to the corresponding limit of 
Eq.~(\ref{highh}) yields\footnote{
The factor $\beta_b m_b v_p^2$
in the exponent of Eq.~(\ref{highh}) restricts the 
contribution of the
$u$-integration to values less than or of the order of
$(\beta_b m_b v_p^2)^{-1}$ which, for large $v_p$,
is much less than one. 
Hence the factor $1-u$ in the logarithm may
be replaced by unity and, moreover,   
the contribution of
the small-particle-mass electronic component 
of the plasma dominates.  Since the exponent damps out
the contributions of large $u$, the upper limit of the
integral may be extended from $u=1$ to $u \to \infty$. 
Writing the last term in the second square brackets in 
Eq.~(\ref{highh}) as 
$
u^{-1/2}= 2 (d/du) \, u^{1/2}
$
and integrating by parts produces no end-point contributions
and cancels many terms save for one involving 
$\beta_e m_p v_p^2$. Finally, changing variables to 
$ z = \beta_e m_e v_p^2 u /2 $
with the aid of the integrals
$$
\int_0^\infty dz \, z^{1/2} \, e^{-z} = \Gamma(3/2)
= \sqrt{\pi} /2 \,,
$$
and   
$$
\int_0^\infty dz \, z^{1/2} \,\ln z \, e^{-z} = \Gamma(3/2) 
\, \psi(3/2) = \Gamma(3/2) \, [ 2 - \ln 4 - \gamma] \,,
$$
gives the limit that leads to the result (\ref{fastt}).}
\begin{eqnarray}
  v_p \to \infty \,  &:&
\nonumber\\
  {d E \over dx} &=&
	  {d E_e \over dx} 
 = {e_p^2 \over 4 \pi }
   \, { \kappa^2_e \over 2\beta_e m_e v_p^2 } \,
  \, \ln \left( { 4 \beta_e m_e m_{pe}^2 v_p^4 
\over \hbar^2 \kappa^2_e} \right) \,.
\label{fastt}
\end{eqnarray}
This formula is simplified and its nature clarified
if we introduce the electron plasma frequency $\omega_e$
defined by
\begin{equation}
\omega_e^2 = {e^2 \,n_e \over m_e } 
	   = {\kappa_e^2 \over \beta_e m_e } \,,
\end{equation}
for we now have
\begin{eqnarray}
  v_p \to \infty \,  &:&
\nonumber\\
  {d E \over dx} &=&
	  {d E_e \over dx} 
 = {e_p^2 \over 4 \pi }
   \, \frac{\omega^2_e}{v_p^2 } \,
  \, \ln\hskip-0.05cm\left( { 2 m_{pe} v_p^2 
\over \hbar \omega_e} \right) \,.
\label{faster}
\end{eqnarray}
This well-known high-velocity limit is valid when the projectile
velocity is much larger than the thermal velocity of the
electrons in the plasma, $v_p \gg \bar v_e$ and, in addition,
when the projectile velocity is sufficiently large that the
quantum Coulomb parameter is small, \hbox{$\eta_p = (e_p e_e/ 
4\pi \hbar v_p) \ll 1$}.  Not that this
high-velocity result is independent of the temperature of the
plasma.

\subsection{Results Relevant for Laser Fusion}

We turn now to examine cases that are relevant to the 
deuterium-tritium (DT) 
plasmas in laser fusion capsules. In an inertial confinement 
fusion (ICF) capsule filled with DT gas, an $\alpha$ particle 
of energy $E_0=3.54\,{\rm MeV}$ is created at threshold 
in the reaction $D+T \to \alpha + n$. This $\alpha$ particle 
slows down and eventually deposits its energy into the plasma, 
if the range is short enough, or exits the ICF capsule entirely, 
if the range is too long. The more energy deposited into the 
plasma by the $\alpha$ particle then the hotter the plasma 
becomes, and this in turn increases the rate of DT fusion. 
Obviously then, the precise value of the $\alpha$ particle
range can have a dramatic impact on ICF performance.

Again we shall assume that all the electrons and  
ions are at a common temperature $T$.  As we shall
see, our results can differ by 20\% or so from those
of Li and Petrasso \cite{li}, results that have been 
used in the description of such laser fusion 
experiments\footnote{N. M. Hoffman and C. L. Lee 
\cite{nels} have used the stopping power computations of 
Li and Petrasso to model the implosion of laser driven
fusion capsules}. 
We first plot in Fig.~{\ref{fig:LPvsBPS.dEdxE.aDT030125}
the total energy loss to all species $dE/dx$ for an 
alpha particle moving through a DT plasma with a 
temperature $T=3$~keV and electron density \hbox{$n_e = 
10^{25}\,{\rm cm}^{-3}$}. We shall also exhibit in
Fig.~{\ref{fig:LPvsBPS.dEdxEei.aDT030125} the separate 
energy losses to the electrons and to the ions that are 
composed of equal numbers of deuterons and tritons. Again 
our results are compared to those of Li and Petrasso.   

\begin{figure}
  \begin{center}
  \epsfxsize=110mm
  \epsfbox{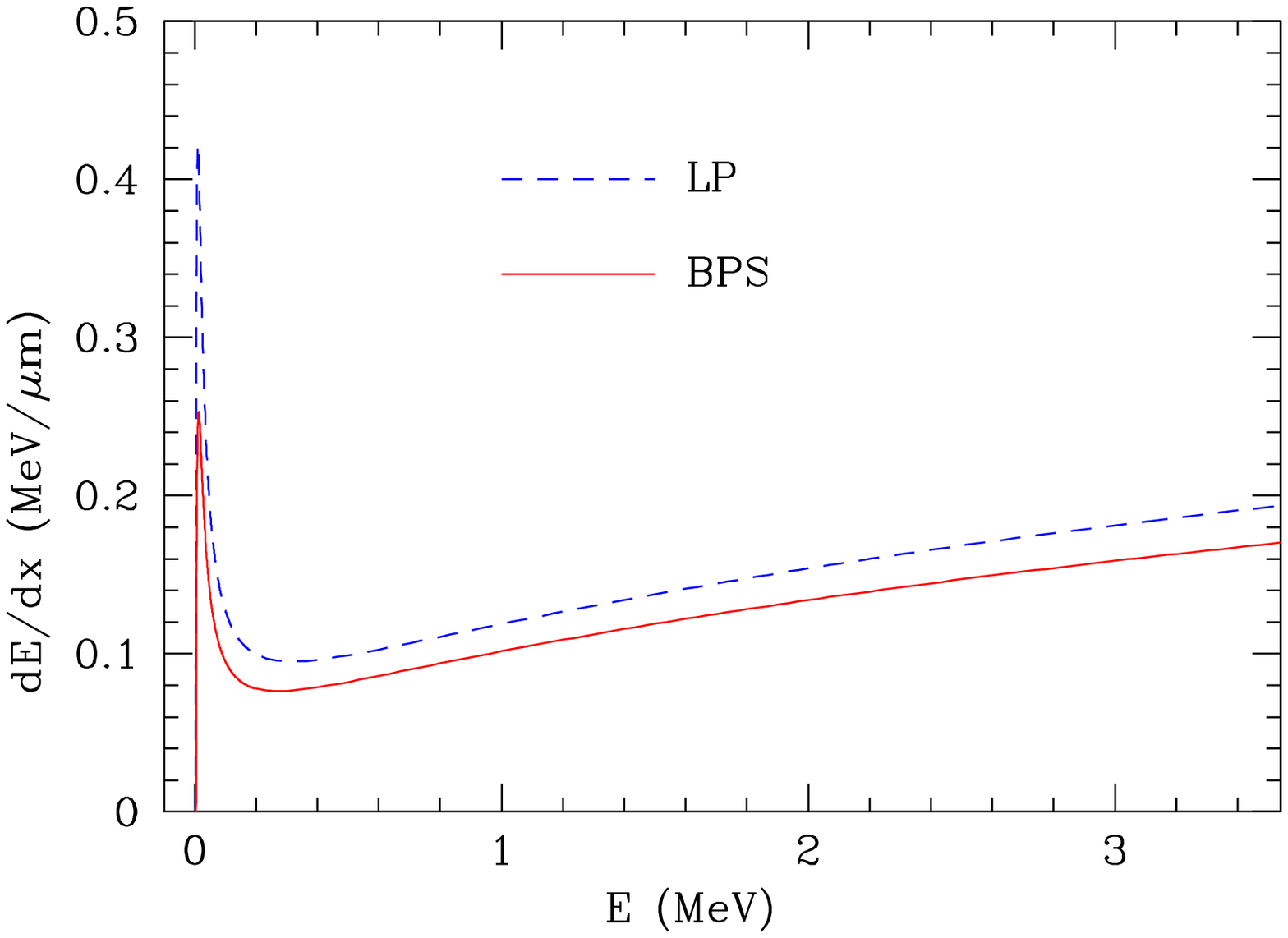} 
  \end{center}
\vskip-1cm 
\caption{
\tenrm
Alpha particle projectile traversing an equal molal DT plasma. 
The stopping power $dE/dx$ (in MeV/$\mu{\rm m}$) is plotted 
{\em vs}.~energy $E$ (in ${\rm Mev}$). The solid line is the
result from this work (BPS), while the dashed line is
the result of Li and Petrasso (LP). The energy domain 
lies between zero and the $\alpha$ particle energy 
$E_0=3.54~{\rm MeV}$ produced in the DT reaction. 
The plasma temperature is $T=3\,{\rm keV}$,  the electron 
number density is $n_e=1.0 \times 10^{25}\, {\rm cm}^{-3}$, 
with deuterium-tritium number densities $n_d=n_t=0.5 \times 
10^{25}\, {\rm cm}^{-3}$ (for charge neutrality). The 
plasma coupling is $g_p=0.011$, 
and the thermal speed of the electron is $\bar v_e=3.98
\times 10^9\,{\rm cm/s} $. The BPS result is essentially 
exact since the plasma coupling is so small. 
}
\label{fig:LPvsBPS.dEdxE.aDT030125}

\vskip1cm 
  \begin{center}
  \epsfxsize=110mm
  \epsfbox{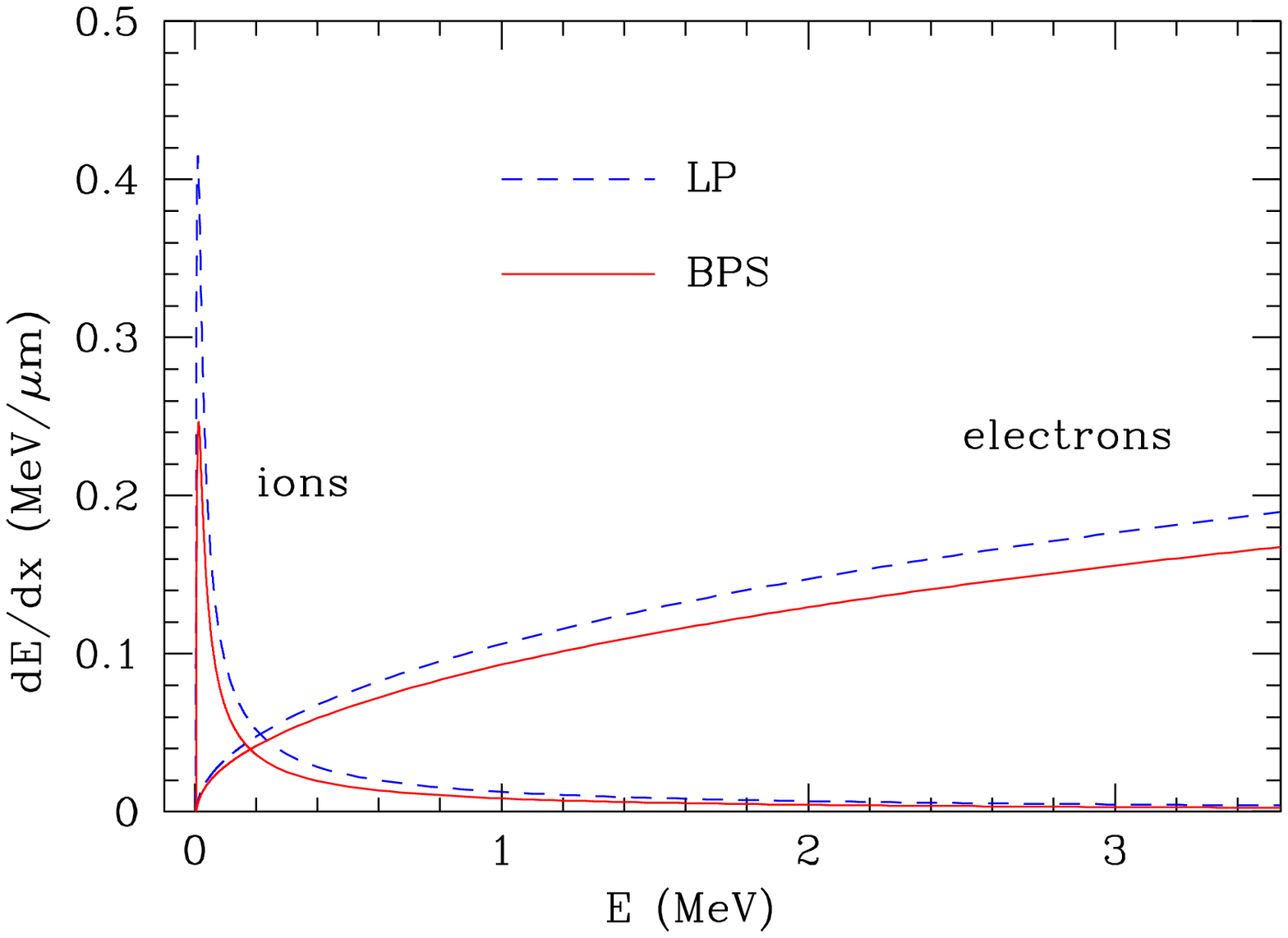} 
  \end{center}
\vskip-1cm 
\caption{
\tenrm
The results of Fig.~\ref{fig:LPvsBPS.dEdxE.aDT030125} 
split into separate ion (peaked curves) and electron 
components (softly increasing curves).
}
\label{fig:LPvsBPS.dEdxEei.aDT030125}
\end{figure}

As shown in Fig.~\ref{fig:LPvsBPS.dEdxEei.aDT030125},
for ion projectiles, the energy loss to the electrons in the plasma
dominates over that to the ions when the projectile energy becomes
sufficiently large on the scale of the temperature $T$.  Here we
provide an estimate of the energy at which this cross over takes
place, an estimate that is valid to logarithmic accuracy, which holds
when the logarithmic term in the energy loss formula --
the ``Coulomb logarithm'' -- is large and dominant.  We do this 
for a plasma whose
various species are all at the same temperature $T$. 
Denoting this logarithmic term by 
$L$, which we treat as a constant since its variation within an
integral is small, the logarithmic contribution to the stopping power
is contained in Eq.~(\ref{wonderclassic}) and reads 
\begin{eqnarray}
   {dE_{b,L} \over dx} &=&
  {e_p^2 \over 4\pi} \, 
	{\kappa^2_b \over m_p \, v_p} \,
  \left( { m_b \over  2\pi \beta  } \right)^{1/2} \,
  \int_0^1 du \, u^{1/2} \,
  \exp\left\{ - {1 \over 2} 
  \beta m_b v^2_p \, u \right\}
\nonumber\\
  && \qquad\qquad
 	L \, 
  \left[  \beta \, \left( m_p + m_b  \right) \, v_p^2
  -  {1\over u} \right] \,.
\label{onlylog}
\end{eqnarray}
As we shall find, near the cross over region the projectile 
energy $E = m_p \, v_p^2 /2 $ is large in comparison with the 
temperature $T$, and the factor 
$\beta m_b v_p^2 $ in the exponent in Eq.~(\ref{onlylog}) 
is large for ions of mass $m_b \sim m_p$. Hence only the small 
$u$ region of the integration makes a significant contribution, 
and the integration region $0<u<1$ can be extended to
$0<u<\infty$ to obtain the leading piece. The resulting Gaussian 
integrals are readily done, and one finds that for an ion
$b$, 
\begin{equation}
{dE_{b,L} \over dx} \simeq
  {e_p^2 \over 4\pi} \, 
	\kappa^2_b \,L \, { 1 \over \beta  m_b  v_p^2} 
	= {e_p^2 \over 4\pi} \, 
	{\omega_b^2 \over v_p^2} \,L  \,.
\label{ionlog}
\end{equation}
Here we have used the ionic plasma frequency, 
$\omega_b^2 = e^2_b \, n_b / m_b $ and 
$\kappa_b^2 = \beta m_b \omega_b^2$, 
in the second equality to emphasize that the result 
is independent of the temperature.
The cross over point is at a projectile energy such
that $\beta m_e v_p^2 \ll 1$, as we shall soon find. Hence for the
electrons in the plasma, when $b=e$, we may approximate the
exponential by unity in Eq.~(\ref{onlylog}) to obtain
\begin{equation}
{dE_{e,L} \over dx} \simeq
  {e_p^2 \over 4\pi} \, 
	\kappa^2_e \,  L \, \, {2\over3} 
	\left( {\beta  m_e  v_p^2 \over 2\pi} \right)^{1/2} \,.
\label{electronlog}
\end{equation}
On comparing these two equations, we find that the electron
contribution dominates over the ionic contribution of species $b$,
\begin{equation}
{dE_{e,L} \over dx}  > {dE_{b,L} \over dx} \,,
\end{equation}
when the projectile energy $E > E_L$, with 
\begin{equation}
E_L \simeq \left[ {9 \pi \over 16} \, {m_p^3 \over m_e \kappa_e^4} \,
\left( \sum_{b \ne e} \, {\kappa_b^2 \over m_b } 
\right)^2 \right]^{1/3} \, T \,.
\label{crossover}
\end{equation}
Note that the expression in square brackets here is independent
of the temperature so that $E_L$ scales linearly with the
temperature.
For the parameters of 
Fig.~\ref{fig:LPvsBPS.dEdxEei.aDT030125}, 
the estimate provided by Eq.~(\ref{crossover}) 
gives the cross over energy
$E_L = 0.10$ MeV compared to the actual 
cross over energy of 0.18 MeV.  

The amount of energy $E_\smI$ that the slowing particle with an
initial energy $E_0$ transfers to the ions may be expressed as
\begin{equation}
E_\smI = \int_0^{E_0} dE \, { {dE_\smI \over dx}(E) \over
       {dE_\smI \over dx}(E) + {dE_e \over dx}(E) } \,,
\label{ratio}
\end{equation}
where
\begin{equation}
{dE_\smI \over dx} = \sum_{\rm{all \,\, ions} \,\, b}
      {dE_b \over dx}   \,.
\end{equation}
The corresponding energy loss to the electrons in the plasma is,
of course, just $E_e = E_0 - E_\smI$.  We can use the rough
logarithmic approximations of the previous paragraph to estimate
the energy transfer.  We use the sum of the approximate ionic
stopping powers (\ref{ionlog}) in the numerator  of
Eq.~(\ref{ratio}) and add the approximate electronic part 
(\ref{electronlog}) to this for the denominator in
Eq.~(\ref{ratio}).  Changing the integration variable $E$ to an
appropriately scaled velocity then yields, in this logarithmic
approximation,  
\begin{equation}
E_{\smI,L} = 2 E_L \, \int_0^{\sqrt{E_0/E_L}} \, {xdx \over 1 + x^3}
\label{fermi}
\,,
\end{equation}
where $E_L$ is the estimate (\ref{crossover}) of the cross over
energy. Since this integral damps out at large $x$ values, the 
simple upper bound to this approximation for the ionic
energy transfer obtained by extending the upper limit of the
integral to infinity should not be too far off.  A glance at
Fig.~\ref{fig:LPvsBPS.dEdxEei.aDT030125}
shows that the ionic energy loss is indeed dominated by small
energies. We use
\begin{equation}
\int_0^\infty \, {xdx \over 1 + x^3} = {2\pi \over 3 \sqrt3} \,,
\end{equation}

\vskip-0.5cm 
\noindent
to obtain
%
\begin{equation}
E_{\smI,L} \, \lesssim \, {4\pi \over 3 \sqrt3} \, E_L \simeq 2.4 \,  E_L \,.
\end{equation}
For the parameters of 
Fig.~\ref{fig:LPvsBPS.dEdxEei.aDT030125}, 
this crude limit gives 0.24 MeV to be compared with the value of
$E_\smI = 0.38$ MeV that comes from a numerical evaluation of
Eq.~(\ref{ratio}) using our complete energy loss formulas
with an initial  energy of $E_0 = 3.54$ MeV [{\it c.f.} 
Fig.~\ref{fig:LPvsBPS.dEdxxei.aDT030125}].  In this case, the energy
loss to the ions is small in comparison to that lost to the
electrons, $E_e = 3.54 -0.38 = 3.16$ Mev. It is interesting to note
that, since the approximate limit scales linearly with the
temperature, if the plasma temperature is increased by an order
of magnitude, from 3 keV to 30 keV, the limit of 0.24 MeV
moves to 2.4 Mev.  A numerical evaluation similar to that
reported in Fig.~\ref{fig:LPvsBPS.dEdxxei.aDT030125} for  this
increased temperature of 30 keV gives $E_\smI = 1.8$ MeV. This is
now comparable to the energy transfer to the electrons, 
$E_e = 3.5 - 1.8 = 1.7$ MeV.  We should
note that, as the temperature is increased with a fixed initial
projectile energy $E_0$, the upper integration limit 
$\sqrt {E_0/E_L} $ in Eq.~(\ref{fermi}) is reduced, and so 
its replacement by the limit $x=\infty$ gives an increasingly 
worse result. 

\vskip-0.2cm 
\begin{figure}[b]
  \begin{center}
  \epsfxsize=110mm
  \epsfbox{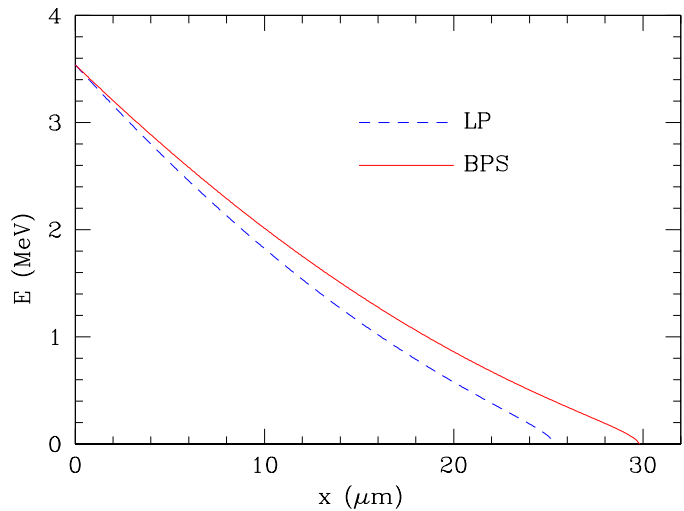} 
  \end{center}
\vskip-0.7cm 
\caption{
\tenrm
The distance $x(E;E_0)$ computed by Eq.~(\ref{slowdown})  
defines an energy $E(x)$ for a particle that has traveled
a distance $x$ starting at $x=0$ with an initial energy
$E(0) = E_0$. An $\alpha$ particle of energy
$E_0=3.54\,{\rm MeV}$ is created from threshold in the 
reaction $D+T \to \alpha + n$. For the DT plasma defined 
in Fig.~\ref{fig:LPvsBPS.dEdxE.aDT030125}, we have plotted
the $\alpha$ particle energy (in MeV) {\em vs}.~the distance
traveled (in $\mu m$). The solid line is the
result from this work (BPS), while the dashed line is
the result of Li and Petrasso (LP). They give the 
respective ranges $R_\smBPS=30\,{\rm \mu m}$ and
$R_\smLP=25\,{\rm \mu m}$ , about a 20\% difference. 
}
\label{fig:LPvsBPS.Ex.aDT030125}
\end{figure} 

From the results shown in Fig.~\ref{fig:LPvsBPS.dEdxE.aDT030125}, 
we can compute the distance $x$ that a projectile, starting with 
energy $E_0$, travels to be slowed down to reach the energy $E$:
\begin{equation}
  x(E;E_0) = \int_E^{E_0} dE \, 
  \left({dE \over dx } \right)^{-1} \,.
\label{slowdown}
\end{equation}
Figure \ref{fig:LPvsBPS.Ex.aDT030125} shows the inverse 
function, $E$ {\em vs}.~$x(E;E_0)$, for an alpha particle with
an initial energy $ E_0 = 3.54$ MeV, 
corresponding to the alpha particle produced
in DT fusion. 
In Fig.~\ref{fig:LPvsBPS.dEdxEei.aDT030125} we illustrated 
the energy dependence of the electron and ion components 
of the stopping power. In 
Fig.~\ref{fig:LPvsBPS.dEdxxei.aDT030125} we plot the 
electron and ion components as a function of the the
distance $x$ that the $\alpha$ particle has traversed,
\begin{eqnarray}
  \frac{dE_e}{dx}(x) = \frac{dE_e}{dx}(E(x)) \ ,
  \hskip1cm 
  \frac{dE_\smI}{dx}(x) = \frac{dE_\smI}{dx}(E(x)) \ .
 \end{eqnarray}

\begin{figure} 
  \begin{center}
  \epsfxsize=110mm
  \epsfbox{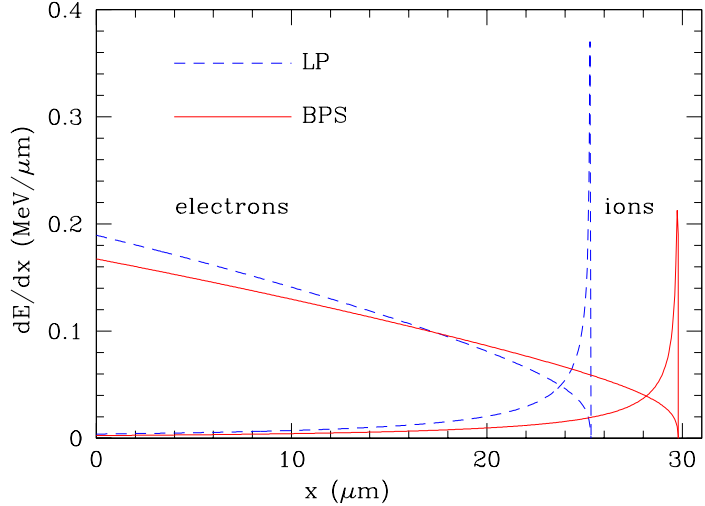} 
  \end{center}
\vskip-0.6cm 
\caption{
\tenrm
The $\alpha$ particle $dE(x)/dx$ (in ${\rm Mev}
/\mu{\rm m}$) {\em vs}.~$x$ (in $\mu m$) split into 
separate ion (peaked curves) and electron components 
(softly decreasing curves). The energy used to compute 
$dE(x)/dx$ is determined from the results shown in 
Fig.~\ref{fig:LPvsBPS.Ex.aDT030125} while the 
corresponding $dE(x)/dx$ is given by the results in
Fig.~\ref{fig:LPvsBPS.dEdxEei.aDT030125}. Again, the 
solid line is the result from this work (BPS), and the 
dashed line is the result of Li and Petrasso (LP). The 
area under each curve gives the corresponding energy 
partition into electrons and ions for this work and that 
of Li and Petrasso. For our results, the total energy 
deposited into electrons is $E_e^\smBPS=3.16$~MeV and 
into ions is $E_\smI^\smBPS=0.38$~MeV, while LP gives 
$E_e^\smLP=3.09$~MeV and $E_\smI^\smLP=0.45$~MeV. Note 
that these energies sum to the initial $\alpha$ particle 
energy of $E_0=3.54\,{\rm MeV}$.  
}
\label{fig:LPvsBPS.dEdxxei.aDT030125}
\vskip0.1cm 
\end{figure}

It is worthwhile comparing the results that we have just
illustrated for DT produced alpha particles moving in a plasma 
at 3.0 keV and electron number density $10^{25}\,{\rm cm^{-3}}$ 
with DT alphas moving in a hotter, more dense plasma, a plasma
at 30 keV and an electron number density $10^{27}\,{\rm cm^{-3}}$;
see 
Figs.~\ref{fig:LPvsBPS.dEdxE.aDT300127}--\ref{fig:LPvsBPS.dEdxxei.aDT300127}.
In the previous case, most of the alpha particle energy was
deposited into electrons. In the new case, much of this energy
is now transfered directly into the ions.  The new, much denser
case clearly has a much shorter alpha particle range. 

\newpage

\begin{figure}[ht]
  \begin{center}
  \epsfxsize=110mm
  \epsfbox{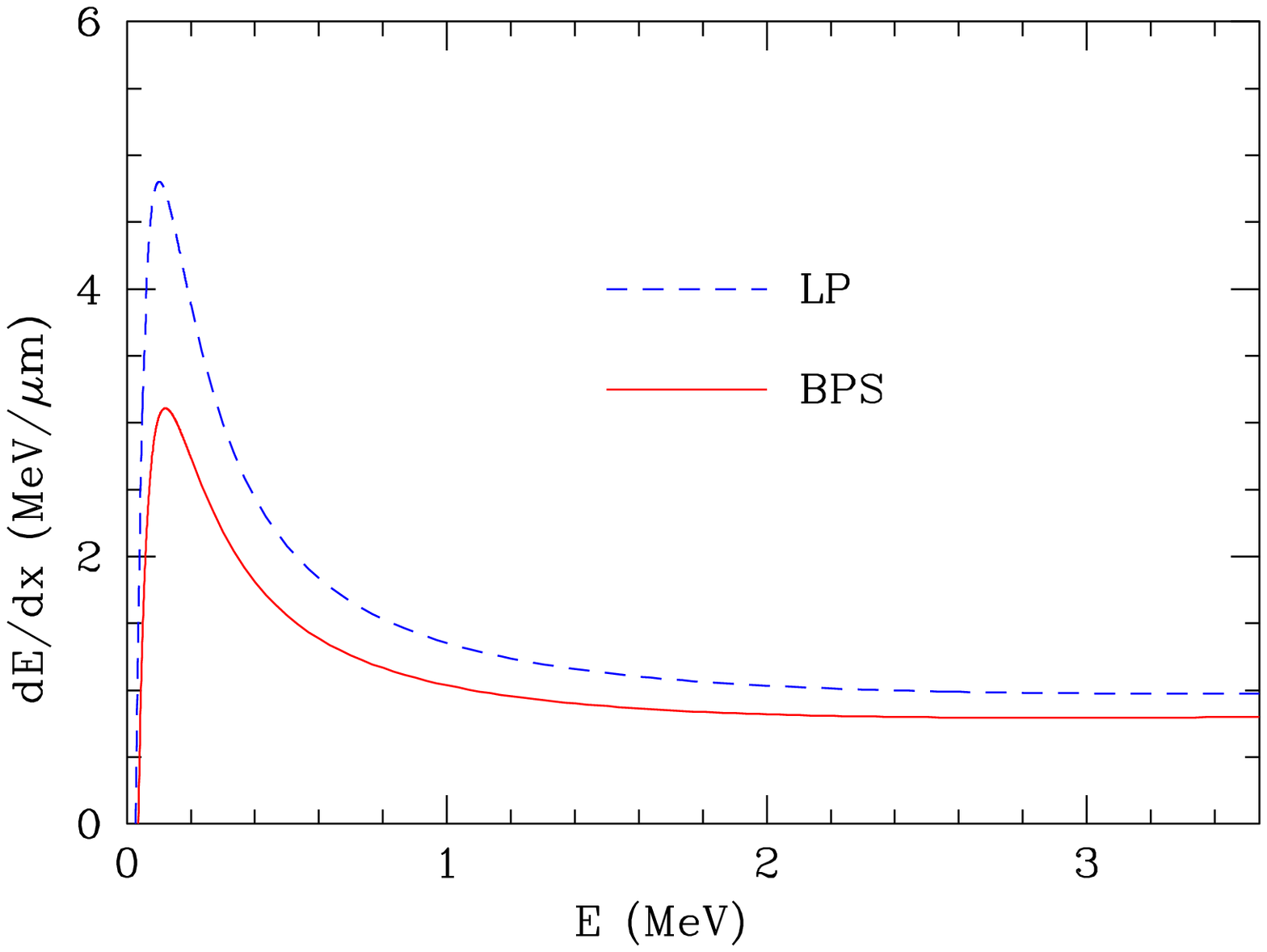} 
  \end{center}
\vskip-1cm 
\caption{
\tenrm
The stopping power $dE/dx$ (in MeV/$\mu{\rm m}$)
of an $\alpha$ particle projectile traversing an 
equal molal DT plasma as a function of the projectile
energy $E$ (in ${\rm MeV}$). The energy domain 
lies between zero and the $\alpha$ particle energy
$E_0=3.54~{\rm MeV}$ produced in DT reaction. The plasma 
temperature is $T=30\, {\rm keV}$ and the electron 
number density is $n_e=1.0 \times 10^{27}\, {\rm 
cm}^{-3}$, which is characteristic of plasmas for
inertial confinement fusion shortly after ignition. 
The deuterium-tritium number densities are $n_d=n_t
=0.5 \times 10^{27}\, {\rm cm}^{-3}$ (for charge 
neutrality).  The solid line is the result from this 
work (BPS), and the dashed line is the result of Li 
and Petrasso (LP).
The plasma coupling is $g_p=0.0033$, 
and the thermal speed of the electron is $\bar v_e=1.26
\times 10^{10}\,{\rm cm/s}$. The BPS result is essentially 
exact since the plasma coupling is so small. 
}
\label{fig:LPvsBPS.dEdxE.aDT300127}

\vskip1cm 

  \begin{center}
  \epsfxsize=110mm
  \epsfbox{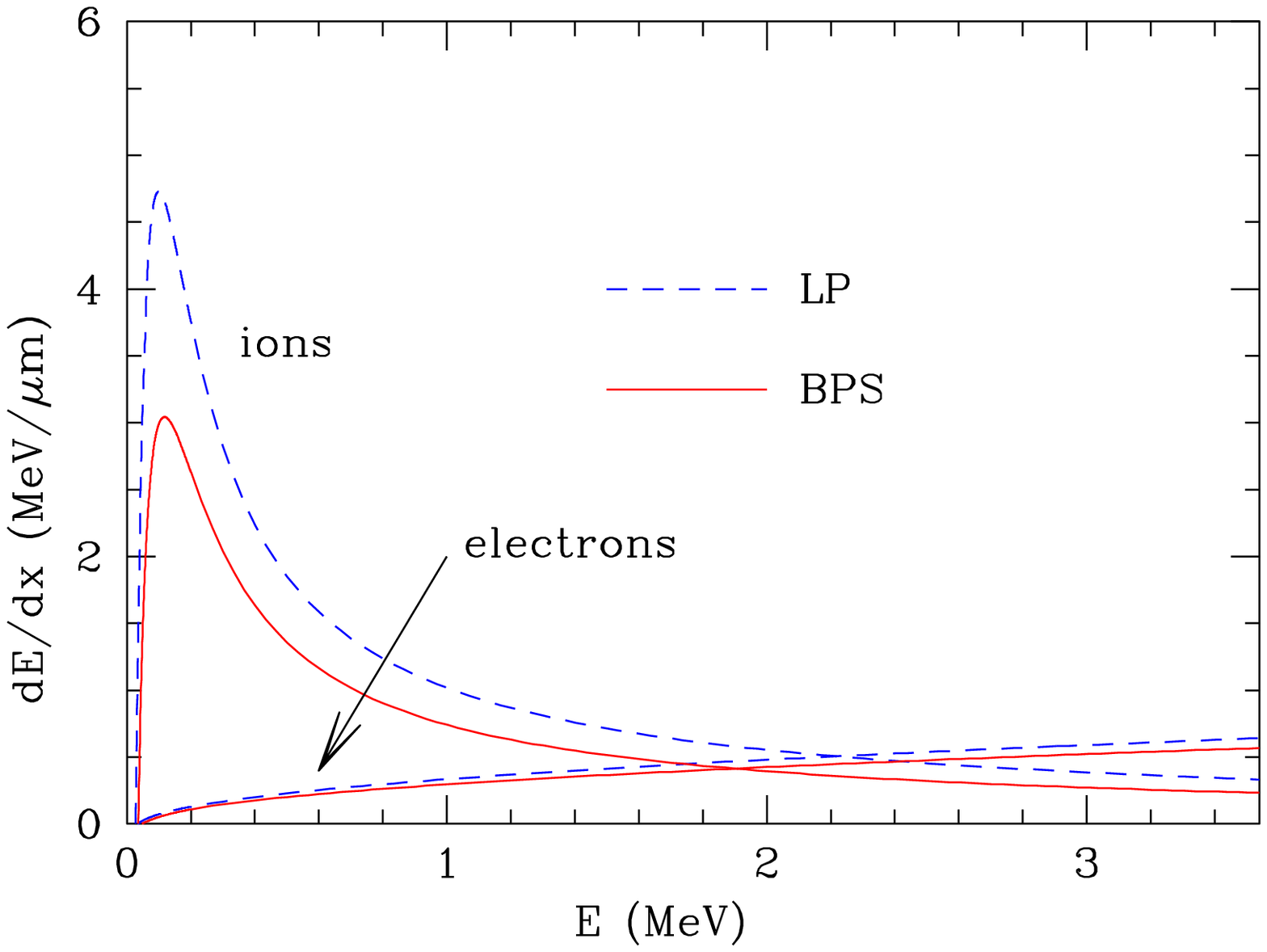} 
  \end{center}
\vskip-1cm 
\caption{
\tenrm
The results of Fig.~\ref{fig:LPvsBPS.dEdxE.aDT300127} 
split into separate ion and electron contributions.
}
\label{fig:LPvsBPS.dEdxEei.aDT300127}
\end{figure}
\newpage

\begin{figure}
  \begin{center}
  \epsfxsize=110mm
  \epsfbox{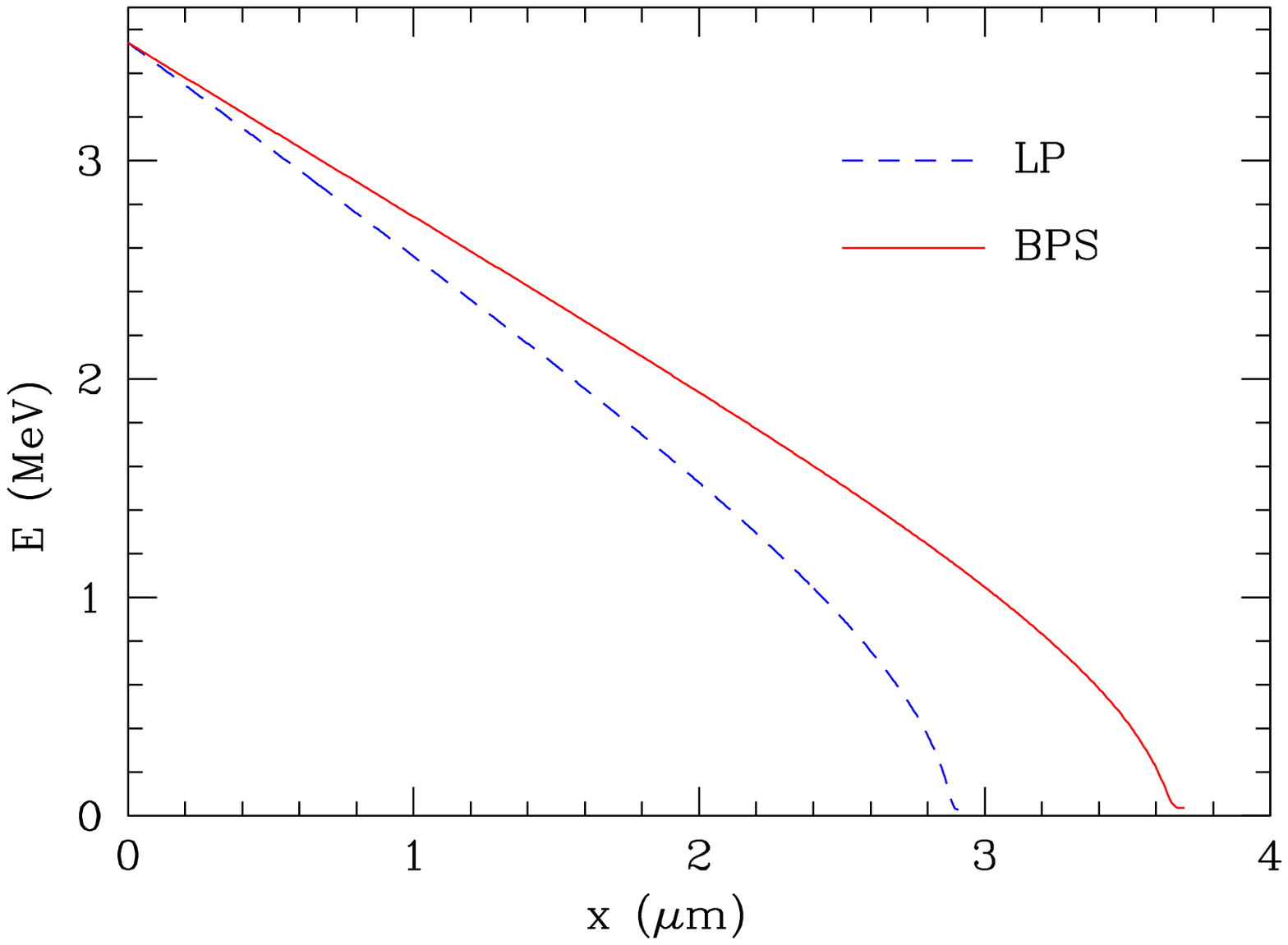} 
  \end{center}
\vskip-1cm 
\caption{
\tenrm
For the DT plasma defined in 
Fig.~\ref{fig:LPvsBPS.dEdxE.aDT300127}, the energy (in MeV) 
as a function of the distance traveled (in ${\rm \mu m}$) 
is shown
for an $\alpha$ particle created at $E_0=3.54\,{\rm MeV}$ 
in the reaction $D+T \to \alpha + n$. The solid line is
the result from this work (BPS), and the dashed line is
the result of Li and Petrasso (LP). They give the respective 
ranges $R_\smBPS=3.7\,{\rm \mu m}$ and $R_\smLP=2.9\,{\rm \mu m}$,
almost a 30\% difference. 
}
\label{fig:LPvsBPS.Ex.aDT300127}
\vskip1cm 
  \begin{center}
  \epsfxsize=110mm
  \epsfbox{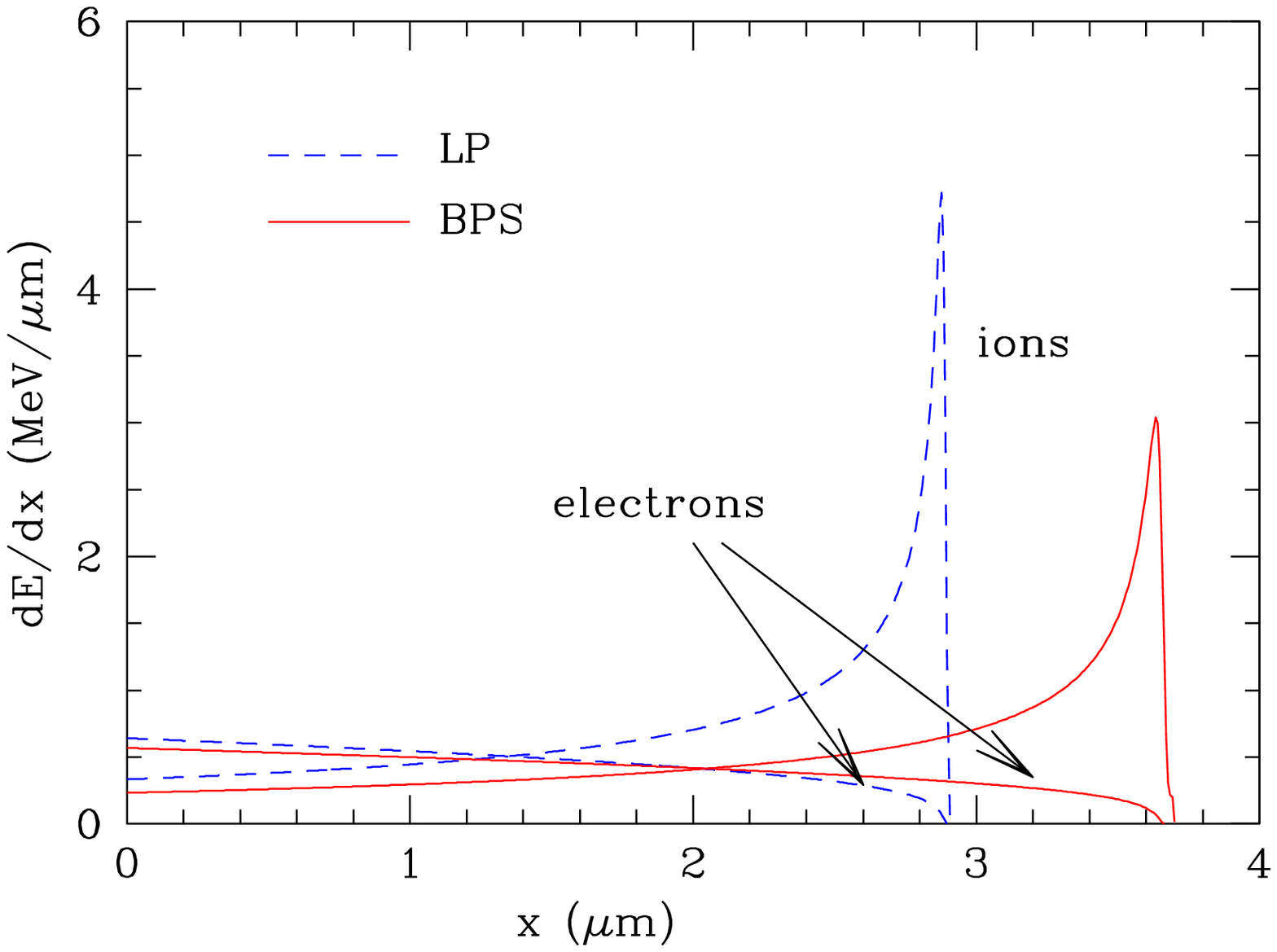} 
  \end{center}
\vskip-1cm 
\caption{
\tenrm
Alpha particle $dE(x)/dx$ (in ${\rm MeV}
/\mu{\rm m}$) {\em vs}.~$x$ (in $\mu m$) split into 
separate ion and electron components. The energy 
used to compute $dE(x)/dx$ is determined from the 
results shown in Fig.~\ref{fig:LPvsBPS.Ex.aDT300127} 
while the corresponding $dE(x)/dx$ is given by 
the results in Fig.~\ref{fig:LPvsBPS.dEdxEei.aDT300127}. 
The solid line is the result from this 
work (BPS), and the dashed line is the result of Li 
and Petrasso (LP). For our result, the energy 
deposited into electrons is $E_e^\smBPS=1.51$~MeV 
and into ions is $E_\smI^\smBPS=2.00$~MeV, while LP 
gives $E_e^\smLP=1.36$~MeV and $E_\smI^\smLP=2.15$~MeV.
Note that both LP and BPS sum to $3.51\,{\rm MeV}$, 
which is within 1\% of the initial $\alpha$ particle 
energy $E_0=3.54\,{\rm MeV}$. 
}
\label{fig:LPvsBPS.dEdxxei.aDT300127}
\vskip0.3cm 
\end{figure}

We conclude this section by plotting similar figures for 
a triton moving through a deuterium plasma with 
$T=0.5$~keV and an electron density $n_e =
10^{24}\,{\rm cm}^{-3}$; see 
Figs.~\ref{fig:LPvsBPS.dEdxE.TD005124}--\ref{fig:LPvsBPS.dEdxxei.TD005124}.
\begin{figure}
  \begin{center}
  \epsfxsize=110mm
  \epsfbox{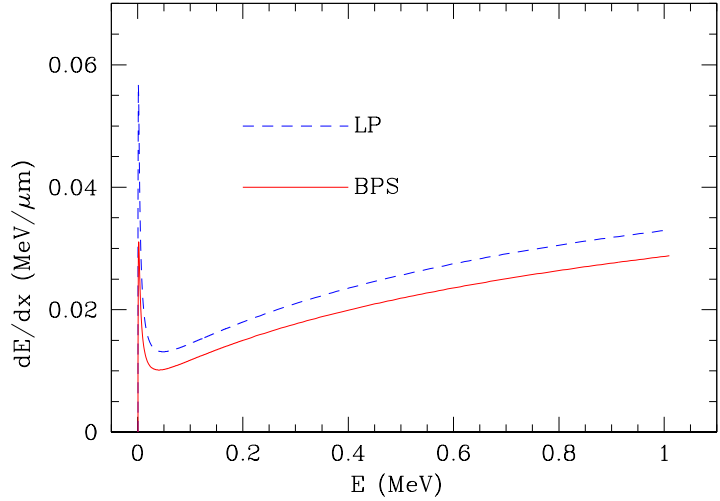} 
  \end{center}
\vskip-1cm 
\caption{
\tenrm
For a triton projectile traversing a deuterium plasma, the 
stopping power $dE/dx$ (in MeV/$\mu{\rm m}$) is plotted
{\em vs}.~energy
$E$ (in ${\rm MeV}$). The solid line is the result from
this work (BPS), and the the dashed line is the result 
of Li and Petrasso~(LP). The energy domain lies between 
zero and the triton energy  $E_0=1.01~{\rm MeV}$ produced
in the DD reaction. The plasma temperature 
is $T=0.5\,{\rm keV}$,  the electron number density is 
$n_e=1.0 \times 10^{24}\, {\rm cm}^{-3}$, with a deuterium
number density $n_d=1.0 \times 10^{24}\, {\rm cm}^{-3}$
(for charge neutrality). The plasma coupling is $g_p=0.025$, 
and the thermal speed of the electron is $\bar v_e = 1.62 
\times 10^9\,{\rm cm/s} $. The BPS result is essentially
exact since the plasma coupling is so small. 
}
\label{fig:LPvsBPS.dEdxE.TD005124}
\end{figure}
\begin{figure}
  \begin{center}
  \epsfxsize=110mm
  \epsfbox{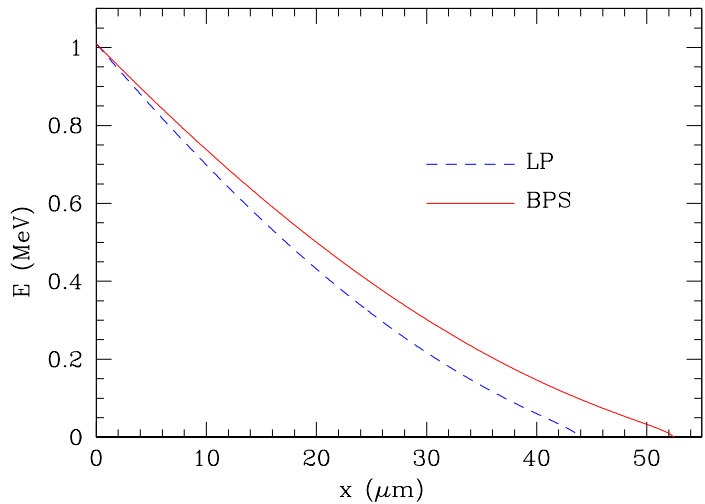} 
  \end{center}
\vskip-1cm
\caption{
\tenrm
For the deuterium plasma defined in 
Fig.~\ref{fig:LPvsBPS.dEdxE.TD005124}, the energy (in MeV) 
as a function of the distance traveled (in ${\rm \mu m}$) 
is shown 
for a triton created at threshold with energy $E_0=1.01\,
{\rm MeV}$  in the reaction $D+D \to T + p$. The solid line 
is the result from this work (BPS) and the dashed line is 
the result of Li and Petrasso (LP). They give the respective 
ranges $R_\smBPS=52\,{\rm \mu m}$ and $R_\smLP=44\,{\rm 
\mu m}$, a 20\% difference. 
}
\label{fig:LPvsBPS.Ex.TD005124}
\end{figure}
\newpage

\begin{figure}
  \begin{center}
  \epsfxsize=110mm
  \epsfbox{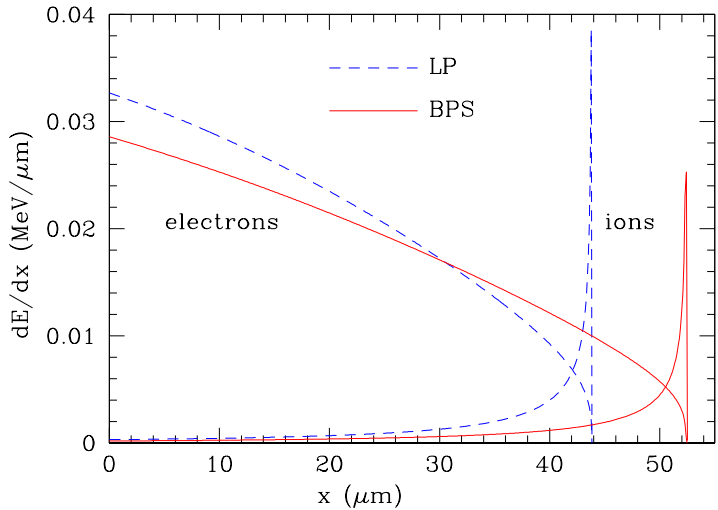} 
  \end{center}
\vskip-1cm
\caption{
\tenrm
Derivative $dE(x)/dx$ of the curve in 
Fig.~\ref{fig:LPvsBPS.Ex.TD005124} split into separate 
ion (peaked curves) and electron components 
(softly decreasing curves). 
This is the triton $dE/dx$ (in ${\rm MeV}
/\mu{\rm m}$) {\em vs}.~$x$ (in $\mu m$).  Again our result 
is the solid curve, and the work of of Li and Petrasso 
is dashed. The area under each curve gives
the corresponding energy partition into
electrons and ions for this work and that 
of Li and Petrasso. For our result, the energy 
deposited into electrons is $E_e^\smBPS=0.95$~MeV 
and into ions is $E_\smI^\smBPS=0.063$~MeV, while LP 
gives $E_e^\smLP=0.93$~MeV and $E_\smI^\smLP=0.075$~MeV.
These energies sum to the initial triton energy
$E_0=1.01\,{\rm MeV}$. 
}
\label{fig:LPvsBPS.dEdxxei.TD005124}
\end{figure}
~~

\subsection{Plasma Temperature Equilibration}

We now present our results for the temperature equilibration 
of a plasma in which different components have different 
temperatures. This is quite common since plasmas may be created 
in ways that more effectively heat one plasma species over 
another; for example, when a plasma experiences a laser pulse 
which preferentially heats the light electrons that have the 
larger scattering cross section. We therefore assume that two 
species $a$ and $b$ are in thermal equilibrium with themselves 
but at two different temperatures $T_a$ and $T_b$. The rate of 
energy exchange between the subsystems $a$ and $b$ is 
\begin{equation}
  \frac{d{\cal E}_{ab}}{dt} = - {\cal C}_{ab}(T_a-T_b) \ ,
\end{equation}
and in this section we present our calculation of the coefficients 
${\cal C}_{ab}$ accurate to order $g^2 \ln C g^2$.

We should remind the reader that there is an  
hierarchy in which the the light electrons first come into 
equilibrium among themselves, then the heaver ions equilibrate 
among themselves, and lastly the electron and ions equilibrate
in temperature. Consider now the typical case of a plasma of light 
electrons and heavy ions, in which the electrons and ions have 
respectively equilibrated with themselves. The electrons will
have a temperature $T_e$, and and suppose all the ions have had 
time to equilibrate to a common ion temperature $T_i$. The
rate of energy exchange between the electrons and the total
ion system is
\begin{equation}
  \frac{d{\cal E}_{e\smI}}{dt} = - {\cal C}_{e\smI}
  (T_e-T_i) \ ,
\end{equation}
where we define ${\cal C}_{e\smI} = {\sum}_i \, {\cal C}_{ei}$.
With electron and ion specific heats per unit volume $c_e$ and 
$c_\smI$ defined by $d {\cal E}_e = c_e\, dT_e$ and $d {\cal 
E}_\smI = c_\smI\, T_i$, the  rate $\Gamma$ is 
given by 
\begin{equation}
  {d \over dt} \left( T_e - T_i \right) = 
  - \Gamma \, \left( T_e - T_i \right) \,,
\end{equation}
with 
\begin{equation}
  \Gamma = {\cal C}_{e \smI} \, 
  \left( { 1 \over c_e} + { 1 \over c_\smI} \right) \ .
\end{equation}
In the regime where the electron and ion temperatures
are not too different, a case of common interest, 
Eq.~(\ref{spit}) gives 
\begin{eqnarray}
T_i m_e &\ll& T_e m_i \,: 
\nonumber\\[10pt]
  {\cal C}_{e\smI} &=& 
  {\kappa_e^2 \over 2\pi } \, \omega_\smI^2 \, 
  \sqrt{ \beta_e m_e \over 2\pi} \, {1\over2} \,
  \left\{ \ln\left( { 8 T_e^2 \over \hbar^2 \omega_e^2} 
  \right)- \gamma - 1 \right\} \,,
\label{spitsum}
\end{eqnarray}
where $\omega^2_\smI = {\sum}_i \, \omega^2_i$ is the sum 
over all the squared ionic plasma frequencies $\omega^2_i=
e^2_i n_i / m_i$. 

The expression for ${\cal C}_{ab}$ in the general case is 
much more complex, and we shall not present it here. We will
only state that it can be written as a classical contribution 
plus a quantum correction ${\cal C}_{ab}=
{\cal C}_{ab}^\smC + {\cal C}_{ab}^{\Delta Q}$, where the
quantum piece is given by Eq.~(\ref{qcorr}) and the classical
piece (\ref{cclassicall}), ${\cal C}_{ab}^\smC=
{\cal C}_{ab,\smS}^\smC + {\cal C}_{ab}^\smlt$, by the
sum of Eqs.~(\ref{wwonderclassicc})~and~(\ref{nnun}).

\section{General Formulation} 
\label{general}

Thus far we have examined only the de-acceleration of charged
particles caused by the stopping power $dE/dx$.  This slowing
down of a particle keeps it moving in a straight line, and
all particles starting from the same place with the same velocity
slow to a thermal velocity at exactly the same final position.
But in fact, the width of a narrow beam of particles will
increase as the particles move through a plasma --- there will be
a sort of Brownian motion in the directions that are transverse to
the beam direction.  A particle will acquire an increasing
average squared transverse velocity as it propagates through a
plasma.  This is transverse diffusion. 
Moreover, the ending positions of a group of particles
with identical starting conditions will be spread out along the
longitudinal, beam direction to a small extent.
This is straggling.  

These random, statistical effects may be accounted for in first
approximation by
describing the charged particle transport by the Fokker-Planck
equation 
\begin{equation}
\left[ {\partial \over \partial t} + {\bf v} \cdot 
  {\mmbf \nabla} \right]
f({\bf r}, {\bf p}, t) =  {\sum}_b 
{\partial \over \partial p^k} \, C^{kl}_b({\bf r},{\bf p},t) \,
\left[ \beta_b v^l + {\partial \over \partial p^l} \right]
f({\bf r}, {\bf p}, t) \ ,
\label{fp}
\end{equation}
where $f({\bf r},{\bf p},t)$ is the phase space number 
density for a swarm of particles injected into the plasma, 
each with a common mass $m$. The momentum derivative 
$\partial/\partial p^k$ acts on everything to its right, 
including the distribution function $f$. An implicit 
summation convention in which repeated indices are summed 
over will be used for vector components, and these indices 
will be denoted by the Latin letters $k$ and $l$. On the 
other hand, sums over the plasma species 
$b$ will always be made explicit. In other words, repeated $b$ 
indices are not summed, while repeated $k$ and $l$ indices are 
summed. These conventions are followed in Eq.~(\ref{fp}). 
For each 
species index $b$, the scattering tensor $C_b^{kl}$ 
is symmetric in $k$ and $l$, as the decomposition (\ref{sstructtt}) 
will illustrate. 

The Boltzmann equation reduces to the Fokker-Planck equation when
the collisions transfer only small momenta in comparison with the
particle momentum. The way in which this limit of the Boltzmann
equation works out is described in detail in 
Appendix\footnote{In this regard, we should note that
  although a ``diffusion approximation'' to the Boltzmann 
  equation is outlined in Lifshitz and Pitaevskii \cite{Lifsss}, 
  that discussion is in the context of dilute heavy particles 
  moving in a gas of light particles.  This is a kinematical 
  restriction that is quite different than the dynamical case of
  sharply peaked forward scattering that we examine in Appendix 
  \ref{fok}.  To make this distinction clear, we also review the 
  work of Lifshitz and Pitaevskii in Appendix \ref{fok}, where we
  show by a simple example that their result is not internally
  consistent unless further restrictions are imposed.} \ref{fok}. 
The Lenard-Balescu equation is of the
form of a Fokker-Planck equation.  Since the right-hand side of
the Fokker-Planck equation entails an overall derivative in
momentum, it conserves the particle number density.  The terms 
in the final square brackets on the right-hand side of the 
equation ensure that this side of the equation vanishes for a 
thermal distribution of particles 
$
f \sim \exp\{- \beta \, m \, {\bf v}^2 /2 \} 
$ 
at inverse temperature $\beta$, {\it provided} the background 
plasma components have the common temperature determined by
the distribution function, namely $\beta_b = \beta$ for
all species $b$. Thus a thermal distribution of particles is 
maintained by the Fokker-Planck equation.  

Here we shall obtain a precise evaluation of the tensor functions
$C^{kl}_b$ that appear in the Fokker-Planck equation.  This we
shall do to the accuracy of the stopping power $dE/dx$ that we
have already discussed.  That is, we shall compute not only the
leading logarithms, but also the constant terms under the
logarithms.  In this way, among other things, we shall give a
precise and unambiguous definition of the Landau Collision
Integral for Coulomb scattering \cite{coll}. We shall define 
the functions $C^{kl}_b$ by requiring that the Fokker-Planck
equation  reproduce the rate of energy and momentum loss to the 
plasma species $b$, quantities that are well defined to the
leading log plus constant order to which we work, and quantities
that we compute using the method of dimensional continuation that
we have described. 

As a charged particle slows down, large angle scattering events
become more important.  Such hard collisions are not described by
the Fokker-Planck equation, and it starts to loose its accuracy of
describing charged particle trajectories, particularly with regard to 
the transverse motion.  We shall obtain quantitative criteria for
the regions where the Fokker-Planck equation ceases to be an
accurate description. 
We shall assess the validity of the
Fokker-Planck description by computing the rate of transverse
energy loss to our order, the order which includes the constant
terms under the leading Coulomb logarithm.  The difference
between this independently calculated quantity and its evaluation
using the Fokker-Planck equation tells us when the Fokker-Planck
description starts to break down. 

\subsection{Energy and Momentum Transfer Rates}

As we have stated, the coefficients $C^{kl}_b$ are constrained to 
produce the energy and momentum exchange between the charged particle
and the background plasma particles of species~$b$. 
To bring this out, we first examine the general case of the
transport of some general quantity $q({\bf p})$. Averaging over
the momentum defines a time dependent spatial density
\begin{equation}
{\cal Q}({\bf r},t) = \int {d^\nu {\bf p} \over (2\pi\hbar)^\nu}
\, q({\bf p}) \, f({\bf r}, {\bf p}, t) \,,
\label{Qdef}
\end{equation}
and flux vector
\begin{equation}
{\cal F}^k({\bf r},t) = \int {d^\nu {\bf p} \over (2\pi\hbar)^\nu}
\, q({\bf p}) \, {p^k \over m} \, 
 f({\bf r}, {\bf p}, t) \,.
\end{equation}
The Fokker-Planck equation (\ref{fp}) then expresses
$
(\partial / \partial t) {\cal Q}
+ {\mmbf \nabla} \cdot  {\mmbf {\cal F}}
$
in terms a momentum integral of functions and derivatives acting
upon the distribution function $f$.  Hence, by partial
integration, we may write the result in the form
\begin{equation}
  {\partial \over \partial t} {\cal Q}({\bf r},t) 
  + {\mmbf \nabla} \cdot {\mmbf {\cal F}}({\bf r},t) = 
  - {\sum}_b \int {d^\nu {\bf p} \over (2\pi\hbar)^\nu}
  \, {d Q_b \over dt}({\bf r},{\bf p},t) \,
  f({\bf r}, {\bf p}, t) \,,
\label{defQdot}
\end{equation}
in which
\begin{equation}
  {d Q_b \over dt}({\bf r},{\bf p},t) = \left[ 
  \beta_b v^l - {\partial \over \partial p^l}
  \right] \, C_b^{kl}({\bf r},{\bf p},t) \, 
  {\partial \over \partial p^k} \, q({\bf p}) \,.
\label{Qgiven}
\end{equation}
The sign convention has been chosen so that $dQ_b/dt$ 
represents the rate at which the quantity flows {\em from} 
the projectile to the plasma medium. 

To bring out the meaning of $d Q_b / dt$, we note that the
Fokker-Planck equation keeps the total particle number
\begin{equation}
N = \int d^\nu {\bf r} \int {d^\nu {\bf p} \over (2\pi\hbar)^\nu}
 \, f({\bf r}, {\bf p}, t) 
\label{fintn}
\end{equation}
constant in time.  Thus
\begin{equation}
\langle q \rangle \, N = 
\int d^\nu {\bf r} 
 \int {d^\nu {\bf p} \over (2\pi\hbar)^\nu}
\, q({\bf p})  \, f({\bf r}, {\bf p}, t)
\label{ave}  
\end{equation}
defines a time-dependent, average value of the property $q$.  
We integrate the local transport equation (\ref{defQdot}) over
all space.  
The spatial divergence of the flux term on the left-hand side 
of Eq.~(\ref{defQdot}) is thus removed, and in view of the definitions 
(\ref{Qdef}) and (\ref{ave}) we have
\begin{equation}
\label{dqavdtn}  
{ d \langle q \rangle \over dt} N = 
- {\sum}_b \int d^\nu {\bf r} 
 \int {d^\nu {\bf p} \over (2\pi\hbar)^\nu}
\, { d Q_b \over dt}({\bf r},{\bf p},t)  \, 
f({\bf r}, {\bf p}, t) \,.
\end{equation}
If we now assume that the distribution function $f$ is
sharply peaked about a point in phase space, at some
definite value (${\bf r}_p$,${\bf p}_p$), then we may
evaluate $dQ_b/dt$ at this phase space point and take
it out of the integral to get
\begin{equation}
{ d \langle q \rangle \over dt}  = 
- {\sum}_b \, { d Q_b \over dt}  \,.
\label{aveqdot}
\end{equation}
Therefore, as stated, our sign convention has been chosen
so that $d Q_b / dt$ is the rate of increase of the quantity 
from the swarm of particles determined by $f$ into the plasma 
species~$b$. 

Let us now apply these general considerations to the projectile
energy and momentum.  As we shall see, we will get 
two constraints that completely determine the scattering tensor 
$C^{kl}_b$.  In this way we obtain a transport equation
that accounts for the secular, long-term build up of the changes
in the velocity of a charged particle moving in a medium, a
transport equation with no long or short distance divergences.  
First we consider the energy density
\begin{equation}
{\cal U} = \int {{d^\nu \bf p} \over (2\pi)^\nu} \,
 { {{\bf p}}^2 \over 2 m} \, f({\bf r}, {\bf p}, t) \,,
\end{equation}
and energy flux
\begin{equation}
{\cal S}^k = \int {{d^\nu \bf p} \over (2\pi)^\nu} \,
 { {{\bf p}}^2 \over 2 m} \, {p^k \over m} 
f({\bf r}, {\bf p}, t) \,,
\end{equation}
with 
\begin{equation}
{\partial \over \partial t}\, {\cal U} + {\mmbf \nabla} \cdot
{\mmbf {\cal S}} = - {\sum}_b 
 \int { d^\nu {\bf p} \over (2\pi\hbar)^\nu} \,
{dE_b \over dt} \, f({\bf r}, {\bf p}, t) \,.
\end{equation}
According to our general discussion, 
$dE_b/dt$ is the rate of energy loss to plasma species $b$
when the charged particle at time $t$ is at the spatial position 
${\bf r}$ with momentum ${\bf p}$.  Similarly, with the
momentum density
\begin{equation}
{\cal P}^k = \int {{d^\nu \bf p} \over (2\pi)^\nu} \,
 p^k \, f({\bf r}, {\bf p}, t) \,.
\end{equation}
and spatial stress
\begin{equation}
{\cal T}^{kl} = \int {{d^\nu \bf p} \over (2\pi)^\nu} \,
 {p^k p^l \over m} f({\bf r}, {\bf p}, t) \,,
\end{equation}
we have
\begin{equation}
{\partial \over \partial t} {\cal P}^k + \nabla^l {\cal T}^{kl} = 
- {\sum}_b 
\int { d^\nu {\bf p} \over (2\pi\hbar)^\nu} \,
{dP^k_b \over dt} \, f({\bf r}, {\bf p}, t) \,.
\end{equation}
The general Fokker-Planck evaluation (\ref{Qgiven}) 
therefore gives

\begin{equation}
 {dE_b \over dt} =
\left[ \beta_b {v}^l - {\partial \over \partial {p}^l} \right]
\left[ C^{kl}_b \,  v^k \right] \,,
\label{sp}
\end{equation}
and 
\begin{equation}
 {dP^k_b \over dt} =
\left[ \beta_b {v}^l - {\partial \over \partial {p}^l} \right]
 C^{kl}_b \ .
\label{pp}
\end{equation}

\noindent
Again, the repeated species index $b$ is not summed, while
the repeated vector indices $k$ and $l$ are summed over.

\subsection{Decomposition of the Collision Tensor}

The collision terms in the Boltzmann or Lenard-Balescu equation
do not involve gradients of spatial variation.  Hence the only
available vector with which the Fokker-Planck collision tensor
$C_b^{kl}$ can be constructed is the particle momentum 
${\bf p} = m {\bf v}$.
This tensor therefore has the general structure\footnote{The
appearance of the tensor $\epsilon^{klm} \, \hat v^m$ is
forbidden by parity invariance.  Note that the unit vector 
$\hat{\bf v}$, the direction of the velocity, is not well-defined
in the limit of vanishing velocity, ${\bf v} \to 0$.  On the
other hand, the tensor $C^{kl}_b$ is well defined in this limit
of small velocity.  Hence a low speed constraint must be obeyed:
$$
v \to 0 \,: \qquad {\cal A}_b = {\cal B}_b \, (\beta_b v)  / 2 \,,
$$
with
$$
v \to 0 \,: \qquad C^{kl}_b = \delta^{kl} \, {\cal B}_b  /2 \,.
$$
Our evaluations satisfy these constraints. 
\label{note}}  
\begin{equation}
  C^{kl}_b = {\cal A}_b \, {\hat v^k \, \hat v^l 
  \over \beta_b \, v} + {\cal B}_b {1 \over 2} 
  \left( \delta^{kl} - \hat v^k \hat v^l
\right) \,,
\label{sstructtt}
\end{equation}
where the additional factors of $1 / \beta_b v$ and $1/2$
multiplying the coefficients ${\cal A}_b$ and ${\cal B}_b$ have been
inserted for later convenience. 
The new scalar coefficients are given by the
projections
\begin{equation}
{\cal A}_b \, { 1 \over \beta _b v } = 
C^{kl}_b \, \hat v^k \, \hat v^l \,,
\label{aproj}
\end{equation}
and
\begin{equation}
{1\over2} (\nu -1) \, {\cal B}_b = C^{kl}_b \,
\left( \delta^{kl} - \hat v^k \hat v^l \right) \,.
\label{bproj}
\end{equation}
Alternatively, 
placing the structure (\ref{sstructtt}) 
in the energy constraint (\ref{sp}) 
produces
\begin{equation}
{dE_b \over dt} =
\left[v - {1 \over \beta_b m } \, 
	{\partial \over \partial v^l} \, \hat v^l \right]
		{\cal A}_b \ .
\label{aaa}
\end{equation}
Likewise, we can relate ${\cal B}_b$ to the momentum change
$d{\bf P}_b/dt$ in the following manner.
Since the plasma is isotropic, $d {\bf P}_b/dt $ must point along
the direction ${\bf v}$.  Hence we need only compute
\begin{equation}
 v^k \, {dP^k_b \over dt} =
\left[ \beta_b {v}^l - {\partial \over \partial {p}^l} \right]
 \left[ C^{kl}_b \, v^k \right] + {1 \over m} \, C_b^{ll} \,,
\end{equation}
where the repeated tensor indices in $C^{ll}_b$ imply a
summation over all the spatial axes. 
In view of Eq.~(\ref{sp}), this can be written as
\begin{equation}
 {1 \over m} \, C_b^{ll} =
 v^k \, {dP^k_b \over dt} -
{ dE_b \over dt} \,.
\label{ppp}
\end{equation}
The remaining function ${\cal B}_b$ now can be obtained from 
Eq.~(\ref{ppp}) together with
\begin{equation}
C_b^{ll} = {\cal A}_b \, {1 \over \beta_b v} +
		{\cal B}_b \, {\nu -1 \over 2} \,.
\label{abc}
\end{equation}

The structure of the Fokker-Planck equation guarantees that a
swarm of particles with a Maxwell-Boltzmann distribution at
temperature $T_b = 1 / \beta_b$  remains in thermal equilibrium
during its interaction
with a plasma species $b$ at this same temperature.  This aspect can
be emphasized if we write Eq.~(\ref{aaa}) as
\begin{equation}
\exp\left\{ -{1\over2} \,\beta_b  m v^2 \right\} \,  
{dE_b \over dt} = - {1 \over \beta_b m } 
{\partial \over \partial v^l} \, \hat v^l \, 
\exp\left\{ -{1\over2} \,\beta_b  m v^2 \right\} \,  
 {\cal A}_b \,.
\label{aaaa}
\end{equation}
We shall find that this formula can be quite convenient for 
the identification of ${\cal A}_b$. Note that the thermal 
average of the rate of energy loss for species $b$ necessarily 
vanishes. Placing the structure (\ref{aaaa}) in the formula 
for this average gives zero, 
\begin{equation}
\label{dedtzero}
  \left\langle { d E_b \over dt } \right\rangle_\smT
  = \left( { \beta_b m \over 2 \pi } \right)^{3/2} 
  \hskip-0.1cm 
  \int \hskip-0.1cm d^3 {\bf v} \,  e^{- {1\over2} \beta_b 
  m v^2 }\, \frac{d E_b}{dt} = 0 \ ,
\end{equation}
since the integral entails a total velocity derivative.
If there is a common temperature $T=1/\beta$ for all
plasma species, then the thermal average of the total
rate of energy loss, 
\begin{equation}
  \frac{dE}{dt} = \sum_b \frac{dE_b}{dt} \ ,
\end{equation}
will also vanish. This should be contrasted to the model 
of Li and Petrasso\cite{li} where the thermal average
energy exchange (\ref{dedtbpsnonzero}) does not vanish. 

\subsection{Sharply Peaked Distributions: Projectiles}

Throughout this work we will often take the distribution 
function $f$ to be sharply peaked in phase space, for example, 
a distribution peaked about a specific momentum value  
\hbox{${\bf p}_p=m_p {\bf v}_p$}. Now, for clarity we write 
the projectile mass as $m_p$ rather than $m$. 
Another useful case is when $f$ is peaked only about the momentum 
direction $\hat {\bf v}_p$ and there is no restriction on the absolute 
value of the momentum itself. For either of these two cases, the 
swarm of particles distributed by $f$ will be called a beam of
{\em projectiles}, and momentum integrals can be performed by
the substitutions ${\bf v} \to {\bf v}_p$ and $\hat{\bf v} \to 
\hat{\bf v}_p$ respectively. Finally, as in the preceding example 
(\ref{fintn})-(\ref{aveqdot}), 
we might also consider the distribution function $f$ to be 
sharply peaked in space as well, about a specific point  
${\bf r}_p$, in which case we can evaluate spatial integrals 
by the substitution ${\bf r} \to {\bf r}_p$.

\subsubsection{Transverse Energy} 

Expression (\ref{aaa}) gives a direct connection between 
the rate of energy transfer and the coefficient ${\cal A}_b$,
in contrast to Eqs.~(\ref{ppp}) and (\ref{abc}) 
which provide an implicit 
relation between ${\cal B}_b$ and the rate of 
momentum transfer. In the case of a sharply peaked
particle beam, however, there is also a simple connection 
between the coefficient ${\cal B}_b$ and the rate of 
{\it transverse} energy flow. As in the second case 
described in the previous paragraph, we consider a 
distribution $f$ that is sharply peaked in the momentum 
or velocity direction $\hat {\bf v}_p$, and define the transverse
energy as
\begin{equation}
E_\perp({\bf p}) = 
{1\over2} \, m_p \left[ {{\bf v}}^2 - \left( {\bf v} \cdot
\hat{\bf v}_p \right)^2 \right] \,. 
\label{te}
\end{equation}
Here our sign convention for this quantity is changed in
that \hbox{$q = - E_\perp$} in Eq.~(\ref{Qdef}). Thus,
the Fokker-Plank evaluation of $dE_{\perp\,b}/dt$ represents 
the rate of transverse energy flow {\it from} the plasma to 
the beam.\footnote{
  To make our sign conventions explicit, we note that
  we have defined $dE/dt$ and $d{\bf P}/dt$ as the time rate
  of energy and momentum transferred to the background plasma
  from the moving projectile.  Thus, as the projectile slows
  down,  $dE/dt$ and ${\bf v} \cdot d{\bf P} / dt$ are {\it
  positive}.  On the other hand, Eq.~(\ref{tc}) defines that
  rate of transverse energy given to the particle by its
  interactions with the plasma.  Thus $dE_\perp / dt$ is 
  {\it positive} as the projectile slows down.  Moreover, 
  Eq.~(\ref{avevdot}) describes the change in the average
  velocity of the projectile and so 
  $\langle {\bf v} \rangle \cdot d \langle {\bf v} \rangle / dt$ 
  is {\it negative} as the projectile slows.} 
Equation~(\ref{Qgiven}) gives
\begin{equation}
  {dE_{\perp\,b} \over dt} = \left[ \beta_b v^l C^{kl}_b -
  \frac{\partial C^{kl}_b}{\partial p^l} \right]  \,
  \frac{\partial E_\perp}{\partial p^k} - C^{kl}_b \,
  \frac{\partial^2 E_\perp}{\partial p^k \partial p^l} \ .
\label{debperp}
\end{equation}
Since the distribution is sharply peaked about the direction 
$\hat{\bf v}_p$, we can substitute $\hat{\bf v} \to \hat 
{\bf v}_p$ in Eq.~(\ref{debperp}). The first term on the right
hand side of Eq.~(\ref{debperp}) involving a single derivative
vanishes, giving a contribution only from the second term 
with two derivatives:
\begin{equation}
\left.  {dE_{\perp\,b} \over dt} \right|_{\rm F-P} =
C^{kl}_b\,\left( \delta^{kl} - \hat v^k_p \hat v^l_p \right)
{1 \over m_p} \,  = { (\nu -1) \over 2 m_p} \, {\cal B}_b \,.
\label{tc}
\end{equation}
Here we have 
placed an F-P designation on the final result because we will
later evaluate this transfer rate exactly within our general
order of calculation.  The comparison of the two results 
provides a signal for the breakdown of the Fokker-Planck equation
when larger angle collisions become important, collisions
that are not accurately described by the Fokker-Planck
approximation. As should be expected, we find in
Section~\ref{valid} that the difference between the Fokker-Planck
evaluation (\ref{tc}) of the rate of transverse energy transfer
and the exact rate to our order of accuracy has no large Coulomb
logarithm.  That is, the difference is of relative order of
one-over the Coulomb logarithm.  We should also hasten to mention
that in general the transverse spreading of a particle beam is a
small effect and so, in general, the transverse error is a small
error in a small effect. 

\subsubsection{Velocity Fluctuations}

We have determined the Fokker-Planck coefficient functions 
$C^{kl}_b$ by the conditions that they correctly describe the
rate of energy and momentum transfer between the background
plasma components $b$ and an arbitrary distribution of test
particles or projectiles.  An alternative approach was
emphasized some time ago by Rosenbluth, MacDonald, and Judd 
\cite{RMJ}.  To make contact with this line of development, 
we consider a distribution function $f$ peaked at the phase 
space point $({\bf r}_p,{\bf p}_p)$, but with finite width. 
Then, for the averages defined by Eq.~(\ref{ave}), 
the momentum transfer rate (\ref{pp}) implies that
\begin{equation}
{d \over dt} \, \langle v^k \rangle  = 
- { 1 \over m_p } \, {\sum}_b 
\left[ \beta_b v_p^l - \frac{1}{m_p}
{\partial \over \partial v_p^l}
                 \right] \, C_b^{kl} \ ,
\label{avevdot} 
\end{equation}
where the scattering tensor $C^{kl}_b$ is evaluated at $({\bf r}_p,
{\bf p}_p)$. This result requires only that the spatial extent of 
the projectile distribution $f$ is small in comparison with the 
scale over which the
plasma properties vary so that $C_b^{kl}$  is adequately
evaluated at the mean position ${\bf r}_p$.  Since our concern
here is with velocity variations, we shall simplify the discussion
by assuming that the background plasma is spatially
uniform so that the spatial coordinate dependence of $C_b^{kl}$ 
can be entirely
neglected.  More to the point, the result requires that the
squared spread in velocity $\Delta v^2$ is small in comparison
with the squared velocity $v^2$ itself, so that $C_b^{kl}$ may be
evaluated at the average $\langle {\bf v} \rangle = {\bf v}_p $.  
Here and in subsequent
work, we neglect the distinction between ${\bf v}_p$ and 
$\langle {\bf v} \rangle$ except when we are specifically
examining the velocity fluctuations.  [Typically, we will have an
equation with $ \langle v^l \cdots \rangle$ on the left and some
function $F(v_p)$ on the right, by which  we implicitly mean 
$F(\langle v \rangle ) $.] 

We define
\begin{equation}
\Delta v^k = v^k - \left\langle v^k \right\rangle \,,
\end{equation}
and next examine
\begin{equation}
\left\langle \Delta v^k \, \Delta v^l \right\rangle
=  \left\langle  v^k \, v^l \right\rangle -
\left\langle v^k \right\rangle \, \left\langle v^l \right\rangle
\,.
\end{equation}
The general relations 
(\ref{Qgiven}) and (\ref{aveqdot}) together with the previous
result (\ref{avevdot}) and a little algebra, show
that\footnote{The the time rates of change of the average
velocity (\ref{avevdot}) and velocity fluctuation
(\ref{avevvdot}) are called ``diffusion coefficients'' by Spitzer
\cite{Spit}.  Our method provides an unambiguous and precise
evaluation of these to the $g^2 [\ln g^2 + C]$ order to which 
we work.}
\begin{equation}
{ d \over dt} \, \langle \Delta v^k \, \Delta v^l \rangle 
= { 2 \over m_p^2} \, {\sum}_b \,
C_b^{kl} \,.
\label{avevvdot}
\end{equation}

Writing the Fokker-Planck equation (\ref{fp}) as
\begin{eqnarray}
\left[ {\partial \over \partial t} + {\bf v} \cdot \nabla \right]
f &=&  {\partial \over \partial p^k} \, 
{\partial \over \partial p^l} \, \left\{ {\sum}_b 
C^{kl}_b \, f \right\}
+ {\partial \over \partial p^k} \, \left\{  {\sum}_b 
\left[ \left( \beta_b v^l - {\partial \over \partial p^l} \right)
C^{kl}_b \right] \, f \right\} \,,
\nonumber\\
&&
\end{eqnarray}
demonstrates that it may be expressed as
\begin{eqnarray}
\left[ {\partial \over \partial t} + {\bf v} \cdot \nabla \right]
f &=&  {\partial \over \partial p^k} \, 
{\partial \over \partial p^l} \, \left\{
\left[ {m_p^2 \over 2} \, 
{ d \over dt} \, \langle \Delta v^k \, \Delta v^l \rangle \right]
 \, f \right\}
- {\partial \over \partial p^k} \, \left\{  
\left[ m_p {d \over dt} \, \left\langle v^k \right\rangle \right]
\, f \right\} \,.
\nonumber\\
&&
\end{eqnarray}
This is of the form advocated by Rosenbluth {\it et al.}
\cite{RMJ}, a form also described by Trubnikov \cite{tru}. 
These authors work with a completely arbitrary
background plasma, a plasma which has no aspects of thermal
equilibrium. Thus in their work, there is no relationship between
the rate of change of the average velocity and the squared
velocity fluctuation. In our
case, however, where we assume that the various plasma species $b$ are 
individually in thermal equilibrium at temperature $T_b = 1 /
\beta_b$, the vector 
$ d \, \left\langle v^k \right\rangle / dt $ 
and the tensor
$ d \, \langle \Delta v^k \, \Delta v^l \rangle / dt $
are, in view of (\ref{avevdot}) and (\ref{avevvdot}),
defined by the same $C_b^{kl}$ coefficients.  When the
different plasma species have the same temperature $T= 1 /
\beta$, then
there is a simple relation between the vector and tensor:
\begin{equation}
  {d \over dt} \, \left\langle v^k \right\rangle  = 
  - {m_p \over 2} \, \left[ \beta v_p^l - \frac{1}{m_p}\,
  {\partial \over \partial v_p^l} \right] \, {d \over dt} \,  
  \langle \Delta v^k \, \Delta v^l \rangle \ .
\label{VT}
\end{equation}

The coefficients $C^{kl}_b$ that appear in the Fokker-Planck
equation could be determined by the contributions of the various
plasma species to the rate of velocity fluctuations
$ d \, \langle \Delta v^k \, \Delta v^l \rangle / dt $ 
rather than by the rate of energy and momentum transfer that we
have chosen.  Such a fixing would give
\begin{eqnarray}
{d E \over dt} &=& - {m_p \over 2} \, \left[ {d \over dt} \,
\left\langle \Delta {\bf v}^2 \right\rangle + 2
\left\langle {\bf v} \right\rangle \cdot {d \over dt} 
\left\langle {\bf v} \right\rangle \right] 
\nonumber\\
&=& -{1 \over m_p} \, {\sum}_b \, C_b^{ll} +
{\sum}_b \, v_p^k \left[ \beta_b v_p^l -
\frac{1}{m_p}\,{\partial \over \partial v_p^l } 
\right] \, C^{kl}_b 
\nonumber\\
&=& {\sum}_b \, \left[ v_p - {1 \over \beta_b m_p} \,
{\partial \over \partial v_p^l } \, \hat v_p^l \right]
\, {\cal A}_b \,,
\label{loser}
\end{eqnarray}
which is precisely our previous determination of the 
${\cal A}_b$ coefficients.  The only change with the 
$C_b^{kl}$ determined by the velocity fluctuations instead of 
the energy and momentum exchange appears in the ${\cal B}_b$
functions.  They would be fixed (in $\nu = 3$ dimensions) by
\begin{eqnarray} 
{\sum}_b \, {\cal B}_b &=& m_p \sum_b { d E_{\perp\,b} \over dt} 
= m_p \, { d E_\perp \over dt} 
\nonumber\\
&=& {1 \over 2} m_p^2 \, 
\left( \delta^{kl} - \hat v_p^k \hat v_p^l \right) \, 
{d \over dt} \,
\left\langle \Delta v^k \Delta v^l \right\rangle \,.
\end{eqnarray}
As will be shown in Sec.~\ref{valid},
this determination differs from the one that we use by terms that
are relatively smaller by one over the large Coulomb logarithm.
Moreover, the ${\cal B}_b$ coefficients describe only very small
corrections to the motion of fast particles. 

We should stress yet again that we are {\em not} working just to
the leading order in the large Coulomb logarithm, but that we
compute exactly the constant terms under this logarithm as
well: We work to the order $ g^2 [\ln g^2 + C]$.  Our method of 
matching the energy and momentum flow
determines the coefficients $C_b^{kl}$ to this order with no
ambiguity.  If instead we would have chosen to match to the 
``diffusion coefficients''  
$d \, \left\langle \Delta v^k \Delta v^l \right\rangle / dt$, 
then we would obtain, to the order to which we work, exactly
the same coefficients ${\cal A}_b$ that determine the rate of
energy flow, but we would obtain slightly different coefficients 
${\cal B}_b$.  The coefficients ${\cal B}_b$ would differ in the
constants $C_b$ under the logarithm.  These different constants 
are determined by the work of Sec.~\ref{valid}. 

\section{Transverse Spreading, Longitudinal Straggling}

We turn now to investigate in more detail the nature and effects
of the spreading in velocity as a projectile moves in the
plasma.  Although the condition of a small
velocity spread can be imposed so that it is obeyed initially, it
may fail at later times. 

In principle, Eq.~(\ref{avevdot}) may be solved to determine 
$\langle {\bf v} \rangle$ as a function of time.  With this
solution inserted in $C_b^{kl}$, Eq.~(\ref{avevvdot}) may then,
in principle, be integrated to determine the fluctuation 
$
\langle v^k v^l \rangle - \langle v^k \rangle \langle v^l \rangle 
$ 
as a function of time.  This procedure remains valid so long as
the fluctuation remains small in comparison with 
$\langle v \rangle^2 $.  

The previous decomposition (\ref{sstructtt}) evaluated at
$\langle v^k \rangle = v_p^k$, 
\begin{equation}
C^{kl}_b = {\cal A}_b \, {\hat v_p^k \, \hat v_p^l \over \beta_b \, v_p} 
+ {\cal B}_b {1 \over 2} \left( \delta^{kl} - \hat v_p^k \hat v_p^l
\right) \,,
\label{decomp}
\end{equation}
shows that the coefficients ${\cal A}_b$ determine the
longitudinal velocity spread --- the `straggling' --- while the
coefficients ${\cal B}_b$ describe the transverse velocity 
spreading,
\begin{equation}
{ d \over dt} \, \left[ 
\langle v^k v^l \rangle - \langle v^k \rangle \langle v^l \rangle 
\right]_L = \hat v_p^k \, \hat v_p^l \, {2 \over m_p^2} \, {\sum}_b 
{\cal A}_b \, {1 \over \beta_b v_p} \,,
\label{longitudinal}
\end{equation}
and
\begin{equation}
{ d \over dt} \, 
\langle v^k v^l \rangle_T 
= \left( \delta^{kl} - \hat v_p^k \, \hat v_p^l \right)
\, {1 \over m_p^2} \, {\sum}_b 
{\cal B}_b  \,,
\label{transverse}
\end{equation}
Here we should note again that by virtue of the isotropy of the
plasma, the average velocity $\langle {\bf v} \rangle = {\bf v}_p$ 
always points along the initial velocity direction $\hat {\bf v}_p$
(while the magnitude of ${\bf v}_p$ changes with time, its direction
remains fixed), and so
the average never has a transverse component.

The transverse spreading (\ref{transverse}) can be expressed as
\begin{equation}
  \frac{d}{dt}\,
  \langle v^k v^l \rangle_T 
= \left( \delta^{kl} - \hat v_p^k \, \hat v_p^l \right)
\, {1 \over m_p} \, { dE_\perp \over dt} \,.
\end{equation}
For {\em very} fast projectiles, the results (\ref{alimmm}) and
(\ref{Bad}) show that 
\begin{eqnarray}
m_p v_p^2 \gg \frac{m_p}{m_e}\,T \,: \qquad\qquad 
{d E_\perp \over dt} 
&\approx& {m_e \over m_p} \,  {dE \over dt} \,,
\end{eqnarray}
and so
\begin{eqnarray}
m_p v_p^2 \gg \frac{m_p}{m_e}\,T\,: \qquad\qquad 
  \langle v_p^k v_p^l \rangle_T 
&\approx& \left( \delta^{kl} - \hat v^k \, \hat v^l \right) \,
{m_e \over m_p} \,  {1 \over m_p} \, \left( E_0 - E \right) \,.
\end{eqnarray}
Thus at high energies, the transverse angular spreading is of order 
$\sqrt { m_e / m_p} $.  This is a small number for ion projectiles,
but of course it is not small for electron projectiles.  When an
ionic projectile slows down to thermal velocities, the transverse
velocity fluctuations must become of order of the thermal
velocity $\bar v_p=\sqrt{3T/m_p}$. However, until thermal velocities 
are reached, the transverse spreading for ions is always small.

To assess the nature of the longitudinal fluctuations, the straggling, 
in a simple way, we shall assume that all the plasma species are at a
common temperature, $\beta^{-1}_b = T$. First we examine  the
motion of a fast projectile so that Eq.~(\ref{loser}) simplifies to
\begin{eqnarray}
m_p v_p^2 \gg T \,: \qquad\qquad
{ d E \over dt}  &=& v_p \, {\sum}_b {\cal A}_b \,.
\end{eqnarray}
Thus, since $ E = m_p v_p^2 /2 $, the rate of longitudinal spreading
(\ref{longitudinal}) can be written as
\begin{eqnarray}
m_p v_p^2 \gg T \,: \qquad\qquad
{ d \over dt} \, \left[ 
\langle v^k v^l \rangle - \langle v^k \rangle \langle v^l \rangle 
\right]_L = - \hat v_p^k \, \hat v_p^l \, {T \over m_p} \, 
{ 1 \over E } {d E \over dt} \,,
\label{toolong}
\end{eqnarray}
which integrates to 
\begin{eqnarray}
m_p v_p^2 \gg T \,: \qquad\qquad
 \left[ 
\langle v^k v^l \rangle - \langle v^k \rangle \langle v^l \rangle 
\right]_L &=&  \hat v_p^k \, \hat v_p^l \, {T \over m_p}  \, 
\ln \left( { E_0 \over E} \right) \,,
\end{eqnarray}
where the $0$ subscript denotes the initial value.  In thermal
equilibrium at temperature $T$, the projectile has a
root-mean-square velocity 
$ \bar v_p = (3 T / m_p)^{1/2}$.  Thus the straggling result may be
written as
\begin{eqnarray}
 v_p \gg \bar v_p \,: \qquad\qquad
 \left[ 
\langle v^k v^l \rangle - \langle v^k \rangle \langle v^l \rangle 
\right]_L &=&  \hat v_p^k \, \hat v_p^l  \, { {\bar v_p}^2 \over 3 } \, 
\ln \left( { E_0 \over E } \right) \,,
\label{stragle}
\end{eqnarray}
As the projectile slows to its thermal velocity $\bar v_p$, its
straggling fluctuations become of order $\bar v_p$ as are those of
all plasma particles.

The total rate of energy loss is given by
\begin{equation}
{d E \over dt} = - {1 \over 2} \, m_p \left[ 
{d \over dt} \langle {\bf v}^2 \rangle_L +
{d \over dt} \langle {\bf v}^2 \rangle_T \right] \,.
\label{reallong}
\end{equation}
The right hand side of Eq.~(\ref{toolong}) can be taken to vanish
in the high energy limit in which terms of order $T/E$ can be
neglected. Hence with the neglect of terms of this order, 
Eq's.~(\ref{toolong}) and (\ref{reallong}) imply
that\footnote{This result also follows from Eq.~(\ref{VT}) as it
must.} (since the
direction of $\langle {\bf v} \rangle$ is constant in time)
\begin{eqnarray}
m_p v_p^2 \gg T \,: \qquad\qquad
{d \over dt} \langle {\bf v} \rangle = - { {\bf v}_p \over m_p 
v_p^2}\, \left[ {d E \over dt} + {d E_\perp \over dt} \right] \,.
\label{veldot} 
\end{eqnarray}
As we have just discussed, for ions $d E_\perp / dt$ 
is of relative order $m_e /m_p $ and can be neglected, so that the
energy loss rate completely determines the slowing down of the
particle.  

Since the logarithm in Eq.~(\ref{stragle}) is a slowly varying 
function, 
as the projectile slows down from very high velocities to
speeds that are more nearly of the order of the thermal velocity,
it acquires velocity fluctuations in the longitudinal  direction 
that are only slightly larger than $\bar v_p$.  This
justifies the integration of the slowing down equations 
(\ref{reallong}) and (\ref{veldot}) from very
high velocities to just above thermal speed using the average
velocity $\langle v \rangle =v_p$, since the
velocity spreading is relatively very small.  However, this
simple picture of the essentially deterministic motion of an
individual particle breaks down when the particle speed
approaches the thermal speed of a particle of its mass.  In this
region, a statistical distribution of particles must be employed
as the proper description.  The Fokker-Planck equation can be
used to describe the time evolution of the phase space density 
$f({\bf r},{\bf p},t)$, and our computation of the coefficients 
$C_b^{kl}$ that enter into the Fokker-Planck equation remain
valid.  What breaks down is the single-particle description of a
particle losing well defined amounts of energy and momentum. 

What we have just said means that whenever $ 1 / \beta_b m_p 
v_p^2$ corrections become important, the notion of a well-defined
projectile trajectory breaks down.  Hence the corrections given
in the second part of Eq.~(\ref{aaa}) are never relevant for
the description of a single particle which, in any relevant
region, is described by 
\begin{eqnarray}
m_p v_p^2 \gg T \,: \qquad\qquad 
{dE_b \over dt} = v_p {\cal A}_b \,,
\label{overk}
\end{eqnarray}
which also determines the complete motion of the particle.
Nonetheless, we have used the form (\ref{overk}) in all of our 
calculations of $dE/dx = (1/v_p)\,dE/dt$.  

We have noted that the transverse spreading of an electron
projectile may be significant, even for fast particles.  To
emphasize this point, we quote the high speed limits for the
total and perpendicular energy loss of an electron that follow
from Eq's.~(\ref{alimmm}) and (\ref{Bad}):
\begin{eqnarray}
m_e v_p^2 \gg T \,: \qquad\qquad 
{dE \over dt} =
  {e^2 \over 4\pi} \,
   { \omega_e^2  \over  v_p } \,
  \ln \left( { m_e v_p^2 \over \hbar \omega_e}
  \right)  \,,
\end{eqnarray}
and
\begin{eqnarray}
m_e v_p^2 \gg T \,: \qquad\qquad 
{d E_\perp \over dt} = {\sum}_b \, {e^2 \over 4 \pi} \,
{\kappa_b^2 \over \beta_b m_e v_p} \, \ln \left( 
{2 m_{eb} v_p \over \hbar \kappa_\smD} \right) \,.
\end{eqnarray}
Here $\omega_e$ is the electron plasma frequency defined by 
$ \omega^2_e = e^2 n_e / m_e$ while 
$ \kappa_b^2 = e_b^2 n_b / T_b $.  The general case is
sufficiently well illustrated by the specific case of a fully
ionized hydrogen plasma with equal numbers of electrons and
protons, $n_e = n_p$, and equal proton-electron temperatures, 
$T_e = T_p = T$.  The reduced masses are given by 
$m_{ee} = m_e /2$ and, with the neglect of the small electron
proton mass ratio, $m_{ep} = m_e$.  Hence
\begin{eqnarray}
m_e v_p^2 \gg T \,: \qquad\qquad &&
\nonumber\\ 
{d E_\perp \over dt} &=&  {e^2 \over 4 \pi} \,
{\omega_e^2 \over  v_p} \, \ln \left( 
{2 m_e^2 v_p^2 \over \hbar^2 \beta e^2 2 n_e } \right) 
\nonumber\\[5pt]
&=&  {e^2 \over 4 \pi} \,
{\omega_e^2 \over  v_p} \, \ln \left( 
{m_e v_p^2 \over \hbar \omega_e } \,  
{ T \over \hbar \omega_e } \right) \,.
\end{eqnarray}
For the plasma parameters that we have used above, 
$ \ln ( T / \hbar \omega_e ) $ 
is not a large (or small) number.  Hence 
$ d E_\perp / dt$ is about the same size as 
$ d E / dt$, and so the the electrons do not slow down along a
straight line.  The electron motion in a plasma
requires the use of a Fokker-Planck description of an ensemble of
particles.

\section{Validity Range of the Transport Equation}
\label{valid}

The exact --- to our order --- rate at which the transverse
energy of a projectile increases is the sum of the 
leading $\nu < 3$ result computed from the Lenard-Balescu equation
plus the leading $\nu > 3$ result computed from the Boltzmann
equation. The difference of this with the evaluation (\ref{tc}) 
given by the Fokker-Planck equation provides a signal for the
breakdown of the Fokker-Planck description.  This measure is
\begin{equation}
\Delta_{b} = 
\left. {d E_{\perp\,b} \over dt } \right|_{\rm exact}
- \left. {d E_{\perp\,b} \over dt } \right|_{F-P} \,.
\label{diffen}
\end{equation}

It is worthwhile providing here the results of this assessment, 
detailed in Sec.~\ref{details} below, so as 
to conclude our general review in a unified manner.  But before 
presenting these results, some general remarks may help clarify 
what we are doing.  We have used the two functions 
$ {\bf v} \cdot d{\bf P}_b / dt $ and $ dE_b / dt $ as inputs to
determine the two scalar coefficients ${\cal A}_b$ and 
${\cal B}_b$ that define the transport tensor $C^{kl}_b$ that 
appears in the Fokker-Planck equation.  Then the time rate of
change of the perpendicular energy, $ dE_\perp / dt $, may be
found from $C^{kl}_b$, a determination that we denote by the
$F-P$ label.  The point is that, within the Fokker-Planck
approximation, only two of the three functions 
$ {\bf v} \cdot d{\bf P}_b / dt $, $ dE_b / dt $, and 
$ dE_\perp / dt $ are independent 
 functions.  However, if
the Fokker-Planck approximation is not made, then these three
functions are linearly independent functions.  The difference 
(\ref{diffen}) is thus a measure of the error in the Fokker-Planck
description. 

Since the $\nu < 3$ contribution to the Fokker-Planck coefficient
$C^{kl}_b$ is the same Lenard-Balescu equation that is used to
evaluate the transverse energy in this region, the difference
defining $\Delta_{b}$ is given by just the $\nu > 3$ parts,
\begin{equation}
\Delta_{b} = 
\left. {d E^\smgt_{\perp\,b} \over dt } \right|_{\rm exact}
- \left. {d E^\smgt_{\perp\,b} \over dt } \right|_{F-P} \,,
\end{equation}
with both terms computed from the scattering cross section
formula that is equivalent to the Boltzmann equation as is
described in Sec.~\ref{big}. This computation is given 
in detail in Sec.~\ref{details}, with the result 
(\ref{TspreadT}) that 
\begin{equation}
\Delta_{b} = - {e_p^2 \over 4\pi} \, { \kappa^2_b \over 2 m_p}
\, \left({ m_b \over 2\pi \beta_b} \right)^{1/2} \,
\int_0^1 {du \over \sqrt u} \, [1 - 3u] \, \
\exp\left\{ - {1\over2} \beta_b m_b v_p^2 \, u \right\} \,.
\label{Tspread}
\end{equation}

In the high speed limit, $ m_b v_p^2 / 2 \gg T_b$, 
the exponential is highly damped.  Hence, in this limit, the
upper integration limit may be extended to $u \to \infty$, and
one finds that
\begin{eqnarray} 
v_p \to \infty \,: &&
\nonumber\\
&&
\Delta_{b} \to - {e_p^2 \over 4\pi} \, 
{ \kappa_b^2 \over 2 \beta_b m_p v_p } \,.
\end{eqnarray}
No ``Coulomb logarithm'' appears here because the difference 
$\Delta_b$ is not sensitive to small angle scattering or,
equivalently, to large distance collisions. 
 
In the low speed limit, the exponential may be expanded in 
powers of $v_p^2$.  The integrations involved in the zeroth 
order term vanish, and one finds that
\begin{eqnarray}
v_p \to 0 \,: &&
\nonumber\\
&&
\Delta_{b} \to - {e_p^2 \over 4\pi} \, 
{2 \kappa_b^2 \over 15 m_p } \, 
\left({ m_b \over 2\pi \beta_b} \right)^{1/2} 
\,\, \beta_b m_b v_p^2  \,.
\label{lowlim}
\end{eqnarray}

The definition of 
$E_{\perp\,b}$ involves the tensor $ \hat v_p^k \hat v_p^l$, 
which is undefined as $v_p \to 0$. Therefore it must be
accompanied an additional factor of $v_p^2$ (as $v_p \hat 
v_p^k = v_p^k$), thereby giving a well defined and quadratically 
vanishing tensor $v_p^k v_p^l$.

These limits can be used to assess the validity of the
Fokker-Planck evaluation
\begin{equation}
\left. { d E_{\perp\,b} \over dt } \right|_{F-P} = 
{1 \over m_p} \, \left[ C^{ll}_b - {\cal A}_b 
	{1 \over \beta_b v_p} \right] =
		{1 \over m_p} \, {\cal B}_b \,.
\label{fpT}
\end{equation}
The high-velocity limit (\ref{Bad}) for 
${\cal B}_b$ gives
\begin{eqnarray}
v_p \to \infty \,: &&
\nonumber\\
&&
\left. { d E_{\perp\,b} \over dt } \right|_{F-P}
 = {e_p^2 \over 4 \pi} \,
{\kappa_b^2 \over \beta_b m_p v_p} \, \ln \left( 
{2 m_{pb} v_p \over \hbar \kappa_\smD} \right) \,.
\end{eqnarray}
As should have been expected, this is larger than the difference 
$\Delta_b$ by the logarithmic factor
$$
2 \, \ln\left( {2 m_{pb} v_p \over \hbar \kappa_\smD} \right) \,.
$$ 
As far as the transverse spreading is concerned, this shows, 
at least at high energy, that the Fokker-Planck description 
is valid only to leading logarithmic order:  The validity of the
transverse spreading given by the Fokker-Planck equation is 
valid to the accuracy to which the Coulomb logarithm is large
in comparison to unity.  It must immediately be remarked,
however, that, in the high-energy limit, the rate of energy loss
\begin{eqnarray}
v_p \to \infty \,: &&
\nonumber\\
&&
{ d E \over dt } = v_p \, {\cal A}_e = 
  {e_p^2 \over 4 \pi} \,
{\kappa_e^2 \over \beta_e m_e v_p} \, \ln \left( 
{2 m_{pe} v_p^2 \over \hbar \omega_e} \right)
\end{eqnarray}
is much larger that the rate of transverse spreading, larger by
the very large ion/electron mass ratio $m_b / m_e $.  Thus the
spreading entails very small angles, and one can tolerate a
rather large error in this small effect. 

In general, the error $\Delta_b$ is smaller than the transverse
energy
spread itself by a factor of one over the Coulomb logarithm.  To
see this, we may, for example, turn to Eq.~(\ref{wonderclassic}).
It contains a logarithm whose factor $K$ in its argument cancels
against that in another contribution to yield the large Coulomb
logarithm [with a quantum or classical cutoff as is appropriate
to the velocity $v_p$.]  Thus the leading Coulomb logarithm 
contribution to the Fokker-Planck transverse energy rate (\ref{fpT}) 
is given by
\begin{equation}
\left. { d E_{\perp\,b} \over dt } \right|_{F-P \, log} = 
{e_p^2 \kappa_b^2 \over 4\pi } \, \left( { m_b \over 2\pi
\beta_b} \right)^{1/2} \, { L_b \over m_p } \, 
\int_0^1 du \, u^{-1/2} \, (1-u) \, 
\exp\left\{ - {1 \over2} \beta_b m_b v_p^2 \, u \right\}
\,,
\end{equation}
in which $L_b$ is the appropriate Coulomb logarithm for this
plasma species $b$.  This is indeed larger than $\Delta_{b}$ 
[Eq.~(\ref{Tspread})] by essentially the factor $L_b$.

The result (\ref{Blim}) gives the small velocity limit
of the Fokker-Planck approximation (\ref{fpT}):
\begin{eqnarray}
  v_p &\to& 0 \,:
\nonumber\\
\left. { d E_{\perp\,b} \over dt } \right|_{F-P}  
&=& -
 {e_p^2 \kappa_b^2\over 4\pi} \, 
  \left({ m_b \over  2\pi\beta_b} \right)^{1/2}
  \, {4 \over 3 m_p } \,   
  \left[  \ln\left(\beta_b  { e_p e_b \over 16 \pi} \,
  \kappa_\smD { m_b \over m_{pb} } \right) + {1\over2} + 
  2\gamma \right] \,.
\label{fpTslow}
\end{eqnarray}
This result is in accord with the comments of footnote
\ref{note}, which explains why
\begin{equation}
v_p \to 0 \,: \qquad C^{kl}_b \to {\rm const} \cdot \delta^{kl} 
= \delta^{kl} \, {\cal B}_b \, / 2 \,.
\label{cklsmallv}
\end{equation} 
Using Eqs.~(\ref{sp}) and (\ref{cklsmallv}), along with
Eq.~(\ref{tc}), in an arbitrary number of spatial dimensions 
$\nu$ we find
\begin{eqnarray}
v_p \to 0 \,: \qquad &&
\nonumber\\
&& { dE_b \over dt} = -{\nu \over 2m_p } \, {\cal B}_b \,,
\qquad\qquad
 { dE_{\perp\, b} \over dt} = {\nu -1 \over 2m_p } \, {\cal B}_b \,,
\end{eqnarray}
and hence 
\begin{equation}
v_p \to 0 \,: \qquad\qquad {d E_{\perp\,b} \over dt} =
-{\nu - 1 \over \nu} \, {dE_b \over dt} \,.
\end{equation}
Our sign conventions, in which $dE_b/dt$ is the rate of
energy {\it loss} of plasma species $b$, while $dE_{\perp\,b}$ 
is rate of transverse energy {\em gain} of species $b$, dictate
the relative minus sign above. 
The low speed limit of $d E_b / dt$ for $\nu = 3$, as computed by
Eq's.~(\ref{aaa}) and (\ref{lessreglimm}), is indeed just 
the factor $(3/2)$ times the result (\ref{fpTslow}) for 
$dE_{\perp\,b} / dt$. On other other hand, the result 
(\ref{fpTslow}) is quite different than the low speed limit 
(\ref{lowlim})  of the error $\Delta_{b}$ which behaves as 
$v^2_p$ when $v_p \to 0$, not as a constant as given in 
Eq.~(\ref{fpTslow}).  Thus, the Fokker-Planck equation gives 
a very accurate description of the transverse spreading of
low velocity particles.  

We turn now to the details of our calculation.

\section{Long Distance Effects Dominate When $\mmbf{\nu <3}$}
\label{small}

When the spatial dimensions are less than three,
long-distance, collective effects are dominant. 
This ``soft physics'' is described to leading 
order in the plasma density by the 
Lenard\cite{len}--Balescu\cite{bal} 
equation.\footnote{
\baselineskip 15pt
Again we note that  Refs.~\cite{clem,dup,nich} contain well 
written expositions.} Indeed, to leading order
in the density, one can prove that the rigorous 
BBGKY hierarchy reduces to the Lenard--Balescu 
equation when \hbox{$\nu < 3$}. This can be 
demonstrated, for example, by carefully examining 
the discussion given in Nicholson\cite{nich} or in 
Clemmow and Dougherty\cite{clem}.  It is significant 
that the proof of the reduction of the BBGKY hierarchy 
to the Lenard--Balescu equation breaks down at 
precisely \hbox{$\nu = 3$}: this happens because 
of the  appearance of short-distance, ultra-violet 
divergences, which are absent in dimensions less than
three.

The Lenard--Balescu equation for the case of interest in which
each background plasma species $b$ is in thermal equilibrium and
described by a Maxwell-Boltzmann distribution at temperature 
$T_b = 1 / \beta_b$ is of the Fokker-Planck\footnote{
\baselineskip 15pt
Although this is a purely classical result, the factor 
of $\hbar^{-\nu}$ in the measure
$d^\nu {\bf p}_b/(2\pi\hbar)^\nu$
 is used to convert the momentum integral of the 
dimensionless phase-space distribution $f_b({\bf p}_b)$ 
into a particle number density.} form (\ref{fp}), using
\begin{eqnarray}
  C^{lm}_b &=& e_p^2 e_b^2 \int { d^\nu{\bf k}
  \over (2\pi)^\nu } {  k^l k^m \over ({\bf k}^2)^2  }
  { \pi \over \left| \epsilon ({\bf k} , {\bf v}_p \cdot
  {\bf k} ) \right|^2 }
  \int { d^\nu{\bf p}_b \over (2\pi\hbar)^\nu } \,
  \delta\left({\bf k} \cdot {\bf v}_p - {\bf k} \cdot
  {\bf v}_b \right) \,  f_b({\bf  p}_b) \,,
\label{lbj}
\end{eqnarray}
with\footnote{We now use the subscript $p$ to distinguish the 
projectile velocity ${\bf v}_p$ (which we previously simply
denoted by the unadorned ${\bf v}$) from the velocities 
${\bf v}_b $ of the background plasma particles of species $b$.}   
 ${\bf v}_p = {\bf p}_p / m_p $ and ${\bf v}_b = 
{\bf p}_b / m_b$  the velocities of the projectile and of
the background
plasma species $b$.  The collective behavior 
of the plasma enters through its dielectric function 
$\epsilon({\bf k},\omega)$. For a dilute plasma, the 
case to which the Lenard-Balescu equation applies, 
the dielectric function is given
by\footnote{See, for example, 
Section~29 of Ref.~\cite{Lifs}.}
\begin{equation}
  \epsilon({\bf k},\omega) = 1 + {\sum}_c \, {e_c^2
  \over k^2 } \int { d^\nu {\bf p}_c \over
  ( 2\pi\hbar )^\nu } { 1 \over \omega - {\bf k} \cdot 
  {\bf v}_c + i \eta}\, {\bf k} \cdot {\partial
  \over \partial {\bf p}_c } f_c({\bf p}_c) \,,
\label{epsilon}
\end{equation}
where the prescription $ \eta \to 0^+ $ is implicit
and defines the correct retarded response. 

\subsection{Projectile Motion in an 
Equilibrium Plasma}

Taking the projection (\ref{aproj}) of Eq.~(\ref{lbj}) 
and setting ${\bf k} \cdot \hat{\bf v}_p = k \cos\theta$, or
alternatively taking the trace of the tensor indices in 
Eq.~(\ref{lbj}),  yields
\begin{eqnarray}
 && \left\{ {\cal A}^\smlt_b \, {1 \over \beta_b v_p} 
\,\,,\,\, C_b^{ll \, \smlt} \right\}
\nonumber\\
&& \qquad
= e_p^2 e_b^2 \, \int { d^\nu{\bf k}
  \over (2\pi)^\nu } { 1 \over k^2  }
  { \pi \over \left| \epsilon ({\bf k} , {\bf v}_p \cdot
  {\bf k} ) \right|^2 }
  \int { d^\nu{\bf p}_b \over (2\pi\hbar)^\nu } \,
  \delta\left({\bf k} \cdot {\bf v}_p - {\bf k} \cdot
  {\bf v}_b \right) \,  f_b({\bf  p}_b) \, \left\{ \cos^2\theta 
\,,\, 1 \right\}  \,.
\nonumber\\
&&
\label{givesa}
\end{eqnarray}
Here the less-than superscripts on 
${\cal A}^\smlt_{b}$ and $C^{ll \, \smlt}_{b}$
are written to make it explicit that 
we are now working in spatial dimensions strictly
less than three.  Using the delta function in Eq.~(\ref{givesa})
to remove the component of ${\bf p}_b = m_b {\bf v}_b $
 along the ${\bf k}$ direction, and then integrating out
the remaining $\nu -1$ components of ${\bf p}_b$ using the
Maxwell-Boltzmann distribution
\begin{eqnarray}
f_b({\bf p}_b) 
= n_b \left(\frac{2\pi \hbar^2 \beta_b}
  {m_b}\right)^{\nu/2} \exp\left\{- {\beta_b \over 2} \, 
  m_b v_b^2 \right\}
\label{M-B}
\end{eqnarray}
reduces Eq.~(\ref{givesa}) to
\begin{eqnarray}
 && \left\{ {\cal A}^\smlt_b \, {1 \over \beta_b v_p} 
\,\,,\,\, C_b^{ll \, \smlt} \right\}
\nonumber\\
&& \qquad
   = e_p^2 \,  
	\int { d^\nu{\bf k} \over (2\pi)^\nu }
 \sqrt{ m_b \over 2\pi \beta_b }  \,
	\exp\left\{ - {\beta_b \over 2} m_b \, 
 v_p^2 \cos^2\theta \right\} \,
{ \kappa_b^2 \, \pi \, k  \over \left| 
k^2 \epsilon({\bf k}, k v_p \cos\theta) 
\right|^2} \left\{ \cos^2\theta \,,\, 1 \right\} \,,
\nonumber\\
&&
\label{lesslesslossy}
\end{eqnarray}
where
\begin{equation}
  \kappa^2_b = \beta_b \, e_b^2 \, n_b
\end{equation}
is the contribution of species $b$ to 
the squared Debye wave number.

To work out this result, we first note that
the structure of the dielectric function 
(\ref{epsilon}) can be simplified.  
We use the explicit Maxwell-Boltzmann 
form for the distribution function $f_c({\bf p}_c)$
to compute the derivative in Eq.~(\ref{epsilon}) 
and then integrate out the momentum components of 
${\bf p}_c$ that are perpendicular to ${\bf k}$. 
This gives the structure 
\begin{equation}
  k^2 \, \epsilon( k , k v_p \cos\theta ) =
  k^2 + F(v_p \cos\theta)  \,.
\label{struct}
\end{equation}
The $F$ function appears in the form of a 
dispersion relation
\begin{equation}
  F(u)  =  - \int_{-\infty}^{+\infty} dv \, 
{ \rho_{\rm total}(v) \over u
  - v + i \eta } \,,
\label{disp}
\end{equation}
with the spectral weight
\begin{equation}
  \rho_{\rm total}(v)  = {\sum}_c \, 
	\rho_c(v) \,,
\label{efrelb}
\end{equation}
where
\begin{equation}
\rho_c(v) = 
\kappa^2_c \, v \, \sqrt{ \beta_c m_c \over 2\pi } 
  \exp\left\{
  -{1 \over 2} \beta_c m_c v^2 \right\} \,.
\label{spectral}
\end{equation}
For future use, we
note that $F$ satisfies the relations
\begin{eqnarray}
\label{fref}
  F(-u) = F^*(u)
\end{eqnarray}
and
\begin{equation}
\label{fmfstar}
{\rm Im} \, F(u) = {1 \over 2i} 
\left[ F(u) - F^*(u) \right] = 
  \pi  \rho_{\rm total}(v) \,.
\label{ImF}
\end{equation}

As a first application of this structure of the 
dielectric function, we note that for a plasma 
all of whose components
have the same temperature, the total energy
loss reads
\begin{equation}
  {d E^\smlt \over dx} = {\sum}_b \,  
	{d E^\smlt_b \over dx} 
=
\left[1 - {T \over m_p v} \, 
	{\partial \over \partial v^l} \, \hat v^l \right]
		{\cal A}^\smlt \,,
\label{ddd}
\end{equation}
where
\begin{equation}
{\cal A}^\smlt = {\sum}_b \, {\cal A}^\smlt_b \,.
\end{equation}
Since
\begin{eqnarray}
&&
{\sum}_b  \sqrt{ \beta_b \, m_b \over 2\pi}  \,
	\exp\left\{ - {\beta_b \over 2} m_b \, 
 v_p^2 \cos^2\theta \right\} \,
{ \kappa_b^2 \, \pi \, k \, v_p \cos\theta 
 \over \left| 
k^2 \epsilon({\bf k}, k v_p \cos\theta) 
\right|^2} 
\nonumber\\[2 pt]
&& \qquad\qquad\qquad =
k \, { {\rm Im} \, F(v_p\cos\theta) \over 
\left| k^2 + F(v_p \cos\theta) \right|^2 }
\nonumber\\[2 pt]
&& \qquad\qquad\qquad =
- {1 \over k} \, {\rm Im} \,
{ 1 \over \epsilon(k, v_p\, k  \cos\theta) } \,,
\end{eqnarray}
we have
\begin{eqnarray}
  {\cal A}^\smlt
   &=& - e_p^2  \, 
\int { d^\nu{\bf k} \over  (2\pi)^\nu } 
 {1 \over k } \,
   \cos\theta \,
  {\rm Im} \, { 1 \over \epsilon( k , v_p k \,
  \cos\theta  ) } \,.
\label{lesslossyyy}
\end{eqnarray}
Except for the term involving the derivative in the energy loss
formula (\ref{ddd}), 
Eq.~(\ref{lesslossyyy}) is just the energy loss to 
Joule heating the plasma,  the energy 
loss obtained 
by using Fourier transform techniques to compute
the volume integral of  $ {\bf j} \cdot {\bf E}$, 
where ${\bf j}$ is the current of a point particle 
moving with velocity ${\bf v}_p$ and ${\bf E}$ is 
the electric field produced by this current. The
additional term involving the derivative provided 
by the correct Lenard-Balescu transport equation 
ensures that the total energy loss vanishes for 
a swarm of particles with a thermal distribution 
of velocities at temperature $ T = 1 / \beta$.

\subsection{Calculating the Coefficients}

We use Eq.~(\ref{struct}) and the results that
follow it to place the energy loss coefficients
(\ref{lesslesslossy}) in the form
\begin{eqnarray}
\left\{ {\cal A}^\smlt_b \, {1 \over \beta_b v_p} 
\,\,,\,\, C_b^{ll \, \smlt} \right\}
   &=& {e_p^2 \, \pi \over \beta_b v_p } \, 
	\int { d^\nu{\bf k} \over (2\pi)^\nu }
 \, { k \, \rho_b(v_p\cos\theta) 
 \over \left| 
k^2 + F( v_p \cos\theta) \right|^2} 
\left\{ \cos\theta \,,\, {1 \over \cos\theta} \right\}
\,,
\end{eqnarray}
The wave number integration may be performed by 
passing to hyper-spherical coordinates. For 
functions depending only upon the radial coordinate 
$k$ and the polar angle $\theta$, we may write 
\begin{equation}
  \int { d^\nu {\bf k} \over (2\pi)^\nu}\,
  f(k,\theta) = { \Omega_{\nu-2} \over 
  (2\pi)^\nu } \, \int_0^\infty k^{\nu-1} 
  dk \int_0^\pi d\theta \,\sin^{\nu-2}
  \hskip-0.08cm \theta \, f(k,\theta) \,,
\end{equation}
where $\Omega_{\nu-2}$ is the solid angle
subtended by a $(\nu-2)$-dimensional 
sphere.\footnote{
\baselineskip 15pt
In general, 
by a ``$d$-dimensional sphere'' we mean a 
sphere whose hyper-surface is of dimension 
$d$, which can be thought of as a sphere 
embedded in $(d+1)$-dimensional 
Euclidean space. Points on such a sphere centered 
at the origin with unit radius satisfy 
$\sum_{\ell=1}^{d+1} x_\ell^2 =1$. The 
solid angle $\Omega_d$ is simply the 
surface area of this unit sphere, and 
it can be expressed as $\Omega_d =2 
\pi^{(d+1)/2}/\Gamma((d+1)/2)$.}
The $k$-integral in Eq.~(\ref{lesslesslossy}) 
is of the form
\begin{eqnarray}
\label{kint}
  I(\nu) \equiv 
  \int_0^\infty dk \, \frac{k^\nu}
  {\vert k^2 + F\vert^2} \,,
\end{eqnarray}
which is finite for $\nu < 3$ and log-divergent
at $\nu=3$. Despite the fact that one thinks in 
terms of integer dimensions, one is nonetheless 
free to perform the integral (\ref{kint}) treating 
$\nu$ as an arbitrary complex number. 
Moreover, the solid angle factor $  \Omega_{\nu-2} $
has an analytic form that extends to arbitrary 
complex dimensionality $\nu$, and the power
$\sin^{\nu-2}$ in the polar angular integration 
can also obviously be extended to arbitrary complex 
$\nu$. The whole expression for 
the energy loss rate can be extended  
to a space of arbitrary complex dimensionality 
$\nu$. The physical dimension  $\nu=3$ is, however, 
a singular point, namely a simple pole.  Nonetheless, we 
can {\em regularize} this infinity (that is, 
render it finite) by {\em formally} treating $\nu$ 
as complex number differing slightly from three. 
In any well-defined physical process, all terms that 
diverge in the $\nu \to 3$ limit must cancel among 
themselves. For the problem at hand, we will show
in the next section that short-distance scattering,
which has not yet been included, produces a divergence 
as $\nu \to 3$ that exactly cancels the aforementioned
divergence. This renders the experimentally
measurable energy and momentum loss finite in three dimensions. 

Let us now evaluate the integral (\ref{kint}). 
It is convenient to add 
and subtract a (well chosen) term so as to express
the integral as a sum of two pieces, the first 
having a  
divergence when $\nu \to 3$ but with no 
$\theta$ dependence, 
the second a finite term which does 
have the rather complicated $\theta$-dependence
of the function $F(v_p\cos\theta)$: 
\begin{eqnarray}
\label{integral}
  \int_0^\infty dk \, 
  { k^\nu \over \left| k^2 + F \right|^2 } 
  &=& 
  \int_0^\infty dk \, 
  { k^\nu \over  k^4 + K^4 }
  +\int_0^\infty k^3\,  dk \, \left[
  {1 \over  (k^2 + F ) (k^2 + F^* ) } - 
  {1 \over   k^4 + K^4 } \right] \,,
\nonumber\\
&&
\end{eqnarray}
where $K$ is an arbitrary ($\theta$-independent)
wave number. Since the final result cannot 
depend upon $K$ (as we have merely added 
and subtracted the same $K$-dependent quantity), 
we can choose its value as a  matter of convenience. 
The first term of Eq.~(\ref{integral}) is 
$\theta$-independent and divergent 
as $\nu \to 3$; the second term is finite 
in this limit (so we have taken $\nu=3$), but 
its $\theta$ dependence is non-trivial. 

The first integral on right hand side of 
Eq.~(\ref{integral}) is straightforward to 
evaluate\footnote{One sets $ k^4 = x \, K^4 $ and 
expresses the resulting integral in terms of a 
contour integral involving the discontinuity of 
$ x^{(\nu -3) /4}$. The contour integral may then
be opened up to enclose only the simple pole at
$x=-1$, which gives the result (\ref{cauchy}).}
\begin{equation}
  \int_0^\infty dk \, 
  { k^\nu \over  k^4 + K^4 } =
K^{\nu - 3} \, {\pi \over 4} \, { 1 \over
\sin\left( \pi { 3 - \nu \over 4} \right) }
= {K^{\nu -3} \over 3 - \nu } + O(\nu -3) \,.
\label{cauchy}
\end{equation}
By partial fractions, the
second integral on the right-hand side of
Eq.~(\ref{integral}) is easily evaluated in 
terms of logarithms. Thus, in the $\nu \to 3$ limit, 
\begin{eqnarray}
  \int_0^\infty dk \, \frac{k^\nu}
  {\vert k^2 + F\vert^2} 
  &=&
  \frac{K^{\nu-3}}{3 -\nu}
  -\frac{1}{2(F-F^*)} \left[
  F\ln\left({F \over K^2} \right) - 
  F^*\ln\left({F^* \over K^2} \right) \right] \ .
\label{intkregfinalb}
\end{eqnarray}
The derivative of Eq.~(\ref{intkregfinalb}) with 
respect to $K$ vanishes when $\nu \to 3$,
hence the $K$-dependence cancels in the physical
limit. With the aid of Eq.~(\ref{intkregfinalb}),
the coefficient functions now appear as 
\begin{eqnarray}
&&\left\{ {\cal A}^\smlt_b \, {1 \over \beta_b v_p} 
\,\,,\,\, C_b^{ll \, \smlt} \right\}
    =
 { e_p^2 \, \pi \over \beta_b v_p } \,
  {\Omega_{\nu-2} \over (2\pi)^\nu} 
  \int_{-1}^{+1} d\cos\theta \, 
	\sin^{\nu-3}\theta \,
  \rho_b(v_p\cos\theta) 
\nonumber\\[5pt]
  && \qquad
  \Bigg\{ {K^{\nu-3}  \over 3 - \nu} - 
  {i \over 4\pi \, \rho_{\rm total}(v_p \cos\theta)}  
\left[ F^*   \ln \left( { F^* \over K^2 } \right)
- F \ln \left( { F \over K^2 } \right) \right]
  \Bigg\} \, \left\{ \cos\theta \,,\, 
	{1 \over \cos\theta} \right\}  \,,
\label{almost}
\end{eqnarray}
where Eq's.~(\ref{spectral}) and (\ref{fmfstar}) 
have been used to simplify the notation. 

The second set of terms in the curly braces 
in Eq.~(\ref{almost}) are obviously finite in the 
$\nu \to 3$ limit.  We take this limit for
these terms, and write 
a decomposition into singular and regular parts,
\begin{equation}
  {\cal A}^\smlt_b  = {\cal A}^\smlt_{b,\smS} +
  {\cal A}^\smlt_{b,\smR}  
\,, \qquad\qquad
C_b^{ll \, \smlt} = C_{b,\smS}^{ll \, \smlt} +
C_{b,\smR}^{ll \, \smlt} \,.
\label{dedxblt}
\end{equation}
The singular part is given by
\begin{eqnarray}
&& 
\left\{ {\cal A}^\smlt_{b,\smS} \, {1 \over \beta_b v_p} 
\,\,,\,\, C_{b,\smS}^{ll \, \smlt} \right\}
\nonumber\\
&& \qquad
   = 
  {e_p^2 \over 4 \pi} \, {\Omega_{\nu-2} \over 
  2\pi} \left({K \over 2\pi}\right)^{\nu-3} 
  { 1 \over 3 - \nu} \, {1 \over \beta_b v_p } \,
  \int_{-1}^{+1} d \, \cos\theta \,
  \sin^{\nu-3}\theta \, \rho_b(v_p \cos\theta) \,
\left\{ \cos\theta \,,\, {1 \over \cos\theta} \right\} \,.
\nonumber\\
&&
\label{lesssing}
\end{eqnarray}
The regular part can be simplified 
slightly by using the reflection property 
$F(-u) = F(u)^*$ noted in Eq.~(\ref{fref}) while
the ratio $\rho_a(u) / \rho_{\rm total}(
u)$ is
even in $u$.  Hence
\begin{eqnarray}
&&
\left\{ {\cal A}^\smlt_{b,\smR} \, {1 \over \beta_b v_p} 
\,\,,\,\, C_{b,\smR}^{ll \, \smlt} \right\} 
\nonumber\\
&& \quad
=
  {e_p^2 \over 4 \pi } {1 \over \beta_b v_p} { i \over 2 \pi }
  \int_{-1}^{+1} d\cos\theta \, 
	{\rho_b(v_p\cos\theta) \over 
	\rho_{\rm total}(v_p\cos\theta) }
 F(v_p \cos\theta) \ln \left( { F(v_p
  \cos\theta) \over K^2 }\right) \, \left\{ \cos\theta \,,\,
		{1 \over \cos\theta } \right\}  \,.
\nonumber\\
&&
\label{lessregg}
\end{eqnarray}
The corresponding coefficient functions summed over all the species in
the plasma are easily obtained for a plasma at a common 
temperature, since one only needs to place
\begin{equation}
{\sum}_b {\rho_b(u) \over \rho_{\rm total}(u)} = 1
\end{equation}
in Eq.~(\ref{lessregg}).  
To write the singular contribution 
(\ref{lesssing}) in a convenient form to 
combine with the $\nu > 3$ result, we 
change variables to $u=\cos^2\theta$ to get
\begin{eqnarray}
&& 
\left\{ {\cal A}^\smlt_{b,\smS} \, {1 \over \beta_b v_p} 
\,\,,\,\, C_{b,\smS}^{ll \, \smlt} \right\}
\nonumber\\
&& \qquad
  =
  \frac{e_p^2}{4\pi}\, 
   {1 \over \beta_b v_p} 
  \frac{\Omega_{\nu -2}}{2\pi}\,
  \left(\frac{K}{2\pi}\right)^{\nu-3}
  \hskip-0.3cm 
  \frac{1}{3-\nu}  \int_{0}^1 du\,
  (1-u)^{(\nu-3)/2} 
  \rho_b(v_p u^{1/2}\,) \, \left\{ 1 \,,\, {1\over u} \right\} \,.
\label{lesssingg}
\end{eqnarray}

\subsection{Asymptotic Results for Large and Small 
Velocity}

The asymptotic forms of our results for large 
and small projectile velocities $v_p$ are of 
interest. Here we shall work out these limits
for the regular terms. The corresponding limits
for the singular terms are much easier to
compute once they are combined with the 
singular terms produced from the $\nu > 3 $
calculation, and so we defer this until later on. 

To obtain the small velocity behavior 
of the dielectric function, we first add and
subtract $v_p\cos\theta/v$ in the numerator of 
the integrand of (\ref{disp}) to get
\begin{equation}
  F(v_p  \cos\theta)  = \kappa^2_\smD  - 
  {\sum}_c \, \kappa^2_c \, 
  \int_{-\infty}^{+\infty} dv \, { v_p 
  \cos\theta  \over v_p \cos\theta + i 
  \eta - v}\sqrt{ \beta_c m_c \over 2\pi } 
  \exp\left\{ - { 1 \over 2} \beta_c m_c v^2
  \right\} \,,
\label{structt}
\end{equation}
where
\begin{equation}
  \kappa^2_\smD = {\sum}_c \, \kappa_c^2
\end{equation}
is the total squared Debye wave number of 
the plasma. We now make use of the relation
\begin{equation}
  {1 \over v_p \cos\theta -v + i \eta} =
  - i \pi \delta(v_p \cos\theta - v ) +
  {\cal P} { 1 \over v_p \cos\theta - v } \,,
\end{equation}
in which ${\cal P}$ denotes the principal part
prescription. Since ${\cal P} (1/x)$ defines 
an odd function, the translation $ u = v - 
v_p \cos\theta$ of the integration variable
gives\footnote{
\baselineskip 15pt
The integral defining the real 
part which appears here may be written in the 
form 
$$
  d(x) = { 1 \over 2} \int_{-\infty}^{+\infty} 
  {dv \over v}\,\sinh 2vx \, e^{-v^2} \,,
$$
since the integrand in an even function.  
Differentiating this with respect to $x$, writing 
out the resulting hyperbolic cosine in exponential 
terms, and completing the square yields two simple 
Gaussian integrals which give $d'(x) = \sqrt{\pi} 
\exp\{x^2\}$. Thus we have the alternative evaluation
$$
  d(x) = \sqrt{\pi} \int_0^x dy \, \exp\{y^2\} \,,
$$
which is essentially Dawson's integral.}
\begin{eqnarray}
  F(v_p  \cos\theta)  &=& \kappa^2_\smD  + 
  \pi \, i \, {\sum}_c \, \rho_c(v_p \cos\theta) 
\nonumber\\
  &&
  - 2 \, v_p \cos\theta \, {\sum}_c \, 
  \kappa^2_c\, \sqrt{ \beta_c m_c \over 2\pi } 
  \exp\left\{ - {1 \over 2} \beta_c m_c \, (v_p 
  \cos\theta)^2\right\}
\nonumber\\
  && \quad 
  \int_0^{\infty}  { du \over u} \, \sinh 
  \left( \beta_c m_c \, u \, v_p \cos\theta 
  \right) \, \exp\left\{-{1 \over 2} \beta_c 
  m_c \, u^2 \right\} \,.
\label{structtt}
\end{eqnarray}
In this form the small $v_p$ limit is reduced 
to the evaluation of elementary Gaussian integrals 
and we have
\begin{eqnarray}
  v_p \to 0 \,: &&
\nonumber\\
  F(v_p  \cos\theta)  &=& \kappa^2_\smD -
  {\sum}_c \, \kappa^2_c \, \beta_c \, m_c \,
   v_p^2 \cos^2\theta + O(v_p^4)  
+ \pi i \, \rho_{\rm total}(v_p\cos\theta) \,,
\nonumber\\
\label{smstruct}
\end{eqnarray}
where we note that $\rho_{\rm total}(u)$ starts 
out at order $u$. 
Placing this result in Eq.~(\ref{lessregg})  produces
\begin{eqnarray}
  v_p &\to& 0 \,:
\nonumber\\
  {\cal A}^\smlt_{b,\smR}  &=&
    -  {e_p^2 \over 4 \pi } \,  
   \kappa_b^2 \,
 \left( {\beta_b  m_b \over 2 \pi } \right)^{1/2} \, v_p
\Bigg\{ \left[ {1\over3} - 
{1\over10} \beta_b m_b v_p^2 \right]
\left[\ln \left({\kappa_\smD^2 \over K^2} \right) + 1 \right]
\nonumber\\
&& \qquad
   - {v_p^2 \over 5} \, {\sum}_c  \, { \kappa_c^2 \over
  \kappa_\smD^2 } \, \beta_c \,  m_c \,
 + { \pi \, v_p^2  \over 60} \, \left[ {\sum}_c \, 
	{\kappa^2_c \over \kappa_\smD^2} 
\left(\beta_c  m_c \right)^{1/2} \right]^2 \Bigg\} \,.
\label{lessreglim}
\end{eqnarray}
and 
\begin{eqnarray}
  v_p &\to& 0 \,:
\nonumber\\
  C^{ll \, \smlt}_{b,\smR}  &=&
    -  {e_p^2 \over 4 \pi } \,  
  { \kappa_b^2 \over \beta_b} \,
 \left( {\beta_b  m_b \over 2 \pi } \right)^{1/2} \,
\Bigg\{ \left[ 1 - 
{1\over6} \beta_b m_b v_p^2 \right]
\left[\ln \left({\kappa_\smD^2 \over K^2} \right) + 1 \right]
\nonumber\\
&& \qquad
   - {v_p^2 \over 3} \, {\sum}_c  \, { \kappa_c^2 \over
  \kappa_\smD^2 } \, \beta_c \,  m_c \,
 + { \pi \, v_p^2  \over 36} \, \left[ {\sum}_c \, 
	{\kappa^2_c \over \kappa_\smD^2} 
\left(\beta_c  m_c \right)^{1/2} \right]^2 \Bigg\} \,.
\label{lesssreglim}
\end{eqnarray}

To obtain the large projectile velocity limit 
of the ${\cal A}_{b,\smR}^\smlt$ coefficient, we first note that 
the numerator and each term in the denominator of
the spectral weight ratio  
$\rho_b(v_p\cos\theta) / 
\rho_{\rm total}(v_p\cos\theta)$ 
[where $\rho_b$ and $\rho_{\rm total}$ are defined in 
Eq's.~(\ref{spectral}) and (\ref{efrelb})] contains 
a factor 
$\exp\{ - \beta_c m_c v_p^2 \cos^2\theta \}$. 
In view 
of the very small electron/ion mass ratio, 
these exponential factors approach 0 much faster for 
ions than for electrons in the $v_p \to \infty$ limit.
Thus the spectral weight ratio is very small  
except for the case in which the index $b$ refers 
to the electron, $b=e$, for which case we have
\begin{equation}
v_p \to \infty\,: \qquad\qquad
{\rho_e(v_p\cos\theta) \over 
\rho_{\rm total}(v_p\cos\theta) } \to 1 \,.
\end{equation}
Thus the sum defining the energy loss to all
plasma particle species is dominated by the
electron contribution in the large projectile
velocity limit:
\begin{eqnarray}
v_p \to \infty\,: \qquad &&
\nonumber\\
 && {\cal A}^\smlt_{\smR}   =
  {\sum}_b   {\cal A}^\smlt_{b,\smR}   \to
	{\cal A}^\smlt_{e,\smR}  \,.
\end{eqnarray}
The asymptotic limit that we are about to obtain
is valid when $\beta_e m_e v_p^2 /2 \gg 1$, or 
when
\begin{equation}
E_p = {1\over2} m_p v_p^2 \gg 
	{m_p \over m_e} \, T_e \,.
\end{equation} 

To obtain this large velocity limit, we first replace 
the spectral
weight ratio by unity in formula
(\ref{lessregg}).  We then write
$\cos\theta = z$ in 
Eq.~(\ref{lessregg}) 
and note that since the integrand is analytic in 
the upper-half $z$ plane, we may deform the original 
integration along the $ -1 < z < +1 $ portion of
the real axis into a semicircle of unit radius in 
the upper-half $z$ plane. Thus (\ref{lessregg}) 
becomes 
\begin{eqnarray}
  {\cal A}^\smlt_\smR   &=&
  {e_p^2 \over 4 \pi }  \int_0^\pi { d \phi \over
  2 \pi} e^{2i \phi } F(v_p e^{i\phi}) \ln \left(
  { F(v_p e^{i\phi}) \over K^2 } \right) \,.
\label{semicir}
\end{eqnarray}
When $u  = \zeta$ is a large complex variable 
in Eq.~(\ref{disp}) we can perform simple Gaussian 
integrals to obtain the limit
\begin{eqnarray}
  |\zeta| \to \infty \, : &&
\nonumber\\
  F(\zeta)  &=& -  { \kappa^2_e \over
  \beta_e m_e \, \zeta^2 } ~+~ {\cal O}(\zeta^{-4}) \ .
\end{eqnarray}
This limit actually entails a sum of terms over
all the species $b$, with the electron mass 
$m_e$ replaced by that of the species $b$,
$m_e \to m_b$.  However, because of the
very large ratio of the ion masses to the 
electron mass, this sum is dominated by the
electron contribution that we have written.
Keeping only the dominant electron terms 
and placing this limiting behavior in 
Eq.~(\ref{semicir}) gives
\begin{eqnarray}
  v_p \to \infty \,  &:&
\nonumber\\
  {\cal A}^\smlt_\smR  &=&
	  {\cal A}^\smlt_{e,\smR}
 = {e_p^2 \over 4 \pi }
   \, { \kappa^2_e \over 2\beta_e m_e v_p^2 } \,
  \ln\hskip-0.1cm \left( {  K^2 \beta_e m_e v_p^2 
  \over \kappa^2_e} \right) \ .
\label{semicirlim}
\end{eqnarray}

The asymptotic behavior of 
$C^{ll \, \smlt}_{b , \smR}$ is quite different.  The final
factor of $1 / \cos\theta$ in Eq.~(\ref{lessregg}) emphasizes the
region about $\cos\theta = 0$.  Hence the large $v_p$
considerations given above for 
$ {\cal A}^\smlt_{b,\smR}$ 
do not hold.  Instead, we must regulate the integrand near 
$\cos\theta = 0$.  Since 
$F(0) = \kappa_\smD^2$, this is done by writing 
Eq.~(\ref{lessregg}) as 
\begin{eqnarray}
 C_{b,\smR}^{ll \, \smlt}
&=&
  {e_p^2 \over 4 \pi } {1 \over \beta_b v_p} { i \over 2 \pi }
  \int_{-1}^{+1} d\cos\theta \, 
	{\rho_b(v_p\cos\theta) \over \cos\theta} 
  { F(v_p \cos\theta) \over \rho_{\rm total}(v_p\cos\theta) }
\nonumber\\
&& \qquad\qquad\qquad\qquad 
\left\{ \ln \left( { \kappa_\smD^2 \over K^2} \right)
 + \ln \left( { F(v_p\cos\theta) \over F(0) }\right) \right\} \,.
\end{eqnarray}
For the integral involving the first constant logarithm, we note
that the overall function multiplying $F(v_p\cos\theta)$ is odd
in $\cos\theta$.  Hence in view of Eq.~(\ref{ImF}), we may
replace
\begin{equation}
F(v_p\cos\theta) \to {1\over2} 
\left[ F(v_p\cos\theta) - F(-v_p\cos\theta) \right]
= \pi i \rho_{\rm total}(v_p\cos\theta)
\end{equation}
in this part.  Since 
$ \ln \left( { F(v_p\cos\theta) / F(0) }\right) $ 
vanishes at $\cos\theta = 0$, the integral involving it may be
treated in the same way as was done previously for 
$ {\cal A}^\smlt_{b,\smR} $.  Only the electrons in the plasma
contribute to this second piece.  Thus, using the Kroenecker delta
function $\delta_{b,e}$ to distinguish the electron contribution, 
the two parts give
\begin{eqnarray}
 C_{b,\smR}^{ll \, \smlt}
&=&
 - {e_p^2 \over 4 \pi } {1 \over \beta_b v_p} { 1 \over 2 }
  \int_{-1}^{+1} d\cos\theta \, 
	{\rho_b(v_p\cos\theta) \over \cos\theta} 
\,  \ln \left( { \kappa_\smD^2 \over K^2} \right)
\nonumber\\
 && \qquad + \, \delta_{b,e} \,
 {e_p^2 \over 4 \pi } {1 \over \beta_b v_p} 
  \int_0^\pi {d\phi \over 2 \pi}  \, 
	F\left( v_p e^{i\phi} \right) \,
  \ln \left( { F(v_p e^{i\phi}) \over F(0) }\right) \,.
\end{eqnarray}
In view of the exponential damping in the definition 
(\ref{spectral}), we may replace the $\cos\theta$ integration
limits of $\pm 1$ by $\pm \infty$ to obtain the asymptotic form
\begin{eqnarray}
v_p &\to& \infty \,: 
\nonumber\\
&& 
\int_{-1}^{+1} d\cos\theta \, 
{\rho_b(v_p\cos\theta) \over \cos\theta} \to
\int_{-\infty}^{+\infty} dz \, \kappa_b^2 v_p \,
\sqrt{ \beta_b m_b \over 2 \pi} \,
\exp\left\{ - {1 \over 2} \beta_b m_b v_p^2 z^2 \right\}
= \kappa_b^2 \,.
\end{eqnarray}
On the other hand, in the large velocity limit,
\begin{eqnarray}
v_p \to \infty \,: \qquad\qquad\qquad\qquad &&
\nonumber\\
\int_0^\pi {d\phi \over 2\pi} \, 
F\left(v_p e^{i\phi}\right) \, \ln \left(
{ F\left( v_p e^{i\phi} \right) \over F(0) }
\right) 
&\to&
\int_0^\pi {d\phi \over 2\pi} \, 
\left( - {\kappa_e^2 \over \beta_e m_e v_p^2 }
	e^{-2i\phi} \right) \,
\ln\left( - {\kappa_e^2 \over \kappa_\smD^2 \beta_e m_e v_p^2 }
	e^{-2i\phi} \right) 
\nonumber\\
&=& 
- {\kappa_e^2 \over \beta_e m_e v_p^2 } \, 
\int_0^\pi {d\phi \over 2\pi} \, 
\left( -2i \phi \right) \, e^{-2i\phi} 
=
- {\kappa_e^2 \over 2 \beta_e m_e v_p^2 } \,.
\end{eqnarray}
This is of relative order $T_e / m_e v_p^2 $ and hence may be neglected in
the asymptotic limit, leaving only
\begin{eqnarray}
v_p &\to& \infty \,:
\nonumber\\
&& \qquad
C^{ll \, \smlt}_{b , \smR} = - {e_p^2 \over 4\pi} \,
{\kappa_b^2 \over 2 \beta_b v_p } \, 
\ln\left( { \kappa_\smD^2 \over K^2 } \right) \,.
\label{different}
\end{eqnarray}

\section{Short Distance Effects Dominant When 
$\mmbf{\nu > 3}$: \\ Classical Case}
\label{big}

To the leading order in the plasma density with 
which we are concerned, the Boltzmann equation 
correctly describes the Coulomb interactions in 
the plasma for spatial dimensions $ \nu $ larger 
than three. Again one can prove that, to leading 
order in the plasma density, the rigorous BBGKY 
hierarchy reduces to the Boltzmann equation when 
the spatial dimension exceeds three.\footnote{
\baselineskip 15pt
This 
can be established by carefully examining, for example, 
the discussion of the derivation of the Boltzmann
equation from the BBGKY hierarchy given in Sec.~3.5 
of Huang\cite{huang}. The derivation breaks down 
at $\nu =3$ because of the long range of the Coulomb 
force.} The Boltzmann equation for the phase-space 
density $f_a({\bf p}_a)$ of species $a$ reads
\begin{equation}
\left[  {\partial \over \partial t} 
+ {\bf v}_a \cdot \nabla \right] f_a({\bf r}, {\bf p}_a , t) 
   ={\sum}_b C_{ab}({\bf r}, {\bf p}_a, t ) \,.
\label{Beq}
\end{equation}
We suppress the common space and time coordinates 
${\bf r}, t$ and write 
the collision term involving species $b$ 
in the form
\begin{eqnarray}
  C_{ab}( {\bf p}_a)\! &=& \!\int {d^\nu{\bf p}'_b
  \over (2\pi\hbar)^\nu} {d^\nu{\bf p}'_a \over
  (2\pi\hbar)^\nu} {d^\nu{\bf p}_b \over (2\pi\hbar)^\nu}
  \left| T(W,q^2) \right|^2 \!(2\pi\hbar)^\nu
\delta^{(\nu)}
  \! \left( {\bf p}_b' + {\bf p}_a' - {\bf p}_b - {\bf p}_a
  \right)
\nonumber\\
  && \qquad\qquad
  (2\pi\hbar) \delta\left( {1\over2} m_b {v'_b}^2 +
  {1\over2} m_a {v'_a}^2  -  {1\over2} m_b  {v_b}^2
  -  {1\over2} m_a {v_a}^2 \right)
\nonumber\\
  && \qquad\qquad\qquad
  \Big[ f_b({\bf p}_b') f_a({\bf p}_a')- f_b({\bf p}_b)
  f_a({\bf p}_a) \Big] \,.
\label{coll}
\end{eqnarray}
Here, although at this stage the scattering process 
is taken to be purely classical, a quantum-mechanical 
notation has been adopted\footnote{
\baselineskip 15pt
The roles of the 
factors of Planck's constant $\hbar$ that appear in 
Eq.~(\ref{coll}) are worth pointing out. It suffices to
consider the factors associated with the first product 
of phase-space densities in the square brackets in 
Eq.~(\ref{coll}).  The two factors of $\hbar^{-\nu}$ that 
appear in the first two integration volume elements 
are the factors of $\hbar$ that appear even in purely
classical statistical mechanics.  They change the 
dimension of $d^\nu{\bf p}'$ to that of an inverse 
volume (in a $\nu$-dimensional space) so that the 
momentum integral of $f({\bf p}')$ becomes a particle 
number density.  The remaining factor of $\hbar^{-\nu}$ 
in the $d^\nu{\bf p}_b$ measure just cancels the 
conventional factor of $\hbar^\nu$  associated with
the total-momentum-conserving delta function. Since 
the dimension of $\delta(x)$ is $x^{-1}$, the single
factor of $\hbar$ associated with the energy-conserving
delta function produces a quantity with the dimensions 
of time.  So far, we have the dimension count $L^{-2\nu} 
\, T$.  The final factor to be examined is the scattering 
amplitude.  To obtain its dimensions and its overall
$\hbar$ dependence, we consider the first Born 
approximation result $T = \hbar^{-1} \, \tilde V({\bf 
q} / \hbar )$, where $\tilde V$ is the Fourier transform 
of the potential and ${\bf q}$ is the momentum transfer 
in the scattering.  Thus the scattering amplitude has 
the dimensions $ T^{-1} \, L^\nu $, and we conclude that 
the collision term $C_{ab}({\bf p}_a)$ has the dimensions 
of a rate, $T^{-1}$, as it must. As we shall later see 
explicitly, all the factors of Planck's constant $\hbar$ 
cancel in the classical limit save for those that convert 
the two momentum integrals of the phase-space densities 
into number densities, the $\hbar$ factors that appear 
even in classical statistical mechanics.}  to describe 
the scattering of the particles of mass $m_a$ and $m_b$,
the scattering from the initial momenta ${\bf p}_a = 
m_a {\bf v}_a ,\, {\bf p}_b = m_b {\bf v}_b $ to the 
final momenta ${\bf p}_a' = m_a {\bf v}_a' ,\, {\bf 
p}_b' = m_b {\bf v}_b'$, with the scattering amplitude
$T(W,q^2)$ depending on the center-of-mass energy $W$ 
and the squared momentum transfer $q^2$. It is convenient 
to employ this quantum-mechanical notation even for 
classical scattering for several reasons. 
It explicitly displays 
the complete kinematical character of a scattering 
process, including the detailed balance symmetry.  
It explicitly shows that the Boltzmann equation may be
generalized to an arbitrary number of spatial
dimensions $\nu$. It connects the collision term explicitly with
the cross section generalized to $\nu$ dimensions as shown in
Eq.~(\ref{sigma}) below.  Finally, it shows that 
the collision term 
(\ref{coll}) vanishes when all the particles are 
in thermal equilibrium with the generic densities 
$f({\bf p}) \sim \exp\{- \beta m v^2 /2 \} $
because of 
the conservation of energy enforced by the 
delta function.

\subsection{Projectile Motion in an 
Equilibrium Plasma}

The transport of energy, momentum, and transverse energy was
discussed at the start of Sec. \ref{general}.  To do this in
terms of the Boltzmann equation, we first review the standard
treatment for the sake of clarity and completeness, and to
establish our notation.  In general, we deal with a
momentum-dependent quantity $q({\bf p})$ which gives a spatial
density for species $a$ as
\begin{equation}
{\cal Q}_a({\bf r},t) = \int {d^\nu {\bf p}_a \over (2\pi\hbar)^\nu}
\, q({\bf p}_a) \, f_a({\bf r}, {\bf p}_a, t) \,,
\end{equation}
and flux vector
\begin{equation}
{\cal F}_a^k({\bf r},t) = \int {d^\nu {\bf p}_a \over (2\pi\hbar)^\nu}
\, q({\bf p}_a) \, {p_a^k \over m_a} \, 
 f_a({\bf r}, {\bf p}_a, t) \,.
\end{equation}
The Boltzmann equation (\ref{Beq}) then gives
\begin{equation}
{\partial \over \partial t} {\cal Q}_a({\bf r},t) 
+ \nabla^k {\cal F}_a^k({\bf r},t) = 
\int {d^\nu {\bf p}_a \over (2\pi\hbar)^\nu}
\, q({\bf p}_a) \,{\sum}_b \, C_{ab}({\bf r}, {\bf p}_a, t) \,.
\label{ccolll}
\end{equation}
The sum involves collisions
with the other particle species in the plasma. Thus,
we have an unambiguous identification of the 
rate of transfer to each species $b$ in the 
plasma.
The scattering amplitude, delta functions, and 
momentum integrations in the collision term are 
symmetrical under the interchange of initial and 
final particles. Hence, we may make the replacement
\begin{equation}
  q({\bf p}_a)  \,  \Big[ f_b({\bf p}_b')
  f_a({\bf p}_a') - f_b({\bf p}_b) f_a({\bf p}_a) \Big]
  \to
   \Big[ q({\bf p}_a') - q({\bf p}_a) \Big]
  f_b({\bf p}_b) f_a({\bf p}_a) \ .
\label{symm}
\end{equation}
The collision term in Eq.~(\ref{ccolll}) can be expressed as
\begin{equation}
\int {d^\nu {\bf p}_a \over (2\pi\hbar)^\nu}
\, q({\bf p}_a) \,{\sum}_b \, C_{ab}({\bf r}, {\bf p}_a, t) 
= - {\sum}_b \, \int {d^\nu {\bf p}_a \over (2\pi\hbar)^\nu}
\, f_a({\bf r}, {\bf p}_a, t) \, {d Q_{ab} \over dt}  \,, 
\end{equation}
where the sign is chosen so that $dQ_{ab} /dt$ gives the rate at
which the property $Q$ is transfered from species $a$ {\em to} 
the plasma species $b$.  We now concentrate on a particular
momentum and position of a particular ``projectile'' particle 
$p$ and identify  $dQ_b / dt$ as the transfer from this projectile
particle to the plasma species $b$. Now, in general,  
the cross section for the scattering of particles 
$p$ and $b$ into a restricted momentum interval 
$\Delta$ is given by
\begin{eqnarray}
  v_{pb} \, \int_\Delta d \sigma_{pb}
  &=& \int_\Delta {d^\nu{\bf p}'_b \over (2\pi\hbar)^\nu}
  {d^\nu{\bf p}'_p \over (2\pi\hbar)^\nu} \left|
  T(W,q^2) \right|^2 (2\pi\hbar)^\nu \delta^{(\nu)}
  \left( {\bf p}_p' + {\bf p}_b' - {\bf p}_p - {\bf
  p}_b \right)
\nonumber\\
  && \quad
  (2\pi\hbar) \,  \delta\left( {1\over2} m_p {v'_p}^2 +
  {1\over2} m_b {v'_b}^2  -  {1\over2} m_p  {v_p}^2  -
  {1\over2} m_b {v_b}^2 \right) \,.
\label{sigma}
\end{eqnarray}
Using this definition, we find that
\begin{eqnarray}
  { d Q^\smgt_b \over dt} &=& -  \int
  {d^\nu{\bf p}_b \over (2\pi\hbar)^\nu}
  f_b({\bf p}_b) \, v_{pb} \, \int d \sigma_{pb} \,
\left[ q({\bf p}_p') - q({\bf p}_p) \right] \,.
\label{qloss}
\end{eqnarray}
Here
\begin{equation}
  {\bf v}_{pb} = {\bf v}_p - {\bf v}_b
\end{equation}
is the relative velocity between the incident 
particle and plasma species $b$, with magnitude
$v_{pb} = |{\bf v}_{pb}| $. 
Henceforth we use the $\smgt$ superscript to emphasize that this
is the leading result for $\nu > 3$.  Except that we work in a
space of arbitrary dimensionality $\nu$, the result (\ref{qloss})
is the familiar one:  The rate of change of a quantity is its
change in a collision times the rate of these collisions ---
 which is given
by the cross section folded over the incident flux values.  
In this form, rather than the change of the quantity brought
about by the scattering into and out of a momentum region 
as described by the Boltzmann collision term, a 
relabeling of variables 
 expresses the rate of change in terms 
of the change in each collision.  The result (\ref{qloss}) 
expresses the rate in an obvious form, but it does 
not make manifest the fact that the rate  
(\ref{qloss}) for a quantity that is conserved in the collision 
vanishes when integrated over a thermal,
Boltzmann distribution with the same temperature of the
plasma.

Some momentum integrations may be performed by passing 
to the center-of-mass coordinates, where the total and 
relative momenta are defined by 
\begin{equation}
  {\bf P} =  m_p {\bf v}_p + m_b {\bf v}_b \,,
\end{equation}
\begin{equation}
  {\bf p} = m_{pb} {\bf v}_{pb} =
  {1 \over M_{pb} } \left( m_b {\bf p}_p - m_p 
  {\bf p}_b \right) \,,
\label{cmp}
\end{equation}
with $M_{pb}$ the total mass and $m_{pb}$ the 
reduced masses of the system,
\begin{equation}
  M_{pb} = m_p + m_b \,, \qquad\quad
  {1 \over m_{pb} } = { 1 \over m_p} + 
  { 1 \over m_b} \ .
\end{equation}
Similar expressions hold for the final state 
variables. We can now write the momentum and
energy conserving delta-functions as 
\begin{eqnarray}
  && \delta^{(\nu)}\left( {\bf p}_p' + {\bf p}_b'
  - {\bf p}_p - {\bf p}_b \right)\, \delta\left(
  {1\over2} m_p {v'_p}^2 + {1\over2} m_b {v'_b}^2
  -  {1\over2} m_p  {v_p}^2 -  {1\over2} m_b {v_b}^2
  \right)
\nonumber\\
  && \qquad\qquad
  = \delta^{(\nu)}\left( {\bf P}' -  {\bf P} \right)
  \delta\left( { {p'}^2 \over 2 m_{pb} } - { {p}^2
  \over 2 m_{pb} } \right) \,,
\end{eqnarray}
and since there is a unit Jacobian in passing to
center-of-mass coordinates, 
$ d^\nu {\bf p }'_b \, d^\nu {\bf p}'_p = 
d^\nu {\bf P}' \, d^\nu  {\bf p}'$, 
we have 
\begin{equation}
  v_{pb}  \int_\Delta d\sigma_{pb}  =
  \int_\Delta {d^\nu{\bf p}' \over (2\pi\hbar)^\nu}
  \, \left| T(W,q^2) \right|^2 \,(2\pi\hbar) \delta\left(
  { {p'}^2 \over 2m_{pb} } -   { p^2 \over 2m_{pb} }
  \right)  \ .
\label{scatt}
\end{equation}
We now note the energy in the center of mass 
is given by
\begin{equation}
  W = { p^2 \over 2 m_{pb} } \,,
\end{equation}
while the momentum transfer in the scattering is
\begin{equation}
  {\bf q} = {\bf p}_p' - {\bf p}_p =
  {\bf p}_b - {\bf p}'_b = {\bf p}' - {\bf p} \,.
\end{equation}

Let us first apply these considerations to the rate of energy
transfer $dE_b / dt$.  This is obtained from the general 
formula (\ref{qloss}) with
\begin{equation}
q({\bf p}_p') - q({\bf p}_p) \to
{1\over2} m_p \, \left[ {{\bf v}_p'}^2 - {\bf v}_p^2 
\right] \,.
\end{equation}
The conservation of momentum ${\bf P}' = {\bf P}$, 
and the energy constraint ${p'}^2 = p^2$, allows
us to write the energy change of the projectile
as 
\begin{equation}
  {1\over2} m_p \left( {v'}_p^2 - v^2_p \right) =
  {1 \over M_{pb} } \, {\bf P} \cdot {\bf q} \,.
\end{equation}
Since the scattering in the center of mass frame
is axially symmetric about the initial momentum 
${\bf p}$, the transverse components of ${\bf q}$ 
average to zero in the scattering process, and so 
we may make the replacement
\begin{equation}
  {\bf P} \cdot {\bf q} \to
  {1 \over p^2} ({\bf P} \cdot {\bf p}) \,\,
  ({\bf p} \cdot {\bf q}) \to - {1 \over 2 p^2}
  {\bf P} \cdot {\bf p} \,\, q^2 \,,
\end{equation}
with the last form following from the energy 
constraint ${p'}^2 = p^2$. Thus we may write 
the energy loss as
\begin{eqnarray}
  { d E^\smgt_b \over dx} &=& {1\over v_p}
\, \int {d^\nu{\bf p}_b \over (2\pi\hbar)^\nu}
  f_b({\bf p}_b) \, { {\bf P} \cdot {\bf p}
  \over 2 p^2 \, M_{pb} } \, v_{pb} \, \int 
  d \sigma_{pb} \,  \, q^2  \,.
\label{eeloss}
\end{eqnarray}

To extract ${\cal A}^\smgt_b$ from Eq.~(\ref{eeloss}),
we use the relation (\ref{aaaa}), which we repeat here 
for convenience: 
\begin{equation}
\exp\left\{ -{1\over2} \,\beta_b  m_p v^2_p \right\} \,  
\beta_b m_p v_p \, {dE_b \over dx} = - 
{\partial \over \partial v_p^l} \, \hat v^l_p \, 
\exp\left\{ -{1\over2} \,\beta_b  m_p v^2_p \right\} \,  
 {\cal A}_b \,.
\label{aaaaa}
\end{equation}
Since ${\bf P} = m_p {\bf v}_p + m_b {\bf v}_b$ and 
${\bf p} = m_{pb} {\bf v}_{pb}$, and since the cross section
integral is only a function of the magnitude of the relative
velocity $v_{pb}$, the integrand of 
Eq.~(\ref{eeloss}), multiplied by the exponential factor 
in this relation, has the velocity dependence of the form
\begin{eqnarray}
&& - 
\beta_b \exp\left\{ -{1\over2} \,\beta_b  m_p v^2_p \right\} \,  
f_b({\bf p}_b) \, {\bf P} \cdot {\bf p} \, X(v_{pb}) 
\nonumber\\
&& \qquad\qquad
\sim
X(v_{pb}) \, m_{pb} {\bf v}_{pb} \cdot \left(
{\partial \over \partial {\bf v}_p} +
{\partial \over \partial {\bf v}_b} \right) 
\exp\left\{ -{1\over2} \,\beta_b  
\left( m_p v^2_p + m_b v_b^2 \right) \right\} \,.  
\end{eqnarray}
Integration by parts in the $p_b$ integral replaces the
action of $ \partial / \partial {\bf v}_b $ on the 
exponential factor by  
 $ 
- (\partial / \partial {\bf v}_b ) \cdot {\bf v}_{pb} \,
X(v_{pb}) 
$. 
Since 
$
{\bf v}_{pb} = {\bf v}_p - {\bf v}_b
$,
this action of 
$
- (\partial / \partial {\bf v}_b ) 
$
is equivalent to that of
$
 (\partial / \partial {\bf v}_p ) 
$,
and so we have, effectively within the ${\bf p}_b$ integral, 
\begin{equation}
- \beta_b \exp\left\{ -{1\over2} \,\beta_b  m_p v^2_p \right\} \,  
f_b({\bf p}_b) \, {\bf P} \cdot {\bf p} \, X(v_{pb}) \to
{\partial \over \partial {\bf v}_p} \cdot 
\exp\left\{ -{1\over2} \,\beta_b  m_p v^2_p \right\} \,  
f_b({\bf p}_b) \, {\bf p} \, X(v_{pb}) \,.
\end{equation}
The only direction produced by the ${\bf p}_b$ integral is that
along ${\bf v}_p$.  Hence we may replace
$
{\bf p} \to \hat{\bf v}_p \, ( \hat{\bf v}_p \cdot {\bf p} ) 
$.
We thus arrive at the structure (\ref{aaaaa}) upon the 
identification
\begin{eqnarray}
  {\cal A}^\smgt_b  &=& {1 \over 2 m_b} \,
\, \int {d^\nu{\bf p}_b \over (2\pi\hbar)^\nu}
  f_b({\bf p}_b) \, \hat{\bf v}_p \cdot \hat{\bf v}_{pb} \,
  \int d \sigma_{pb} \,  \, q^2  \,.
\label{aloss}
\end{eqnarray}
Here we have simplified an overall factor by using
\begin{equation}
{m_p \over M_{pb} \, m_{pb} } = { 1 \over m_b} \,.
\end{equation}

The remaining independent coefficient may be taken to be 
$C^{ll}_b$ which, according to Eq.~(\ref{ppp}), is given for 
$\nu > 3$, by 
\begin{equation}
 {1 \over m_p} \, C_b^{ll \, \smgt} =
 v_p^k \, {dP^{k \, \smgt}_b \over dt} -
{ dE^\smgt_b \over dt} \,.
\end{equation}
This difference involves
\begin{equation}
{\bf v}_p \cdot m_p \left( {\bf v}_p' - {\bf v}_p \right)
- {1\over2} \, m_p \, \left( {{\bf v}_p'}^2 - {\bf v}_p^2 \right)
= - {1\over2} m_p \, \left( {\bf v}_p' - {\bf v}_p \right)^2 
= - {1 \over 2 m_p} q^2 \,.
\end{equation}
Thus we rather quickly find that
\begin{equation}
C^{ll \, \smgt}_b = {1\over2} \int {d^\nu{\bf p}_b \over (2\pi\hbar)^\nu}
\, f_b({\bf p}_b) \, v_{pb} \, \int d \sigma_{pb} \, q^2 \,.
\end{equation}

In summary, we may write the results as
\begin{eqnarray}
\left\{ {\cal A}^\smgt_b {1 \over \beta_b v_p} \,\,,\,\,
C^{ll \, \smgt}_b \right\}
  &=& {1 \over 2} \,
\, \int {d^\nu{\bf p}_b \over (2\pi\hbar)^\nu}
  f_b({\bf p}_b) \, v_{pb} \,
  \int d \sigma_{pb} \,  \, q^2 \,
\left\{ { \hat{\bf v}_p \cdot \hat{\bf v}_{pb} \over 
\beta_b m_b v_p v_{pb} } \,,\, 1 \right\} \,.
\label{eitherloss}
\end{eqnarray}

\subsection{Classical Coulomb Scattering}

For the remainder of this Section we shall treat
only the case of classical scattering.  We defer
the discussion of quantum-mechanical corrections 
to Section~10.

\subsubsection{Cross Section Integral}

We now apply the energy loss formula to the case of classical Coulomb
scattering which results from the reduction of the classical BBGKY
hierarchy. In $\nu$ dimensions, the element of differential classical
cross section is given by
\begin{equation}
d \sigma_{pb}^\smC = \Omega_{\nu -2} \,  b^{\nu -2} db \,,
\end{equation}
where $\Omega_{\nu -2} $ is the area of the unit $\nu -2$ sphere and $b$
is the classical impact parameter. 
Hence
\begin{equation}
  \int d \sigma_{pb}^\smC \, q^2 = \Omega_{\nu -2} \int^\infty_0 db \,  
  b^{\nu -2} q^2 (b) \,,
\label{weightq}
\end{equation}
with the momentum transfer related to the scattering angle by
\begin{equation}
q^2(b) = 4p^2 \sin^2 \left( \theta (b) / 2 \right) \, ,
\label{q}
\end{equation}
where $ \theta (b) $ is the scattering angle as a function of 
$ b $ in $ \nu $ dimensions. The classical planar trajectory  of
a particle moving in a central potential is independent
of the spatial dimensionality $\nu$. Thus the familiar formula 
for the scattering angle
\begin{equation}
\theta(b) = \pi - 2 b \, \int_{r_{\rm min}}^\infty \, 
{ dr \over r^2 } \, \left[ 1 - \left( {b^2 \over r^2} \right)
	- \left( { 2 m_{pb} \over p^2} \right) V(r)
\right]^{-1/2} \,,
\end{equation}
holds for arbitrary spatial dimensionality $\nu$. Here $r_{\rm min}$ is
the lower turning radius at which the angular brackets in the
integrand vanish. The Coulomb potential energy in $\nu$ dimensions is
given by
\begin{equation}
V(r) = e_p \phi_b(r) = \frac {e_p e_b \, \Gamma (\nu / 2) } 
  {2 \, (\nu - 2) \, \pi^{\nu / 2} } \,{1 \over  r^{\nu-2}} \,,
\end{equation}
which follows from Gauss's law in 
$ \nu $ dimensions, 
\begin{equation}
 e_b  = \int d^{\nu}r \, \nabla \cdot {\bf E}_b = - \Omega_{\nu - 1} \, 
r^{\nu - 1} \, {d \phi_b(r) \over dr} \,.
\end{equation}  
Changing the integration variable to $u = 1/r$ now gives
\begin{equation}
 \theta (b) = \pi - 2 \, \int^{u_0}_0 \frac { du } 
 {\sqrt {1 - 2 \, \xi (b) \, u^{\nu - 2} - u^2} } \,,
\label{thetaofb}
\end{equation}
where 
\begin{equation}
 \xi (b) = \frac {e_p e_b \, \Gamma (\nu / 2) \,  m_{pb} } 
 {2 \, (\nu  - 2) \, \pi^{\nu / 2} \, p^2 \, b^{\nu-2} } \,,
\label{xi}
\end{equation}
and the turning point $u_0$ is now described by the positive root of
\begin{equation}
 1 - 2 \, \xi \, u^{\nu - 2}_0 - u^2_0 = 0 \,.
\end{equation}

At infinite impact parameter, the scattering angle vanishes and so
does $\xi(b)$. Thus for large impact parameters, the  
integral in Eq.~(\ref{thetaofb}) can be expanded in $ \xi $, giving
\begin{eqnarray}
 \theta(b) &=& 2 \, \xi(b) \, \int^1_0 du \frac {1-u^{\nu - 2}} 
 {(1-u^2)^{3/2}} + {\cal O}(\xi^2) = \sqrt{\pi} \, \frac { (\nu - 2) \, 
 \Gamma[ (\nu - 1) / 2] } {\Gamma(\nu /2)} \, \xi(b) + 
 {\cal O}(\xi^2) \,. 
\end{eqnarray}
Making use of Eq.~(\ref{xi}), we see that at large impact
parameters $b$, the momentum transfer (\ref{q}) is given by 
\begin{equation}
q^2 = 4 \, p^2 \, { b_0^{2\nu -4} \over b^{2\nu -4} } \,,
\label{q2atnu}
\end{equation}
in which
\begin{eqnarray}
b_0^{2\nu -4} &=&
\left({ e_p e_b \over 4 \pi } \right)^2 \,\left(
{ m_{pb} \over p^2 } \right)^2  \, 
     { \Gamma \left( { \nu -1 \over 2 } \right)^2 \over \pi^{\nu -3} }
\nonumber\\
&=& \left({ e_p e_b \over 4 \pi } \right)^2 \,\left(
{ m_{pb} \over p^2 } \right)^2  \, 
     \pi^{3-\nu} \left[ 1 - \gamma ( \nu - 3 ) + \cdots \right] \,.
\label{min}
\end{eqnarray}

For finite impact parameters, no divergence appears if we simply
set $\nu = 3$.  In this case, the integral in Eq.~(\ref{thetaofb})
gives
$
\theta(b) = 2 \tan^{-1} \xi(b) 
$,
and with $\nu = 3$ in the definition (\ref{xi}) of $\xi(b)$, one
finds the well-known result for Rutherford scattering, 
\begin{eqnarray}
\nu = 3 \, : &&
\nonumber\\
&&
q^2(b) = { 4 m_{pb}^2 \over p^2 } \, 
         { \left(e_p e_b / 4\pi \right)^2  
\over b^2 + ( e_p e_b \, m_{pb} / 4\pi \,  p^2 )^2 } \,.
\label{q2at3}
\end{eqnarray}

Taking into account our results for $q^2$ at $\nu =3$ and
arbitrary but finite $b$, Eq.~(\ref{q2at3}), 
and at large $b$ but arbitrary
$\nu$, Eq.~(\ref{q2atnu}), we arrive at an interpolation formula 
\begin{equation}
q^2_I(b) =  4 \, p^2  
   \, \,  { b_0^{2\nu -4} \over  b^{2 \nu -4} + b_0^{2\nu -4} } 
\label{lope}
\end{equation}
which is valid for $\nu$ slightly above $3$. 
Since 
$
[q^2(b) - q^2_I(b)]
$
vanishes in the limit $\nu \to 3$ for finite values of $b$, 
while the integral of 
$
b^{\nu -2} \, [q^2(b) - q^2_I(b)]
$
over very large $b$ values vanishes, we conclude that, in the
limit $\nu \to 3$, we may replace the exact $q^2(b)$ by the
interpolation function $q^2_I(b)$ in computing the momentum
transfer integral (\ref{weightq}), with
\begin{eqnarray}
\nu \to 3^+ \,: &&
\nonumber\\
&&
v_{pb} \, \int d\sigma_{pb}^\smC \, q^2 = { p \over m_{pb} } \, 
\Omega_{\nu-2}\, 4\, p^2 \, I \,,
\label{classq}
\end{eqnarray}
in which
\begin{equation}
I = \int_0^\infty db \, b^{\nu -2}
      { b_0^{2\nu -4} \over b^{2\nu -4} + b_0^{2\nu -4} } \,.
\end{equation}
This integral is akin to the previous integral (\ref{cauchy}),
and a similar evaluation 
gives\footnote{Namely, one sets $b^{2\nu -4} = 
b^{2\nu -4}_0 \, x$ to express $I$ as a contour integral that
gives the discontinuity of $x^{(3-\nu) / (2\nu -4)}$. The contour
may be then opened up to enclose only the simple pole at 
$x = -1$, which gives the result (\ref{clever}).}
\begin{equation}
I = {\pi \, b_0^{\nu -1} \over 2\nu - 4} \, 
{1 \over \sin \left( \pi { \nu - 3 \over 2\nu - 4} \right) }
= 
{ b_0^{\nu -1} \over \nu - 3} + O( \nu - 3) \,.
\label{clever}
\end{equation}
Placing this result in Eq.~(\ref{classq}) and using 
the definition (\ref{min}) of $b_0$, we obtain, after a little
algebra, the $\nu \to 3$ limit
\begin{equation}
v_{pb} \, \int d\sigma_{pb}^\smC \, q^2 = 
	{ (e_p e_b)^2 \over 2\pi } \, { m_{pb} \over p } \, 
\left( { e_p e_b \, m_{pb} \over 
 4 \,  p^2 } \right)^{\left({3-\nu \over \nu -2}\right)}
   \, { \Omega_{\nu -2} \over 2\pi}  
	\left[ { 1 \over \nu - 3 } -  \gamma \right] \,. 
\label{classic}
\end{equation}
The pole at $\nu = 3$ that appears here reflects the long-distance,
infra-red divergence that appears when $\nu$ approaches 3 from above. 
Note that for simplicity, we have written the result for like
charges, $e_p e_b > 0$; otherwise this product should be replaced
by $|e_pe_b|$.

\subsubsection{Classical Coefficients}

We turn now to compute the transport coefficients 
when the classical cross 
section (\ref{classic}) is placed in the general 
result (\ref{eitherloss}).  
To do this, it is convenient 
to use the velocity variables ${\bf v}_p$ and ${\bf 
v}_b$, with ${\bf p} = m_{pb} \, 
{\bf v}_{pb} $ and $ {\bf v}_{pb} = {\bf v}_p - {\bf 
v}_b$.  The classical cross section (\ref{classic}) produces 
the factor $p^{-1}{p^2}^{(\nu-3)/(\nu -2)}$ or
\begin{equation}
v_{pb} \, \int d\sigma_{pb}^\smC \, q^2 \sim
v_{pb}^{\left( { \nu -4 \over \nu -2} \right) } \,.
\end{equation}
As Eq.~(\ref{eitherloss}) shows, the ${\cal A}_b^\smgt$
coefficient involves $\hat{\bf v}_{pb} / v_{pb}$ times this factor.
Since 
\begin{equation}
{ \hat{\bf v}_{pb} \over v_{pb} } \,\,
v_{pb}^{\left( { \nu -4 \over \nu -2} \right) } =
- { \nu -2 \over \nu -4} \, {\partial \over \partial {\bf v}_b } 
\,\, v_{pb}^{\left( { \nu -4 \over \nu -2} \right) } \,,
\end{equation}
and an integration by parts makes the derivative act upon the
distribution function $f_b({\bf p}_b)$, we have, effectively, 
\begin{equation}
 {\partial \over \partial {\bf v}_b } \to 
\beta_b m_b \, {\bf v}_b \,.
\end{equation}
Hence, for classical scattering, Eq.~(\ref{eitherloss}) may be
expressed as 
\begin{equation}
\left\{ {\cal A}^\smgt_{b,\smC}{1\over \beta_b v_p}  \,\,,\,\, 
	C^{ll \smgt}_{b,\smC}  \right\} =
{1\over2} \int {d^\nu {\bf p}_b \over (2\pi\hbar)^\nu }
\, f_b({\bf p}_b) \, v_{pb} \int d\sigma^\smC_{pb} \, q^2
\, \left\{ - {\nu-2 \over\nu-4} \, 
  { {\bf v}_p \cdot {\bf v}_b  \over v_p^2 } \,,\, 1 \right\} \,.
\end{equation}

To reduce this expression, we use the integral representation
\begin{eqnarray} 
  v_{pb}^{\left( {\nu -4 \over  \nu - 2} \right) } 
&=& \left( {\beta_b m_b \over 2}
  \right)^{{4 - \nu \over 2\nu-4} } 
{ 1 \over 
\Gamma\left( {4 - \nu \over  2\nu - 4 } \right) } 
\, \int_0^\infty {ds \over s}  \,
  s^{{4 - \nu \over  2\nu -4 } }  \, 
\exp\left\{ - {1 \over 2}  \beta_b
  \, m_b \, v^2_{pb} \, s \right\} \,.
\end{eqnarray}
With
\begin{equation}
  \left( { m_b \over 2 \pi \hbar} \right)^\nu \,
  f_b({\bf p}_b) =n_b \, \left( { \beta_b m_b \over
  2\pi } \right)^{\nu / 2} \exp\left\{ - {1 \over 2}
  \beta_b \, m_b \, v_b^2 \right\} \,,
\end{equation}
and with ${\bf v}_{pb} = {\bf v}_p - {\bf v}_b $,   
we may interchange the integrals and complete the 
square in the Gaussian integral to get 
\begin{eqnarray}
  &&
 \int {d^\nu{\bf p}_b \over (2\pi\hbar)^\nu} \,
  f_b({\bf p}_b) \, 
  v_{pb}^{\left( {\nu -4 \over \nu - 2} \right) } \,
\left\{ - { \nu -2 \over \nu -4 } \, 
{ {\bf v}_p \cdot {\bf v}_b \over v_p^2 } \,,\, 1 \right\}
\nonumber\\
  && \qquad =  n_b \, 
\left({\beta_b m_b\over 2} \right)^{4 - \nu \over 2\nu -4 }
 \,   
{ 1 \over \Gamma \left( {4 - \nu \over  2\nu - 4 } \right) } 
\, \int_0^\infty {ds \over s} \, s^{4 - \nu \over 2\nu - 4} \, 
(1 + s)^{-\nu/2}
\nonumber\\
  && \qquad\qquad\qquad\qquad
  \exp\left\{ - {1 \over 2} \beta_b m_b \, v^2_p \,
  { s \over 1 + s } \right\} 
\, 
\left\{-{\nu -2 \over \nu -4} \, {s \over 1 + s} 
\,,\, 1 \right \} \,. 
\end{eqnarray}
We use this expression to evaluate the general 
formula (\ref{eitherloss})  using the classical cross 
section (\ref{classic}). The result is simplified 
by the variable change $ u = s \, ( 1 + s )^{-1}$, 
and its nature clarified by introducing the squared
Debye wave numbers for the particles of species $b$,
\begin{equation}
  \kappa_b^2 = \beta_b \, e^2_b \, n_b \,.
\end{equation}
Thus
\begin{eqnarray}
&& \left\{ {\cal A}^\smgt_{b, \smC} {1 \over \beta_b v_p}
 \,\,,\,\, C^{ll \smgt}_{b,\smC} \right\} 
= {e_p^2 \over 4\pi} \,
  { \Omega_{\nu -2} \over 2 \pi} \,\left[ { 1 \over
  \nu -3 } - \gamma \right] \, \kappa_b^2 
  \,  \left( { m_b \over 2 \beta_b } \right)^{1/2} \,
  \left[ { e_p e_b \, \beta_b m_b \over 8 m_{pb} }
  \right]^{\left( {3-\nu \over \nu-2} \right) }
{1\over \Gamma \left({4 - \nu \over 2\nu -4}  \right) } 
\nonumber\\
  && \qquad
\int_0^1 du \, u^{-1/2} \,  u^{(3 - \nu)/(\nu -2)} \,
  (1 - u)^{\nu (\nu-3) / (2\nu -4) }
  \exp\left\{ - {1 \over 2}
  \beta_b \, m_b \, v^2_p \, u \right\} 
\left\{ - {\nu -2 \over  \nu -4} \, u \,,\, 1 \right\} \,.
\nonumber\\
&&
\label{classical}
\end{eqnarray}

To facilitate the comparison of this $\nu > 3$ 
result with the result (\ref{lesssingg}) for
$\nu < 3$, we first note that, in the limit in
which $\nu$ approaches three,
\begin{eqnarray}
  \nu \to 3 \, : \qquad\qquad\qquad &&
\nonumber\\
  { 1 \over \Gamma \left( {4 - \nu \over 2\nu -4 }
  \right) }&=& { 1 \over \Gamma \left( {1\over2}
  \right) } \left[ 1 + \psi\left( { 1\over2}
  \right) \, (\nu -3) \right] ~+~
  {\cal O}(\nu-3)^2
\nonumber\\[5pt]
  &=& { 1 \over \sqrt \pi} \left[1 - ( \ln 4
  + \gamma ) \, (\nu -3) \right]  ~+~
  {\cal O}(\nu-3)^2 \,,
\label{gfact}
\end{eqnarray}
with
\begin{equation}
  \nu \to 3 \, : \quad\qquad
  { 1 \over \nu -3 } - \ln 4 = {1 \over \nu -3 } \,
  4^{(3-\nu)} ~+~ {\cal O}(\nu-3)\,.
\end{equation}
Using
\begin{equation}
 (1 - u)^{\nu (\nu-3) / (2\nu -4) } =
(1-u)^{(\nu-3)/2} \, 
(1-u)^{(\nu-3)/(\nu-2)} \,,
\end{equation}
and
\begin{equation}
- { \nu -2 \over \nu - 4} = 1 + 2 \, (\nu -3) ~+~
	{\cal O}(\nu -3)^2 \,,
\end{equation}
we may now cast Eq.~(\ref{classical}) in the form
\begin{eqnarray}
&& \left\{ {\cal A}^\smgt_{b, \smC} {1 \over \beta_b v_p}
 \,\,,\,\, C^{ll \smgt}_{b,\smC} \right\} 
\nonumber\\
&& \qquad\qquad   = {e_p^2 \over 4\pi} \,
  { \Omega_{\nu -2} \over 2 \pi} \, \kappa_b^2 \, 
  \, \left[ { 1 \over \nu -3 }  - 2\gamma \right] \,
  \left( { m_b \over
  2 \pi \beta_b  } \right)^{1/2} \,
  \int_0^1 du \, u^{-1/2} \, (1-u)^{(\nu-3)/2} 
\nonumber\\
&& \qquad\qquad\qquad 
\exp
  \left\{ - {1 \over 2} \beta_b m_b v^2_p \, u \right\} \,
  \left[ { e_p e_b \, \beta_b m_b \over 2 \, m_{pb} } \,
  { u \over 1-u} \right]^{\left( {3-\nu \over \nu-2}
  \right) } \, \left\{ \left[ 1 + 2 (\nu -3) \right] \, u
\,,\, 1 \right\} \,.
\label{goodclassic}
\end{eqnarray}

\section{Classical Results}
\label{cclassic}

The $\nu >3$ contributions (\ref{goodclassic}) we have just 
computed and the
$\nu < 3$ contributions of Eq.~(\ref{dedxblt}) are each
separately divergent in the $\nu \to 3$ spatial
limit. However, it follows from the general principle 
of our dimensional continuation that their sum is
well defined in the limit\footnote{
\baselineskip 15pt
In this regard, it 
is worth noting that as the spatial dimensions 
change, the physical dimensions of a charge
$e^2$ change. This is made explicit if we 
replace $e^2 \to e^2 \,\mu^{(\nu-3)}$.  Now 
the wave number $\mu$ carries the dimensional
change in the original definition of the 
squared charge. The fact that the singular 
pole terms cancel implies that the complete
result is independent of what precise numerical 
value is taken for the arbitrary wave number 
$\mu$; the result is  insensitive to the way 
in which the squared charge is extrapolated 
away from three dimensions. This is the analog, 
within our dimensional continuation method, to 
the renormalization group invariance of quantum
field theory.}, combinations we denote by 
\begin{eqnarray}
  {\cal A}^\smC_b  &=&  
  \lim_{\nu \to 3}\left\{
  {\cal A}^\smgt_{b,\smC}  +  
  {\cal A}^\smlt_b  \right\} \,,
\label{dedxcca}
\end{eqnarray}
and
\begin{eqnarray}
  C^{ll \smC}_b  &=&  
  \lim_{\nu \to 3}\left\{
  C^{ll \smgt}_{b,\smC}  +  
  C^{ll \smlt}_b  \right\} \,.
\label{dedxccc}
\end{eqnarray}

It is worthwhile emphasizing again that this
 provides the leading and next-to-leading 
contributions to the classical stopping power
in the plasma coupling $g$.   
The lower-dimensional contributions ($<$) 
have  several pieces,  
Eqs.~(\ref{dedxblt})--(\ref{lessregg}). They 
consist of the sum of 
 singular terms, Eq.~(\ref{lesssing}), and 
regular pieces, Eq.~(\ref{lessregg}). 
Note that it is 
redundant to employ a classical subscript in the 
$\nu < 3$ case, as 
we did for the $\nu > 3$ piece, since all 
contributions in dimensions less than three 
are purely classical. 

Since the regular piece is finite and has already
been reduced to its simplest form, let us concentrate 
on the sum of the singular pieces, and write 
\begin{equation}
  {\cal A}^\smC_{b,\smS}   =
  \lim_{\nu \to 3}    \left\{
  {\cal A}^\smgt_{b,\smC} +
  {\cal A}^\smlt_{b,\smS}  \right\} \,,
\qquad\qquad
  C^{ll \smC}_{b,\smS}   =
  \lim_{\nu \to 3}    \left\{
  C^{ll \smgt}_{b,\smC} +
  C^{ll \smlt}_{b,\smS}  \right\} \,.
\label{dedxcsing}
\end{equation}
The subscript S on the left-hand side of 
Eq.~(\ref{dedxcsing}) should not be taken 
to indicate that the limit of the sum is 
singular, but only that the result comes
from the well-defined sum of individually singular pieces.
The $\nu \to 3$ limit is well defined,
as the respective pole terms cancel. 
The first term was derived in (\ref{goodclassic}),
and is dominant for $\nu > 3$, while the
second term is dominant for $\nu < 3$
and is given by (\ref{lesssingg}), which
we repeat here for convenience:
\begin{eqnarray}
&& 
\left\{ {\cal A}^\smlt_{b,\smS} \, {1 \over \beta_b v_p} 
\,\,,\,\, C_{b,\smS}^{ll \, \smlt} \right\}
\nonumber\\
&& \qquad
  =
  \frac{e_p^2}{4\pi}\, 
   {1 \over \beta_b v_p} 
  \frac{\Omega_{\nu -2}}{2\pi}\,
  \left(\frac{K}{2\pi}\right)^{\nu-3}
  \hskip-0.3cm 
  \frac{1}{3-\nu}  \int_{0}^1 du\,
  (1-u)^{(\nu-3)/2} 
  \rho_b(v_p u^{1/2}\,) \, \left\{ 1 \,,\, {1\over u} \right\} \,.
\label{convenient}
\end{eqnarray}
Some computation shows that the $\nu \to 3$ limit 
of the sum of Eq's.~(\ref{goodclassic}) and (\ref{convenient}) 
gives
\begin{eqnarray}
&& 
\left\{ {\cal A}^\smC_{b,\smS} \, {1 \over \beta_b v_p} 
\,\,,\,\, C_{b,\smS}^{ll \, \smC} \right\} = 
  {e_p^2 \over 4\pi} \, 
	\kappa^2_b \, 
  \left( { m_b \over  2\pi \beta_b } \right)^{1/2} \,
  \int_0^1 du \, u^{-1/2} \,
  \exp\left\{ - {1 \over 2} 
  \beta_b m_b v^2_p \, u \right\}
\nonumber\\
  && \qquad\qquad
  \left[ - \left\{ \ln \left(\beta_b  { e_p e_b  \over 4 \pi} \,
K \, { m_b \over m_{pb} } \, { u \over 1-u} \right) 
+ 2 \gamma \right\} 
\, \Big\{ u \,,\, 1 \Big\} 
+ \Big\{ 2  \, u \,,\, 0 \Big\} \right] \,.
\label{wonderclassic}
\end{eqnarray}

Since the classical energy loss functions 
${\cal A}^\smC_{b,\smS}$ and $C^{ll \smC}_{b,\smS}$ 
contain the complete contributions for $\nu>3$, the 
only additional finite part is that which comes 
from $\nu < 3$, and so the complete energy loss
functions for the species $b$ plasma particles 
in the classical case are given by
\begin{equation}
  {\cal A}^\smC_b = {\cal A}^\smC_{b,\smS} +
  {\cal A}^\smlt_{b,\smR} \,, \qquad
  C^{ll \smC}_b = C^{ll\smC}_{b,\smS} +
  C^{ll\smlt}_{b,\smR} \,,
\label{classicall}
\end{equation}
in which the two contributions to each function 
are given by
the results (\ref{wonderclassic}) that we have
just dealt with and the previous results
(\ref{lessregg}), which we repeat here for
convenience: 
\begin{eqnarray}
&&
\left\{ {\cal A}^\smlt_{b,\smR} \, {1 \over \beta_b v_p} 
\,\,,\,\, C_{b,\smR}^{ll \, \smlt} \right\} 
\nonumber\\
&& \quad
=
  {e_p^2 \over 4 \pi } {1 \over \beta_b v_p} { i \over 2 \pi }
  \int_{-1}^{+1} d \cos\theta \, 
	{\rho_b(v_p\cos\theta) \over 
	\rho_{\rm total}(v_p\cos\theta) }
 F(v_p \cos\theta) \ln \left( { F(v_p
  \cos\theta) \over K^2 }\right) \, \left\{ \cos\theta \,,\,
		{1 \over \cos\theta } \right\}  \,.
\nonumber\\
&&
\label{nun}
\end{eqnarray}
Following the methods used to show that the result 
(\ref{doneatlast}) is independent of $K$, it is 
straightforward to verify that the sum 
(\ref{classicall}) 
of Eq's.~(\ref{wonderclassic}) 
and (\ref{nun}) is also 
independent of the particular value of $K$.
Moreover, the reflection symmetry 
$F(-u) = F^*(u)$ 
guarantees that the result (\ref{nun}) is real.  

The classical result applies to the $v_p \to 0$ 
limit of the energy loss (large velocities are
inconsistent with the classical limit).
This limit of (\ref{wonderclassic}) 
entails elementary $u$ integrals\footnote{
\baselineskip 15pt
$$
  \int_0^1 du \, u^{-1/2} = 2 \,, \qquad\qquad
  \int_0^1 du \, u^{-1/2} \ln\left( { 1 -u \over 
  u} \right) = 4 \ln 2 \,,
$$
$$
  \int_0^1 du \, u^{1/2} = {2\over3} \,, \qquad\qquad
  \int_0^1 du \, u^{1/2} \ln\left( { 1 -u \over 
  u} \right) = {4\over3} \left( \ln 2 - 1 \right) \,,
$$
$$
  \int_0^1 du \, u^{3/2} = {2\over5} \,, \qquad\qquad
  \int_0^1 du \, u^{3/2} \ln\left( { 1 -u \over 
  u} \right) = {4\over5} \left( \ln 2 - {4\over3} \right) \,.
$$
}, and one finds that
\begin{eqnarray}
  v_p &\to& 0 \, :
\nonumber\\
{\cal A}^\smC_{b,\smS}   
&=&
  - {e_p^2 \over 4\pi} 
  \, \kappa_b^2  \, v_p \,  
  \left( { \beta_b m_b \over  2\pi } \right)^{1/2} 
  \Bigg\{
    \left( {2\over3} - {1\over5} \beta_b m_b v_p^2  \right)
  \left[  \ln\left(\beta_b  { e_p e_b \over 16 \pi} K
  { m_b \over m_{pb} } \right)  +  2\gamma \right]
\nonumber\\
&& \qquad\qquad\qquad\qquad\qquad
  - {2\over15} \, \beta_b m_b v_p^2 \Bigg\} ~+~
  {\cal O}(v_p^5) \,.
\label{wonderclassiclim}
\end{eqnarray}
This result, added to the small velocity limit 
of the regular part (\ref{lessreglim}) derived 
in Sec.~\ref{small}, 
produces
\begin{eqnarray}
  v_p &\to& 0 \,:
\nonumber\\
  {\cal A}^\smC_b   &=& -
 {e_p^2 \over 4\pi} \, \kappa_b^2 \, v_p \,
  \left({\beta_b m_b \over  2\pi} \right)^{1/2}
  \Bigg\{
   \left( {2\over3} - {1\over5} \beta_b m_b v_p^2  \right)
  \left[  \ln\left(\beta_b  { e_p e_b \over 16 \pi} \,
  \kappa_\smD { m_b \over m_{pb} } \right) + {1\over2} + 
  2\gamma \right] 
\nonumber\\[5pt]
  && - {2\over15} \, \beta_b m_b v_p^2 -
     {v_p^2 \over 5} \, {\sum}_c  \, { \kappa_c^2 \over
  \kappa_\smD^2 } \, \beta_c m_c 
 + { \pi \, v_p^2 \over 60} \, \left[ {\sum}_c \, 
	{\kappa^2_c \over \kappa_\smD^2} 
 \left(\beta_c  m_c \right)^{1/2} \right]^2 \Bigg\} 
 ~+~ {\cal O}(v_p^5) \,.
\label{lessreglimm}
\end{eqnarray}
Note that the arguments of the logarithms that 
appear here 
involve a small factor that is essentially 
the plasma coupling parameter $g \sim\beta e^2
\kappa_\smD/4\pi$.  Hence these logarithms are 
negative numbers that are large in magnitude. 

In a similar fashion, we compute
\begin{eqnarray}
  v_p &\to& 0 \, :
\nonumber\\
C^{ll \smC}_{b,\smS}   
&=&
  - {e_p^2 \kappa_b^2 \over 4\pi} 
  \,   \left( {m_b \over  2\pi\beta_b } \right)^{1/2} 
  \Bigg\{
    \left( 2 - {1\over3} \beta_b m_b v_p^2  \right)
  \left[  \ln\left(\beta_b  { e_p e_b \over 16 \pi} K
  { m_b \over m_{pb} } \right)  +  2\gamma \right]
\nonumber\\
&& \qquad\qquad\qquad\qquad\qquad
  - {2\over3} \, \beta_b m_b v_p^2 \Bigg\} ~+~
  {\cal O}(v_p^4) \,,
\label{wonderclassiclimc}
\end{eqnarray}
which, added to the small velocity limit 
of the regular part (\ref{lesssreglim}), 
produces
\begin{eqnarray}
  v_p &\to& 0 \,:
\nonumber\\
  C^{ll \smC}_b   &=& -
 {e_p^2 \kappa_b^2\over 4\pi} \, 
  \left({ m_b \over  2\pi\beta_b} \right)^{1/2}
  \Bigg\{
   \left( 2 - {1\over3} \beta_b m_b v_p^2  \right)
  \left[  \ln\left(\beta_b  { e_p e_b \over 16 \pi} \,
  \kappa_\smD { m_b \over m_{pb} } \right) + {1\over2} + 
  2\gamma \right] 
\nonumber\\[5pt]
  && - {2\over3} \, \beta_b m_b v_p^2 -
     {v_p^2 \over 3} \, {\sum}_c  \, { \kappa_c^2 \over
  \kappa_\smD^2 } \, \beta_c m_c 
 + { \pi \, v_p^2 \over 36} \, \left[ {\sum}_c \, 
	{\kappa^2_c \over \kappa_\smD^2} 
 \left(\beta_c  m_c \right)^{1/2} \right]^2 \Bigg\} 
 ~+~ {\cal O}(v_p^5) \,.
\label{lesssreglimm}
\end{eqnarray}
Finally, we note that 
$
{\cal B}_b = C^{ll \smC}_b - 
{\cal A}_b^\smC (1/ \beta_b v_p ) $
has the leading small velocity limit 
\begin{eqnarray}
  v_p &\to& 0 \,:
\nonumber\\
  {\cal B}^\smC_b   &=& -
 {e_p^2 \kappa_b^2\over 4\pi} \, 
  \left({ m_b \over  2\pi\beta_b} \right)^{1/2}
  \, {4\over3} \,   
  \left[  \ln\left(\beta_b  { e_p e_b \over 16 \pi} \,
  \kappa_\smD { m_b \over m_{pb} } \right) + {1\over2} + 
  2\gamma \right] \,.
\label{Blim}
\end{eqnarray}

\section{Quantum Corrections}

\subsection{Quantum Scattering in the Born 
Approximation}

No dimensionless parameter can be formed from  the 
basic quantities $e_p e_b \,, m_{pb} \,,$ and 
$ v_{pb}$  
that describe the classical scattering. Hence 
dimensional analysis determines all of the 
classical result (\ref{classic}) except for 
the purely numerical factors that it contains.  
Quantum-mechanical scattering, on the other hand,  
is richer in that it involves the dimensionless
parameter 
\begin{equation}
  \eta_{pb} = { e_p e_b \over 
	4\pi \, \hbar \, v_{pb}} \,
  \left( { p \over \pi \hbar } \right)^{\nu -3}\,,
\label{dimeta}
\end{equation}
where $ v_{pb} = p / m_{pb} $ is the velocity of 
the projectile relative to a particle of species 
$b$ in the plasma. We have previously made use of 
this parameter in the $\nu = 3$ limit. The factor 
$( p / \pi \hbar)^{(\nu-3)}$ that disappears when
$\nu =3$  is introduced to make
$\eta_{pb}$ dimensionless when the spatial 
dimensionality $\nu$ is extended away from 
$\nu = 3$.\footnote{Apart from three dimensions, 
the Coulomb potential, being the $\nu$-dimensional
Fourier transform of $1/k^2$, behaves as $1 /
r^{(\nu -2)}$. Hence $e^2 / r^{(\nu -2)}$ has the 
dimensions of energy.  Since Planck's constant
$\hbar$ has dimensions of momentum times distance
or, equivalently energy times time, an inverse distance
has the dimensions $p / \hbar$ while $ \hbar v$ has
the dimensions of energy times distance.  We conclude
that Eq.~(\ref{dimeta}) does indeed define a dimensionless
parameter for arbitrary spatial dimensionality $\nu$.
The additional factor of $\pi$ in 
$ ( p / \pi \hbar)^{(\nu -3)}$
is introduced for later convenience.}  
The quantum-mechanical extension of the
classical squared momentum-transfer cross section 
(\ref{classic}) involves a dimensionless function 
of the dimensionless parameter $\eta_{pb}$. This 
parameter, which describes short-distance, 
quantum-mechanical effects, cannot appear to 
leading order in the long-distance physics which 
is involved in the $\nu < 3$ process evaluated 
in Sec.~\ref{small}. It can, however, enter
into and correct the short-distance scattering 
process with which we are now concerned.  Although 
the quantum-mechanical description that we now 
turn to is strictly outside of the derivation of 
the Boltzmann equation from the classical BBGKY 
hierarchy, it is clear from physical grounds that 
quantum mechanics must be employed when $\eta_{pb}$ 
is small, which formally corresponds to a limit in 
which Planck's constant $\hbar$ becomes large. In 
the limit in which the projectile velocity $v_p$ 
becomes small and the $\eta$ parameter is large, 
the previous classical limit must be employed. 
Otherwise, a quantum-mechanical treatment of the 
scattering in $\nu > 3$ must be made.

We turn now to  evaluate the scattering factor
(\ref{scatt}) multiplied by $q^2$ when $\eta_{pb}$ 
is small and the quantum-mechanical Born
approximation result is appropriate.  This gives 
the extreme quantum-mechanical limit that applies 
for very high projectile velocities. We shall soon 
bridge the gap between the quantum Born and the
classical results.  In the Born approximation,
\begin{equation}
  T = { \hbar e_p e_b \over q^2 } \,,
\end{equation}
and so
\begin{equation}
  v_{pb} \int d\sigma_{pb}^\smB \, q^2 = \int{
  d^\nu {\bf p}' \over (2\pi\hbar)^\nu }\,
  2 \pi \hbar \, \delta \left( { {p'}^2 \over
  2m_{pb}} - { p^2 \over 2m_{pb} } \right)
  \left( { \hbar e_p e_b \over q^2 } \right)^2
  \, q^2\,.
\end{equation}
We use
\begin{equation}
  q^2 = 4 \, p^2 \, \sin^2\theta/2  \,,
\end{equation}
and express the momentum integration volume in
(hyper-)spherical coordinates, with an implicit 
integration over all the angles save for the 
polar angle $\theta$, to write
\begin{equation}
  d^\nu {\bf p}' = m_{pb} \, {p'}^{(\nu-2)} \,
  d ( {p'}^2 /2m_{pb} ) \, \Omega_{\nu-2} \,
  \sin^{\nu -2} \theta \, d\theta \,,
\end{equation}
where $\Omega_{\nu -2}$ is the solid angle 
of a $\nu -2$ dimensional sphere. Setting 
$\theta = 2\chi$ gives
\begin{equation}
  v_{pb} \int d\sigma_{pb}^\smB \, q^2 =  { (e_p e_b)^2
  \over 2\pi}  {m_{pb} \over p } \left( { p^2 \over
  \pi^2 \hbar^2 } \right)^{(\nu -3)/2} { \Omega_{\nu -2}
  \over 2 \pi }\int_0^{\pi/2} d\chi \, \cos^{\nu -2}\chi
  \, \sin^{\nu-4}\chi  \,.
\label{q2}
\end{equation}
The integral which appears here has the value 
$(\nu -3)^{-1} + O(\nu- 3)$, as one can show by 
dividing it into two parts with a suitable partial 
integration, or by expressing it in terms of the
standard integral representation of the beta 
function. Hence
\begin{equation}
  v_{pb} \int d\sigma_{pb}^\smB \, q^2 = {(e_p e_b)^2
  \over 2\pi}  {m_{pb} \over p } \left( { p^2 \over
  \pi^2 \hbar^2 } \right)^{(\nu -3)/2} { \Omega_{\nu -2}
  \over 2\pi } \, { 1 \over \nu - 3 }  \,.
\label{bornn}
\end{equation}
Again the pole which appears here reflects the 
infrared divergence when $\nu$ approaches $3$ 
from above.

\subsection{Full Quantum Correction}

To fill in the region of arbitrary $\eta_{ab}$ 
values, we consider the weighting of the squared 
momentum transfer with the difference between the 
complete and first Born approximation cross
sections,
\begin{equation}
  \int \left( d\sigma - d\sigma^\smB \right) \, q^2 \,.
\label{difff}
\end{equation}
This integral of a cross section difference is 
well behaved in the limit $\nu \to 3 $. The pole 
at $\nu = 3$ produced by the integral involving 
the full cross section is canceled by an identical 
pole in the integral involving the Born approximation. 
This is because these poles come from soft, infrared 
physics corresponding to long distances where the 
potential is weak. The divergence behavior leading 
to the poles is produced entirely by the first Born 
approximation term of the full cross section
$d\sigma$ which is then canceled by the subtraction 
of its Born approximation $d\sigma^\smB$. However, we 
cannot simply set the spatial dimension $\nu = 3$ 
in the cross section difference that is the integrand
in the integral (\ref{difff}) 
because in three dimensions the Born and 
full Coulomb cross section elements are identical, 
$d\sigma = d\sigma^\smB$. 
The $\nu \to 3$ limit of the integral is not the 
integral of the $\nu \to 3$ limit of the integrand.
The integral does not converge uniformly at small
scattering angles, and the order of the limits cannot 
be interchanged. 
What we shall do is to 
implicitly assume that the spatial dimensionality
is slightly greater than three so as to regulate 
the theory\footnote{
\baselineskip 15pt
Other infrared regularizations 
may be used, such as the replacement of the Coulomb 
potential with a screened Debye potential, since
only a potential logarithmic divergence is to be
avoided.\label{avoid}}. Then we shall make a 
(implicitly generalized) partial wave expansion 
for the cross section difference (\ref{difff}).  
High partial waves with $l\gg 1$ correspond to 
large impact parameter scattering where the effect 
of the potential 
is weak and the first Born approximation becomes
exact\footnote{At large angular momentum, the effective 
centrifugal potential is much larger than the Coulomb 
potential, justifying the treatment of the Coulomb potential
by the first Born approximation.}.  
Thus the subtraction of the first Born 
approximation within a partial wave decomposition 
yields a partial wave sum that converges at large $l$ 
values and hence gives no pole at 
$\nu = 3$ when the physical limit of three dimensions 
is taken. Thus all we really need do is to express 
everything in terms of partial waves in three 
dimensions and subtract the Born approximation 
in the partial wave summand. In this way we may 
exploit some clever mathematics of Lindhard and  
Sorensen \cite{lind2}, but in a manner which
justifies its use.  It should be emphasized that 
the Born approximation must be subtracted in the 
partial wave summand before the sum is performed --- 
the separate sums do not converge at large $l$. 

Although we always have in mind that the Born 
term is to be subtracted, for simplicity we shall 
omit this explicit subtraction in some intermediate 
steps. We use the standard partial wave decomposition 
of the scattering amplitude,
\begin{equation}
  d \sigma = d \Omega_2 \, \left| f(\theta) 
  \right|^2 \,,
\end{equation}
with
\begin{equation}
  f(\theta) = { 1 \over 2 i p} 
  \sum_{l=0}^\infty (2 l + 1 ) 
  \left( e^{2i\delta_l} - 1 \right) 
P_l(\cos\theta) \,.
\end{equation}
The cross section weighted integral of 
$q^2 = 2 p^2 \, (1-\cos\theta )$ may now be 
evaluated using
\begin{equation}
(2l + 1 ) \, \cos\theta  P_l(\cos\theta) =
  (l+1) \, P_{l+1}(\cos\theta) + l \, P_{l-1}
  (\cos\theta) \,,
\end{equation}
and the orthogonality relation
\begin{equation}
  \int d\Omega_2 \, (2l + 1) P_{l'}(\cos\theta)
  P_l(\cos\theta)= 4\pi \, \delta_{l , l'} \,.
\end{equation}
These give\footnote{
\baselineskip 15pt
Appendix \ref{csavtcl} explains
how the cross section averaged momentum 
transfer (\ref{wave}) is simply related to 
the classical limit.}
\begin{equation}
  \int d\sigma  \, q^2 = 2 \pi \hbar^2
  \sum_{l=0}^\infty (l+1)\left\{ 2 - e^{2i
  [\delta_l - \delta_{(l+1)}] } -e^{- 2i
  [\delta_l - \delta_{(l+1)}] } \right\} \,.
\label{wave}
\end{equation}
For the Coulomb potential\footnote{This formula for the Coulomb
partial wave phase shift is derived in many graduate level quantum
mechanics texts.  See, for example, Gottfried \cite{Gott}, Sec.~17,
Landau and Lifshitz \cite{LLQ}, Sec.~36, or Schwinger \cite{S},
Sec.~9.2.  The connection of the Coulomb phase function to that of the
Debye potential, and the recovery of the Coulomb result in the
infinite screening radius limit, has been presented by Brown \cite{B}
using determintal methods and Jost functions.},
\begin{equation}
  e^{2i\delta_l} = { \Gamma(l+1+i\eta) \over
  \Gamma(l+1-i\eta) } e^{i\phi} \,,
\end{equation}
where the phase $\phi$ is independent of $l$, and
$\eta$ is the generic quantum parameter. For the
specific $p$-$b$ system that we consider, $\eta
\to \eta_{pb}$, where 
\begin{equation}
  \eta_{pb} = { e_p e_b \, m_{pb} \over 4 \pi
  \hbar p } = { e_p e_b \, \over 4 \pi \hbar
  v_{pb} } \,.
\end{equation}
Using $\Gamma(z+1) = z\Gamma(z)$, a little 
algebra, and subtracting the Born approximation,
we find that\cite{bloch}
\begin{eqnarray}
  \int \left( d\sigma_{pb} - d\sigma_{pb}^\smB 
  \right) q^2 &=&
  4 \pi \eta_{pb}^2 \hbar^2 \sum_{l=0}^\infty \left[ {1
  \over l+1+i\eta_{pb}} + { 1 \over l+1-i\eta_{pb} } - { 2
  \over l+1} \right]
\nonumber\\
  &=& -  {(e_p e_b)^2 \over 4\pi } \, { 1  \over
  v^2_{pb} }\,  2 \left[ {\rm Re} \, \psi(1+i\eta_{pb})
  + \gamma \right] \,,
\label{diff}
\end{eqnarray}
where $\psi(z)$ is the logarithmic derivative 
of the gamma function, $\psi(z) = \Gamma'(z)/ 
\Gamma(z) $, and Re denotes the real part.
Note that, by combining denominators, we 
may write
\begin{equation}
  {\rm Re} \, \psi(1+i\eta)  = 
  \sum_{k=1}^\infty{ 1 \over k}
  { \eta^2 \over k^2 + \eta^2 }  
  -\gamma \,.
\label{combd}
\end{equation}

To check the partial wave method that
has been employed, we consider the limit in 
which $\eta_{pb}$ becomes large, in which
case the full quantum cross section becomes 
equal to its classical limit, 
\begin{equation}
  v_{pb} \int \left( d\sigma_{pb} -
  d\sigma_{pb}^\smB \right) q^2 \to
  v_{pb} \int \left( d\sigma_{pb}^\smC -
  d\sigma_{pb}^\smB \right) q^2 \ .
\end{equation} 
Indeed, using
\begin{equation}
  |z| \to \infty \,: \qquad \psi(1+z) = \ln z
  + O(z^{-1}) 
\label{psilim}
\end{equation}
on the right-hand side of Eq.~(\ref{diff}), we 
find that
\begin{equation}
  v_{pb} \int \left( d\sigma_{pb}^\smC -
  d\sigma_{pb}^\smB \right) q^2 = - {(e_p e_b)^2
  \over 4\pi } \, { 1 \over v_{pb} }\,\left[
  \ln \left( {e_p e_b  \over 4\pi \hbar v_{pb}}
  \right)^2+ 2 \gamma \right] \,.
\label{cbdiff}
\end{equation}
This is precisely the $\nu \to 3$ limit of 
the difference of Eq's.~(\ref{classic}) and 
(\ref{bornn}), 
confirming the validity of our use of the partial
wave expansion\footnote{
\baselineskip 15pt
Although we are
not able to explicitly compute the cross section 
difference (\ref{diff}) for $\nu>3$, it easy to 
make a {\it model} of the mathematical expression
for this case, a model that gives the essence of 
the $\nu>3$ behavior and which reduces to the 
correct $\nu \to 3$ limit. We set
$$
  v_{pb} \int \left( d\sigma_{pb} - d\sigma_{pb}^\smB 
  \right) q^2 = {(e_p e_b)^2 \over 2\pi v_{pb} } \,
  \left( {p \over \pi \hbar } \right)^{(\nu -3)} \,
  {\Omega_{\nu -2} \over 2\pi } \, F(\eta_{pb}, \nu ) \,.
$$
The overall factors here have the correct dimensionality
of velocity times squared momentum times length to the
power $\nu -1$.   
We combine denominators as in Eq.~(\ref{combd}) with a
fractional power of $1/k$ chosen, as we shall see, 
to give correspondence with previous results. Thus
we choose the model function
\begin{eqnarray}
  F(\eta,\nu) &=& - \sum_{k=1}^\infty \left(
  {1 \over k} \right)^{\left({2\nu-5 \over \nu
  -2} \right) } \, { \eta^2 \over k^2 + \eta^2 }
\nonumber\\
  &=&- \sum_{k=1}^\infty \left( {1 \over k}
  \right)^{\left({2\nu-5 \over \nu -2 } \right)}
  + \sum_{k=1}^\infty \left( {1 \over k} \right)^{
  \left({2\nu-5 \over \nu -2 } \right) } \, { k^2
  \over k^2 + \eta^2 } \,.
\nonumber
\end{eqnarray}
These formulae reduce to Eq.~(\ref{diff}) in the limit 
$\nu \to 3$.  Of more interest is the character of 
the $\eta \to \infty$ limit. To obtain this limit 
for $\nu$ slightly greater than 3, we note that the
first sum in the second equality above defines a zeta
function, and with $\zeta(s) = (s-1)^{-1} + \gamma 
+ O(s-1)$, we have, for $\nu \to 3$,
$$
  \sum_{k=1}^\infty \left( {1 \over k} \right)^{
  \left({2\nu-5 \over \nu -2 } \right) } = { 1 
  \over \nu - 3} + 1 + \gamma \,.
$$
The second sum, in the large $\eta$ limit with $\nu$
slightly above 3, may be replaced by the integral 
that provides the asymptotic value
$$
  \int_1^\infty dk \, \left( {1 \over k} 
  \right)^{\left({2\nu-5 \over \nu -2 } 
  \right) } \, { k^2 \over k^2 + \eta^2 }
  \sim {1 \over \nu -3} \eta^{\left({3 - \nu 
  \over \nu -2 } \right) } + 1 \,.
$$
Hence, in the $\eta \to \infty$ limit,
$$
  F(\eta,\nu) = {1 \over \nu -3} \left\{ 
  \eta^{\left({3 - \nu \over \nu -2 } 
  \right) } - 1\right\}- \gamma \,,
$$
and we see that, using the definition 
(\ref{dimeta}) of $\eta$, our mathematical 
model reproduces, in this limit, the difference 
of the result (\ref{classic}) for the classical 
scattering integral and the result (\ref{bornn})
for the Born approximation scattering integral 
(with the neglect of terms that vanish when $\nu 
\to 3$).}. 

It is convenient to refer the total cross section 
integral of the squared momentum transfer to the 
classical cross section, not the quantum Born 
approximation. Thus, we subtract Eq.~(\ref{cbdiff}) from 
Eq.~(\ref{diff}) to obtain the purely quantum mechanical 
correction
\begin{eqnarray}
  v_{pb} \,\int \left( d\sigma_{pb} -
  d\sigma^\smC_{pb} \right) q^2 &=& -
  {(e_p e_b)^2 \over 4\pi \, v_{pb} }\,
  \left\{ 2 \, {\rm Re} \, \psi(1+i\eta_{pb})
  -\ln \, \eta_{pb}^2\right\} \,.
\label{diffcl}
\end{eqnarray}

Equation (\ref{eitherloss}) gives the explicit form 
for the energy loss functions for the 
plasma species $b$ in the $\nu > 3$ region that
we are now considering.  For convenience, we
repeat this formula here:  
\begin{eqnarray}
\left\{ {\cal A}^\smgt_b {1 \over \beta_b v_p} \,\,,\,\,
C^{ll \, \smgt}_b \right\}
  &=& {1 \over 2} \,
\, \int {d^\nu{\bf p}_b \over (2\pi\hbar)^\nu}
  f_b({\bf p}_b) \, v_{pb} \,
  \int d \sigma_{pb} \,  \, q^2 \,
\left\{ { \hat{\bf v}_p \cdot \hat{\bf v}_{pb} \over 
\beta_b m_b v_p v_{pb} } \,,\, 1 \right\} \,.
\label{genbig}
\end{eqnarray} 
In view of this general formula and the 
expression (\ref{diffcl}) for the difference
of the complete quantum cross section and its
classical limit, we may write the $\nu > 3$
result for the energy loss functions in the 
general case as
\begin{equation}
{\cal A}^\smgt_b = {\cal A}^\smgt_{b,\smC} 
  + {\cal A}^{\Delta Q}_b \,, \qquad
C^{ll\smgt}_b = C^{ll\smgt}_{b,\smC} 
  + C^{ll\Delta Q}_b \,,
\label{dedxgt}
\end{equation}
where $ {\cal A}^\smgt_{b,\smC} $ and
$C^{ll\smgt}_{b,\smC} $ are the $\nu>3$
classical results given in Eq.~(\ref{goodclassic}),
while $ {\cal A}^{\Delta Q}_b $ and
$ C^{ll\Delta Q}_b $ are the quantum mechanical 
corrections to this classical result, 
the results given by inserting
the correction (\ref{diffcl}) into the general
formula (\ref{genbig}). Explicitly,
\begin{eqnarray}
 \left\{  {\cal A}^{\Delta Q}_b {1 \over \beta_b v_p} 
\,,\, C^{ll \Delta Q}_b \right\}  &=&  - {1\over 2} 
\int {d^3{\bf p}_b
  \over (2\pi\hbar)^3} f_b({\bf p}_b) \, 
{ e_p^2 e_b^2 \over 4 \pi \, v_{pb}}
 \left\{ 2 \, {\rm Re} \,
  \psi \left( 1 + i \eta_{pb} \right) - \ln \eta^2_{pb}
  \right\} 
\nonumber\\
&& \qquad\qquad\qquad
\left\{ { \hat{\bf v}_p \cdot \hat{\bf v}_{pb} \over 
\beta_b m_b v_p v_{pb} } \,,\, 1 \right\} \,.
\label{remains}
\end{eqnarray}

In accordance with our general principle of
dimensional continuation, to find the leading
and next-to-leading contributions of the
stopping power, we must take the $\nu \to 3$ 
limit of the sum of the $\nu > 3$ piece (as
calculated above) and the $\nu < 3$ piece. The 
complete quantum result in three dimensions is 
therefore provided by 
\begin{equation}
  {\cal A}_b  = 
  \lim_{\nu\to 3}\left\{
  {\cal A}^\smgt_b + {\cal A}^\smlt_b  
  \right\}
=
  {\cal A}^\smC_b + {\cal A}^{\Delta Q}_b  \,,
\label{fulla}
\end{equation}
and
\begin{equation}
  C^{ll}_b  = 
  \lim_{\nu\to 3}\left\{
  C^{ll\smgt}_b + C^{ll\smlt}_b  
  \right\}
=
  C^{ll\smC}_b + C^{ll \Delta Q}_b  \,,
\label{fullll}
\end{equation}
which, with the aid of Eq.'s~(\ref{dedxgt}) and 
(\ref{dedxcca}) -- (\ref{dedxccc}), 
we have written in the form of a purely classical piece 
plus a quantum correction. We have already calculated
the classical term (\ref{classicall}) and found that the
potential divergences cancel, so we turn to simplifying 
the quantum piece (\ref{remains}). Namely, the 
integration variable can be changed to the
relative velocity, and the angular integrations 
performed. This gives
\begin{eqnarray}
 \left\{  {\cal A}^{\Delta Q}_b {1 \over \beta_b v_p} 
\,,\, C^{ll \Delta Q}_b \right\}  && = 
 -  { e_p^2 \kappa_b^2
  \over 4 \pi }  { 1 \over 2 \beta_b v_p }
  \left( { \beta_b m_b \over 2\pi } \right)^{1/2}
  \int_0^\infty dv_{pb}
  \bigg\{ 2\, {\rm Re} \, \psi \left( 1 + i \eta_{pb}
  \right) - \ln \eta^2_{pb}  \bigg\}
\nonumber\\
  &&
  \Bigg[ \exp\left\{ - {1 \over 2} \beta_b
  m_b \left( v_p - v_{pb} \right)^2\right\} 
\left\{ { 1 \over \beta_b m_b v_p v_{pb} } 
 \left( 1 - { 1 \over \beta_b m_b v_p v_{pb} } \right) 
	\,,\, 1 \right\}	
\nonumber\\
  && 
  + \exp\left\{ - {1 \over 2} \beta_b
  m_b \left( v_p + v_{pb} \right)^2\right\} 
\left\{ { 1 \over \beta_b m_b v_p v_{pb} } 
 \left( 1 + { 1 \over \beta_b m_b v_p v_{pb} } \right) 
	\,,\, - 1 \right\} \Bigg]	\,.
\nonumber\\
&&
\label{regb}
\end{eqnarray}
This expression provides a small correction  
when the integration is dominated 
by regions in which the quantum Coulomb parameter 
$\eta_{pb}$ is large so that the scattering is 
nearly classical.  When the effective $\eta_{pb}$
values are of order unity, then a detailed evaluation 
of Eq.~(\ref{regb}) is called for. But there are some 
limits in which Eq.~(\ref{regb}) simplifies.

\subsection{Simplifications and Asymptotic Limits}

One simplification appears for cold plasmas, that is,
plasmas for which  
the thermal speed $v_\smT = \sqrt { 3 / 
\beta_b m_b}$ can be neglected. This is described 
by the formal limit $\beta_b \to \infty$. In this limit
the first exponential in Eq.~(\ref{regb}) sets 
$v_{pb} = v_p$ in the factors that multiply 
it, the second exponential gives a negligible 
contribution, and so Eq.~(\ref{regb}) becomes,
\begin{eqnarray}
 v_b \to \infty \,: \qquad\qquad\qquad &&
\nonumber\\
 \left\{  {\cal A}^{\Delta Q}_b {1 \over \beta_b v_p} 
\,,\, C^{ll \Delta Q}_b \right\}  &=& 
 -  { e_p^2 \kappa_b^2
  \over 4 \pi }  { 1 \over 2 \beta_b v_p }
  \left( { \beta_b m_b \over 2\pi } \right)^{1/2}
  \int_{-\infty}^\infty dv
  \bigg\{ 2\, {\rm Re} \, \psi \left( 1 + i \eta_{p}
  \right) - \ln \eta^2_{p}  \bigg\}
\nonumber\\
  && \qquad\qquad
  \exp\left\{ - {1 \over 2} \beta_b m_b v^2\right\} 
\left\{ { 1 \over \beta_b m_b v_p^2 } 
	\,,\, 1 \right\}	
\nonumber\\
&=& 
-  { e_p^2 \kappa_b^2
  \over 4 \pi }  { 1 \over 2 \beta_b v_p }
\,  \bigg\{ 2\, {\rm Re} \, \psi \left( 1 + i \eta_{p}
  \right) - \ln \eta^2_{p}  \bigg\}
\left\{ { 1 \over \beta_b m_b v_p^2 } 
	\,,\, 1 \right\} \,,
\nonumber\\
&&
\end{eqnarray}
where
\begin{equation}
  \eta_p = {e_p e_b \over 4 \pi \hbar v_p } \,.
\end{equation}

The other simplification appears when the thermal velocity 
$v_\smT$ or $v_p$ or both are large in comparison with $e_b 
e_p / 4 \pi \hbar $.  In these cases we may use 
the small $\eta_{pb}$ limit which Eq.~(\ref{combd})
reveals to be\footnote{
\baselineskip 15pt
Because of the small 
$v_{pb}$ integration region in Eq.~(\ref{regb}), 
the formal order $\eta^2$ error in this limit 
is actually of order $\eta^2 \, \ln \eta^2 $.}
\begin{eqnarray}
  \eta \to 0: &&
\nonumber\\
  && {\rm Re} \psi \left( 1 + i \eta \right) =
  - \gamma + O(\eta^2) \,.
\end{eqnarray}
It is difficult to implement this limit directly 
in Eq.~(\ref{regb}). It is much easier to return to 
the starting point (\ref{remains}) and evaluate 
it in the manner of the evaluation of the classical
energy loss functions.  Since $p^2 = 
m_{pb}^2 v_{pb}^2$, the starting point 
Eq.~(\ref{remains}) involves
\begin{eqnarray}
  &&
  \left\{ 2\, {\rm Re} \, \psi \left( 1 + i \eta_{pb}
  \right) - \ln \eta^2_{pb}  \right\} \, 
\left\{ {1\over v_{pb}^3} \,,\, {1\over v_{pb}} \right\}
\nonumber\\
  && \qquad \to 
  \left\{ \ln \left[ { 1 \over 2} \beta_b m_b
  v_{pb}^2 \right]- \ln \left[ { 1 \over 2}
  \beta_b m_b \left( { e_p e_b \over 4 \pi \hbar}
  \right)^2 \right]- 2 \gamma \right\} \,
\left\{ {1\over v_{pb}^3} \,,\, {1\over v_{pb}} \right\}
\,.
\end{eqnarray}
We first express
\begin{equation}
  \ln \left[ { 1 \over 2} \beta_b m_b v_{pb}^2
  \right] =\lim_{\epsilon \to 0} {1 \over
  \epsilon} \left[\left( { 1 \over 2} \beta_b
  m_b v_{pb}^2 \right)^\epsilon- 1 \right] \,,
\end{equation}
and then exponentiate the terms involving 
$v_{pb}$ using
\begin{equation}
   \left( { 1 \over 2} \beta_b m_b v_{pb}^2
  \right)^{-p} = { 1 \over \Gamma(p) } \int_0^\infty
  {ds \over s} \, s^p \,\exp\left\{ - { 1 \over 2}
  \beta_b m_b v_{pb}^2 \, s \right\} \ ,
\end{equation}
with $p=3/2-\epsilon$, $p=3/2$, and $p=1/2 - \epsilon$,
$p=1/2$. 
Writing Eq.~(\ref{gfact}) in the form
\begin{equation}
  \frac{1}{\Gamma(1/2-\epsilon)} =
  \frac{1}{\sqrt{\pi}}\, \bigg[\,
  1 - \left(\log 4 + \gamma \right)
  \epsilon   \bigg] ~+~ {\cal O}( \epsilon^2) \ ,
\end{equation}
and using $z\Gamma(z) = \Gamma(z+1)$ to also obtain
\begin{equation}
  \frac{1}{\Gamma(3/2-\epsilon)} =
  \frac{2}{\sqrt{\pi}}\, \bigg[\,
  1 + \left(2 - \log 4 -\gamma \right)
  \epsilon   \bigg] ~+~ {\cal O}(
  \epsilon^2) \ ,
\end{equation}
we find that
\begin{eqnarray}
  \eta_{pb} &\to& 0: 
\nonumber\\
  && {1\over v_{pb}^3} \,
  \left\{ 2\, {\rm Re} \, \psi \left( 1 +
  i \eta_{pb} \right) - \ln \eta^2_{pb}
  \right\} \, 
\left\{ 1 \,,\, \beta_b m_b v_{pb}^2  \right\}
\nonumber\\
  &&
  = - \left( {1\over2} \beta_b m_b \right)^{3/2} \,
  {2\over \sqrt\pi}\, \int_0^\infty ds \, s^{1/2}
  \exp\left\{ - {1\over2} \beta_b m_b v_{pb}^2 s
  \right\}
\nonumber\\
  && \quad\qquad\qquad
  \left[ \left\{ 3\gamma + \ln\hskip-0.1cm \left[ 
  2 s \beta_b m_b
  \left( {e_p e_b \over 4 \pi \hbar} \right)^2
  \right]\, \right\} \, \Big\{ 1 \,,\, {1 \over s} \Big\}
   - \Big\{ 2 \,,\, 0 \Big\} \right] \,.
\end{eqnarray}
Placing this representation in Eq.~(\ref{remains}),
interchanging integrals, performing the resulting 
Gaussian integration, and making the variable change 
previously used,
\begin{equation}
  s = { u \over 1 - u } \,,
\end{equation}
yields
\begin{eqnarray}
\eta_{pb} \to 0 \,: \, &&
\nonumber\\
 \left\{  {\cal A}^{\Delta Q}_b {1 \over \beta_b v_p} 
\,,\, C^{ll \Delta Q}_b \right\}  &=& 
  {e_p^2 \kappa_b^2 \over 4\pi} \,
  \, \left( { m_b \over 2 \pi \beta_b } \right)^{1/2} 
\, \int_0^1 du \, u^{-1/2}
  \exp\left\{ - {1\over2} \beta_b m_b v_p^2 u \right\}
\nonumber\\
  &&
  \left[  {1\over 2} \left\{ \ln\hskip-0.1cm \left( 
  2 \beta_b m_b 
  \left( { e_p e_b \over 4 \pi \hbar}
  \right)^2 {u \over 1-u}\right) + 3 \, 
  \gamma \right\} \, \Big\{ u \,,\, 1 \Big\} 
 - \Big\{ u \,,\, 0 \Big\} \right] \,.
\nonumber\\
&&
\label{wonderquantum}
\end{eqnarray}

The $\eta_{pb} \to 0$ limit is formally the large $\hbar$ limit.
This is the limit in which quantum uncertainty rather than a
classical turning point sets a minimum distance scale.  This may
be brought out explicitly if we replace the previous combination
(\ref{dedxcsing}) of potentially singular parts by 
\begin{equation}
{\cal A}^{\smQ}_{b,\smS} = {\cal A}^{\smC}_{b,\smS} + 
{\cal A}^{\Delta Q}_b 
\,, \qquad\qquad
 C^{ll \smQ}_{b,\smS} = C^{ll \smC}_{b,\smS} + 
C^{ll \Delta Q}_b \,.
\end{equation}
The complete functions defined by Eq's.~(\ref{fulla}) and
(\ref{fullll}) now read, in view of Eq.~(\ref{classicall}),  
\begin{equation}
{\cal A}_b = {\cal A}^{\smQ}_{b,\smS} + 
{\cal A}^{\smlt}_{b,\smR} 
\,, \qquad\qquad
 C^{ll}_b = C^{ll \smQ}_{b,\smS} + C^{ll \smlt}_{b,\smR} \,,
\end{equation}
where the regular parts coming from the $\nu < 3$ contribution
were defined in Eq.~(\ref{nun}).  Adding
Eq's.~(\ref{wonderclassic}) and (\ref{wonderquantum}) gives the
quantum regime limit
\begin{eqnarray}
\eta_{pb} \to 0 \,: \qquad\qquad &&
\nonumber\\
 \left\{  {\cal A}^Q_{b,\smS} {1 \over \beta_b v_p} 
\,,\, C^{ll Q}_{b,\smS} \right\}  &=& 
  {e_p^2 \kappa_b^2 \over 4\pi} \,
  \, \left( { m_b \over 2 \pi \beta_b } \right)^{1/2} 
\, \int_0^1 du \, u^{-1/2}
  \exp\left\{ - {1\over2} \beta_b m_b v_p^2 u \right\}
\nonumber\\
  &&
  \left[  {1\over 2} \left\{ - \ln\hskip-0.1cm \left( 
  {m_b \over m_{pb} } \,  
   { \beta_b \hbar^2 K^2 \over 2 m_{pb} }
   {u \over 1-u}\right) - 
  \gamma \right\} \, \Big\{ u \,,\, 1 \Big\} 
 + \Big\{ u \,,\, 0 \Big\} \right] \,.
\nonumber\\
&&
\label{qlim}
\end{eqnarray}
To simply compare quantum formula (\ref{qlim}) with the classical
formula (\ref{wonderclassic}), we neglect mass ratios.  Then we
see that the classical cutoff length $\beta_b \, (e_p e_b / 4\pi) $
in Eq.~(\ref{wonderclassic}) is replaced by the quantum length
$ \hbar \, \sqrt{ \beta_b / 2 m_{pb} } $ here in Eq.~(\ref{qlim}).

Equation (\ref{qlim}) 
is the high velocity limit in the quantum-mechanical sense
that \hbox{$|\eta_{pb}| \ll 1$}, But this limit entails no 
restriction on
the kinetic energy $m_b v_p^2 / 2$ relative to the temperature
$ T_b = 1 / \beta_b$.  When the kinetic energy is also large in
comparison with the temperature, a limit that we shall simply
denote as $v_p \to \infty$, the exponential damping in the
integrand of the integral (\ref{qlim}), which emphasizes the $u=0$
region, allows us to set $1-u \to 1$ in the logarithm and extend
the upper integration limit to $u = \infty$.  In this way, we
obtain 
\begin{eqnarray}
v_p \to \infty \,: \qquad\qquad\qquad &&
\nonumber\\
 \left\{  {\cal A}^Q_{b,\smS} {1 \over \beta_b v_p} 
\,,\, C^{ll Q}_{b,\smS} \right\}  &=& 
  {e_p^2 \kappa_b^2 \over 4\pi} \,
   { 1 \over \beta_b v_p } \,
  \ln\hskip-0.1cm  
  \left( { 2 m_{pb} v_p \over \hbar K}
  \right) 
 \, \Big\{ {1 \over \beta_b m_b v_p^2} \,,\, 1 \Big\} 
 \,.
\label{vlim}
\end{eqnarray}
This is to be combined with the large velocity limits
(\ref{semicirlim}) and (\ref{different}) of 
${\cal A}^\smlt_{b,\smR}$ and $C^{ll \smlt}_{b,\smR}$ .  
The coefficients ${\cal A}_b$ in Eq's.~(\ref{vlim}) and 
(\ref{semicirlim}) are both dominated
by the electron contribution, a contribution $m_b / m_e$ larger
than that of an ion $b$. To bring out the nature of the result,
it is convenient to use the squared electron plasma frequency
$ \omega^2_e = \kappa_e^2 / \beta_e m_e$, and we have
\begin{eqnarray}
v_p \to \infty \,: \, &&
\nonumber\\
  {\cal A}_e &=& 
  {e_p^2 \over 4\pi} \,
   { \omega_e^2  \over  v_p^2 } \,
  \ln \left( { 2 m_{pe} v_p^2 \over \hbar \omega_e}
  \right)  \,.
\label{alimmm}
\end{eqnarray}
On the other hand, all plasma species contribute to 
\begin{eqnarray}
v_p \to \infty \,: \, &&
\nonumber\\
 C^{ll}_b  &=& 
  {e_p^2 \kappa_b^2 \over 4\pi} \,
   { 1 \over \beta_b v_p } \,
 \ln \left( { 2 m_{pb} v_p \over \hbar \kappa_\smD}
  \right) \,.
\label{clllim}
\end{eqnarray}
Since $ {\cal A}_b (1 / \beta_b v_p)$ behaves as 
$1 / v_p^3$, it vanishes more rapidly than 
$C^{ll}_b$ for large $v_p$ 
[of relative order $ 1 / \beta_b m_p v_p^2 $] 
and so the general connection 
$ {\cal B}_b = C^{ll}_b - {\cal A}_b (1 / \beta_b v_p) $ 
gives
\begin{eqnarray}
v_p \to \infty \,: &&
\nonumber\\
&&
{\cal B}_b = {e_p^2 \kappa_b^2 \over 4 \pi} \,
{1 \over \beta_b v_p} \, \ln \left( 
{2 m_{pb} v_p \over \hbar \kappa_\smD} \right) \,.
\label{Bad}
\end{eqnarray}

The limit that we have just described applies to the situation in
which the projectile velocity $v_p$ is so big that 
$ m_e v_p^2 \gg T_e $. 
For the case where the projectile is an ion, this limit implies 
that the projectile kinetic energy 
$ E_p = m_p v_p^2 / 2 $ 
is even greater than a typical plasma temperature $T$ by the
additional large factor of 
$ m_p / m_e $. The limit in which $ E_p \gg T $ but yet
$ m_e v_p^2 $ is not large in comparison with $T$ is also of
interest.  In this case, so long as 
$ ( E_p / T)^3 \, ( m_e / m_p ) > 1 $, 
${\cal A}_e$ still dominates over the ionic contributions to
${\cal A}_b$, but a detailed evaluation of ${\cal A}_e$ is
required.  In this intermediate case, the limit (\ref{Bad}) still
holds for the contribution of the ions in the plasma, while the
contribution of the electrons in the plasma is of relative order 
$ ( m_e v_p^2 / T )^{1/2} $ and thus may be neglected.

\section{Transport Equation Validity Details}
\label{details}

Here we provide the detailed computation of the result
(\ref{Tspread}) used in Sec.~\ref{valid} for the error of the
Fokker-Planck equation as measured by different evaluations of 
the increase in transverse energy. Again we note that   
since the $\nu < 3$ contribution to the Fokker-Planck coefficient
$C^{kl}_b$ is the same Lenard-Balescu equation that is used to
evaluate the transverse energy in this region, the difference
defining $\Delta_{b}$ is given by just the $\nu > 3$ parts,
\begin{equation}
\Delta_{b} = 
\left. {d E^\smgt_{\perp\,b} \over dt } \right|_{\rm exact}
- \left. {d E^\smgt_{\perp\,b} \over dt } \right|_{F-P} \,,
\end{equation}
with both terms computed from the scattering cross section
formula that is equivalent to the Boltzmann equation as is
described in Sec.~\ref{big}.  This is the computation to
which we now turn.

The exact transverse energy change in a scattering with the 
initial and final projectile momenta 
$ {\bf p}_p$ and $ {\bf p}'_p$, is given by
\begin{equation}
\Delta E_\perp = { {p'_{p \perp}}^2 \over 2m_p } 
= { 1 \over 2m_p \, p_p^2 } \,
\left[ {{\bf p}_p'}^2 \, {\bf p}_p^2 - 
({\bf p}_p' \cdot {\bf p}_p )^2 \right]
\,,
\end{equation}
where we now append a subscript and write the projectile mass as 
$m_p$ to avoid possible confusion. Here 
$ {\bf p}_p' = {\bf q} + {\bf p}_p $, 
where ${\bf q}$ is the Galilean invariant momentum transfer, and so 
\begin{equation}
\Delta E_\perp 
= { 1 \over 2m_p \, p_p^2 } \,
\left[ {{\bf p}_p}^2 \, {\bf q}^2 - ({\bf p}_p \cdot {\bf q} )^2 \right]
\,.
\label{sidewise}
\end{equation}
Since the relative velocity ${\bf v}_{pb}$ is the only vector
available to describe the Galilean invariant cross section, 
the tensor
$$
\int d\sigma_{pb} \,\, q^k \, q^l 
$$
can only involve the tensors $v^k_{pb} \, v^l_{pb}$ and $\delta^{kl}$. 
From energy conservation in the center-of-mass system,
\begin{eqnarray}
{\bf v}_{pb} \cdot {\bf q} &=& {1\over2} 
({\bf v}_{pb}' + {\bf v}_{pb}) \cdot m_{pb} 
( {\bf v}_{pb}' - {\bf v}_{pb})
- {1\over2}  ({\bf v}_{pb}' - {\bf v}_{pb}) \cdot 
m_{pb} ({\bf v}_{pb}' - {\bf v}_{pb})
\nonumber\\[3pt]
&=& {1 \over 2}\, m_{pb}\left({\bf v}_{pb}^{\prime\,2}-
{\bf v}_{pb}^2 \right) - 
{1 \over 2}\, m_{pb}\left({\bf v}_{pb}^\prime-
{\bf v}_{pb}\right)^2 
\nonumber\\
&=&
0 - {q^2 \over 2 m_{pb} } \,,
\end{eqnarray}

\noindent
where
\begin{equation}
{1 \over m_{pb} } = { 1 \over m_p} + { 1 \over m_b} 
\end{equation}
defines the reduced mass $m_{pb}$. 
Thus, by contracting with $\delta^{kl}$ and with 
$v^k_{pb} v^l_{pb}$, it is easy to verify that, weighted by the cross
section, we have, effectively, 
\begin{equation}
q^k q^l \to \left( \delta^{kl} - \hat v_{pb}^k \hat v_{pb}^l
\right) { q^2 \over \nu -1 } 
+ \left(\, \nu \, \hat v_{pb}^k \hat v_{pb}^l - \delta^{kl} \right)
\, { \left( q^2 \right)^2 \over 4 (\nu -1) m_{pb}^2 v_{pb}^2 }
\,.
\end{equation}
Accordingly, we may write the transverse energy change 
(\ref{sidewise}) for scattering off particles of plasma
species $b$ as, effectively,
\begin{eqnarray}
\Delta E_{\perp\,b} &\to& { q^2 \over 2 m_p p_p^2 } \,
\left[ {\bf p}_p^2 - {1 \over \nu -1} \, \left( {\bf p}_p^2 -
   ({\bf p}_p \cdot \hat{\bf v}_{pb} )^2 \right) \right] 
\nonumber\\
&& \qquad\qquad
- { q^2 \over 2 m_p p_p^2 } \, { q^2 \over 4 m_{pb}^2 v_{pb}^2 }
\left[ {\bf p}_p^2 - {\nu \over \nu -1} \, \left( {\bf p}_p^2 -
   ({\bf p}_p \cdot \hat{\bf v}_{pb} )^2 \right) \right] 
\,.
\end{eqnarray} 

This appears in the general formula (\ref{qloss}) which entails
an integral involving
\begin{eqnarray}
\int d^\nu {\bf p}_b \, f_b({\bf p}_b) &\sim&
\int d^\nu {\bf v}_{pb} \, \exp\left\{ - {1\over2} \beta_b m_b
\left( {\bf v}_p - {\bf v}_{pb} \right)^2 \right\} 
\nonumber\\
&\sim&
\int_0^\pi \sin^{\nu-2}\theta \, d\theta \, \exp\left\{ \, \beta_b
m_b \, v_p v_{pb} \, \cos\theta \, \right\} \,,
\end{eqnarray}
with
\begin{equation}
{\bf p}_p^2 - \left( {\bf p}_p \cdot \hat{\bf v}_{pb} \right)^2
		= p_p^2 \, \sin^2 \theta \,,
\end{equation}
and the remainder of the integrand independent of the angle 
$\theta$ between $\hat v_{pb}$ and $\hat{\bf v}_p$.
Hence we encounter
\begin{eqnarray}
  \int_0^\pi d\theta \, \sin^2\theta \, \sin^{\nu-2}\theta \, 
  e^{\beta_b m_b \, v_p v_{pb} \, \cos\theta }
  &=& 
  - \int_0^\pi \sin^{\nu-1}\theta \, {1 \over \beta_b m_b \, v_p v_{pb} } 
 \, d \, e^{\beta_b m_b \, v_p v_{pb} \, \cos\theta }
\nonumber\\
&=&
(\nu -1) \,
\int_0^\pi \sin^{\nu-2}\theta \, d\theta 
{\cos\theta \over \beta_b m_b \, v_p v_{pb} } \,
e^{\beta_b m_b \, v_p v_{pb} \, \cos\theta } \,,
\nonumber \\
&&
\end{eqnarray}
with the last line following by partial integration.  We
thus have, effectively, 
\begin{equation}
  \sin^2\theta \to \frac{(\nu-1) \cos\theta}
  {\beta_b m_b v_p v_{pb}}  ,
\end{equation}
or
\begin{equation}
{\bf p}_p^2 - ({\bf p}_p \cdot \hat{\bf v}_{pb} )^2 
\to (\nu -1) \, p_p^2 \, { \hat{\bf v}_p \cdot \hat{\bf v}_{pb} \over 
\beta_b m_b \, v_p v_{pb} } \,,
\end{equation}
and 
\begin{equation}
\Delta E_{\perp\,b} \to { q^2 \over 2 m_p } \, \left[ 1 -
{ \hat{\bf v}_p \cdot \hat{\bf v}_{pb} \over 
\beta_b m_b \, v_p v_{pb} } \right] 
- { (q^2)^2 \over 8 m_p \, m_{pb}^2 \, v_{pb}^2 } \,
\left[ 1 - \nu \, { \hat{\bf v}_p \cdot \hat{\bf v}_{pb} \over 
\beta_b m_b \, v_p v_{pb} } \right] \,.
\end{equation}

Recalling the relationship (\ref{tc}) of the Fokker-Planck
approximation to the rate of transverse energy increase and the
connection (\ref{abc}) amongst the ${\cal A}_b$, ${\cal B}_b$, and 
$C^{ll}_b$ coefficients, we have
\begin{equation}
\left. {d E^\smgt_{\perp\,b} \over dt} \right|_{F-P} 
= {1 \over m_p} \, \left[ C_b^{ll \smgt} - 
   {\cal A}_b^\smgt {1 \over \beta_b v_p } \right] \,.
\end{equation}
Making use of the formula (\ref{eitherloss}) for the 
$C^{ll \smgt}_b$ and ${\cal A}_b^\smgt$ coefficients, and
referring to the general formula (\ref{qloss}), we see that
the previously defined error measure is given by
\begin{equation}
\Delta_{b} = \int {d^\nu {\bf p}_b \over (2\pi\hbar)^\nu} \,
f_b({\bf p}_b) \, v_{pb} \, \int d\sigma_{pb} \left\{
\Delta E_\perp - {q^2 \over 2 m_p} + 
{ \hat{\bf v}_p \cdot \hat{\bf v}_{pb} \over \beta_b m_b \, 
v_p v_{pb} } \, { q^2 \over 2 m_p } \right\} \,.
\end{equation} 
All the potential infrared singular terms, the terms involving a
single power of $q^2$, cancel, and there remains, in the limit
$\nu \to 3$ which now may be taken, 
\begin{equation}
\Delta_{b} = - \int {d^3 {\bf p}_b \over (2\pi\hbar)^3} \,
f_b({\bf p}_b) \, {1 \over 8 m_p \, m_{pb}^2 v_{pb} } \,
\left[ 1 - 3 
{ \hat{\bf v}_p \cdot \hat{\bf v}_{pb} \over \beta_b m_b \, 
v_p v_{pb} } \right] \, 
\int d\sigma_{pb} \, ( q^2)^2 \,.
\end{equation} 
The cross section weighted integral of $(q^2)^2$ that appears
here may be evaluated, for example, by inserting an extra factor
of $q^2 = 4 \, p_p^2 \, \sin^2 \theta/2 $ in Eq.~(\ref{q2})
restricted to $\nu =3$.  The result is that 
\begin{equation}
\int d \sigma_{pb} \, (q^2)^2 = { (e_p e_b)^2 \over \pi} \,
		m_{pb}^2 \,.
\end{equation}
We write
\begin{equation}
{\hat{\bf v}_{pb} \over v_{pb}^2 } = {\partial \over 
\partial {\bf v}_b } \, {1 \over v_{pb} } \,,
\end{equation}
and integrate the velocity derivative by parts so that it
acts on the distribution function $f_b({\bf p}_b)$, giving,
effectively, 
\begin{equation}
{\partial \over \partial {\bf v}_b } \to - \beta_b m_b
		{\bf v}_b \,.
\end{equation}
Hence,  
\begin{equation}
\Delta_{b} = - { (e_p e_b)^2 \over 4\pi} \, { 1 \over 2m_p} \,
\int {d^3 {\bf p}_b \over (2\pi\hbar)^3} \,
f_b({\bf p}_b) \, {1 \over v_{pb} } \,
\left[ 1 - 3 
{ \hat{\bf v}_p \cdot {\bf v}_b \over v_p } \right] \,. 
\end{equation} 
We write the factor $1 / v_{pb}$ in terms of a Gaussian integral,
interchange integrals, complete the square, and change variables
as in the computation of Eq.~(\ref{classical}).
This gives
\begin{equation}
\Delta_{b} = - {e_p^2 \over 4\pi} \, { \kappa^2_b \over 2 m_p}
\, \left({ m_b \over 2\pi \beta_b} \right)^{1/2} \,
\int_0^1 {du \over \sqrt u} \, [1 - 3u] \, \
\exp\left\{ - {1\over2} \beta_b m_b v_p^2 \, u \right\} \,,
\label{TspreadT}
\end{equation}
which is the result previously quoted in Eq.~(\ref{Tspread}) in
Sec.~\ref{valid}.

\section{Rate at Which Different Species Come Into Equilibrium}

Plasmas may be created that contain different species which are
at different temperatures.  This happens, for example, when a
plasma experiences a laser pulse which preferentially heats the
light electrons that have the larger scattering cross section.
Here we use the methods that we have developed to compute the
rate at which the various plasma species come into thermal 
equilibrium.

\subsection{Introduction and Summary}

We shall assume that the particles of two species $a$ and $b$ in the
plasma are individually in thermal equilibrium, but at different
temperatures $T_a$ and $T_b$. We shall compute the leading and
subleading orders, as we have done throughout, of the rate 
$d {\cal E}_{ab} /dt$ at which the energy
density ${\cal E}_a$ of species $a$ changes because of its interaction
with species $b$. Since
\begin{equation}
{\cal E}_a = \int {d^3{\bf p}_a \over (2\pi)^3} \, 
{ {\bf p}_a^2 \over 2 m_a} \, f_a({\bf p}_a) \,,
\end{equation}
where $  f_a({\bf p}_a) $ is a spatially homogeneous thermal
distribution at temperature $T_a = \beta_a^{-1}$, the Fokker-Planck
equation (\ref{fp}) gives
\begin{equation} 
{ d {\cal E}_{ab} \over dt} = - {\cal C}_{ab} \, 
	\left( T_a - T_b \right) \,.
\end{equation}
in which
\begin{equation}
{\cal C}_{ab} = \int { d^3 {\bf p}_a \over (2\pi)^3 }
f_a({\bf p}_a) \, \beta_a v_a \, {\cal A}_b({\bf p}_a)   \,.
\label{ccoeff}
\end{equation}
Since the energy loss of one plasma species is another's gain,
the rate of energy density transfer is skew-symmetric,
\begin{equation} 
{ d {\cal E}_{ab} \over dt} = - { d {\cal E}_{ba} \over dt} 
\,;	
\end{equation}
whence 
\begin{equation} 
{\cal C}_{ab} = {\cal C}_{ba} 
\end{equation}
are symmetric coefficients.

A plasma consists of 
light electrons of mass $m_e = m$ and heavy ions of mass $m_i \gg
m$ which we shall generically denote by $M$.
Before plunging into the details of our computations, we review
the well-known justification for assuming that the electrons and
ions in the plasma are themselves in internal thermal 
equilibrium at the separate temperatures $T_e$ and $T_i$. To do this, 
we note that, as shown for example in Eq.~(\ref{classicdone})
in the results below, 
the mass dependence of the rate appears predominately in
\begin{equation}
{\cal C}_{ab} =
 - { \sqrt{ m_a m_b }  \over 
\left( m_a T_b + m_b T_a \right)^{3/2} } \, \cdots \,.
\end{equation}
Let us now use this result to compute the
rate at which the electrons come into equilibrium with themselves. For
this purpose, we imagine the very simple situation in which the
electrons are in two pieces, one with temperature $T_a$, the other
with temperature $T_b$. Then the rate at which these two pieces come
into equilibrium is controlled by the factor
\begin{equation}
{ 1 \over ( T_a + T_b )^{3/2} \, m^{1/2} } \,.
\end{equation}
If a  similar partition of the ions into two parts at different
temperatures were made, the parts would come into equilibrium at a
rate controlled by the factor
\begin{equation}
{ 1 \over ( T_a + T_b )^{3/2} \, M^{1/2} }  \,.
\end{equation}
Thus the rate at which the ions come into equilibrium amongst
themselves is a factor $\sqrt{m/M}$ slower than the corresponding rate
for the electrons. Now going back to our original problem of electrons
and ions at different temperatures $T_e$ and $T_i$, we see that if the
temperatures are not greatly different, the rate at which the two
species come into thermal equilibrium is controlled by the factor
\begin{equation}
{ m^{1/2} \over ( T_e )^{3/2} \, M }  \,,
\end{equation}
which is a factor of $m/M $ smaller than the rate at which the
electrons come into equilibrium amongst themselves and a factor 
$\sqrt{m/M}$ smaller than the equilibrium rate for the ions alone.
Thus our work which treats the electrons and ions as being in separate
thermal equilibrium but with two different temperatures and
concentrating on computing the rate at which the ions and electrons
come into thermal equilibrium is justified by the very large ion --
electron mass ratio. 

Let us consider the case in which the electrons have come to
temperature $T_e$ and all ions species have equilibrated to
a common ion temperature $T_i$. With electron and ion specific 
heats per unit volume $c_e$ and $c_\smI$ defined by $d {\cal E}_e 
= c_e dT_e$ and $d {\cal E}_\smI = \sum_i {\cal E}_i  =c_\smI dT_i$, 
we define the the rate $\Gamma$ at which the ionic and electronic 
temperatures come into equilibrium by
\begin{equation}
{d \over dt} \left( T_e - T_i \right) = 
		- \Gamma \, \left( T_e - T_i \right) \,,
\end{equation}
with 
\begin{equation}
\Gamma = {\cal C}_{e \smI} \, 
		\left( { 1 \over c_e} + { 1 \over c_\smI} \right) \ .
\end{equation}
The sum of Eq's.~(\ref{smart}) and (\ref{qei}) [evaluated for 
$ K = \kappa_e$ as Eq.~(\ref{smart}) requires] give the limit 
\begin{eqnarray}
T_i m_e &\ll& T_e m_i \,: 
\nonumber\\
{\cal C}_{e\smI} &=& {\sum}_i \, {\cal C}_{ei} = 
{\kappa_e^2 \over 2\pi } \, \omega_\smI^2 \, 
\sqrt{ \beta_e m_e \over 2\pi} \, {1\over2} \,
\left\{ \ln\left( { 8 T_e^2 \over \hbar^2 \omega_e^2} \right)
- \gamma - 1 \right\} \,,
\label{spit}
\end{eqnarray}
where 
\begin{equation}
\omega^2_\smI = {\sum}_i \, \omega^2_i = {\sum}_i \, 
	{e^2_i n_i \over m_i } 
\end{equation}
is the sum over all the squared ionic plasma frequencies.  
The overall coefficient of the Coulomb logarithm 
in the energy transfer rate (\ref{spit})
was obtained long ago by Spitzer \cite{spit1} as described in his
book \cite{spit2}.  However, our determination, as always, gives
not only this coefficient, but also a precise definition of the 
value of the Coulomb logarithm, the constant under the Coulomb
logarithm.

The development in the subsequent sections follows the order
that we have previously used in the stopping power work. 

\subsection{Classical Results}

The decomposition (\ref{classicall}) previously given for the
classical contributions to the ``S'' and  ``R'' contributions 
to the ${\cal A}_b$ functions
gives a corresponding division of the temperature equilibrium 
coefficients,
\begin{equation}
  {\cal C}^\smC_{ab} = {\cal C}^\smC_{ab,\smS} +
  {\cal C}^\smlt_{ab,\smR} \,.
\label{cclassicall}
\end{equation}

The result (\ref{wonderclassic}) for 
${\cal A}^\smC_{b, \smS}$ placed in Eq.~(\ref{ccoeff}) gives
\begin{eqnarray}
{\cal C}^\smC_{ab,\smS} 
 &=& 
-  {\beta_a e_a^2 \kappa_b^2 \over 4\pi} \, 
  \left( { \beta_b m_b \over  2\pi } \right)^{1/2} \,
  \int {d^3{\bf p}_a \over (2\pi\hbar)^3 } \, f_a({\bf p}_a) \, 
 v_a^2 \,  \int_0^1 du \, u^{1/2} \,
  \exp\left\{ - {1 \over 2} 
  \beta_b m_b v^2_a \, u \right\}
\nonumber\\
  && \qquad\qquad\qquad\qquad
  \left[   \ln \left( { e_a e_b  \over 4 \pi} \,
K \, { \beta_b m_b \over m_{ab} } \, { u \over 1-u} \right) 
+ 2 \gamma -2 \right] \,.
\end{eqnarray}
The particle number density may be expressed as 
\begin{equation}
{ d^3 {\bf p} \over (2\pi\hbar)^3 } \, f({\bf p}) =
n \, \left({\beta m \over 2 \pi} \right)^{3/2} 
d^3 {\bf v} \, \exp\left\{ - {1 \over 2} \, \beta m v^2 \right\} \,.
\label{partno}
\end{equation}
The resulting Gaussian integration yields
\begin{eqnarray}
{\cal C}^\smC_{ab,\smS} 
 &=& 
-  {\kappa_a^2 \kappa_b^2 \over 4\pi} \, 
  \left( { \beta_b m_b \over  2\pi } \right)^{1/2} \,
 \int_0^1 du \, u^{1/2} \,
 { 3 \, (\beta_a m_a)^{3/2} \over 
\left( \beta_a m_a + \beta_b m_b u \right)^{5/2} }
\nonumber\\
  && \qquad\qquad\qquad\qquad
  \left[   \ln \left( { e_a e_b  \over 4 \pi} \,
K \, { \beta_b m_b \over m_{ab} } \, { u \over 1-u} \right) 
+ 2 \gamma -2 \right] \,.
\label{wwonderclassic} 
\end{eqnarray}
To place this in a form that exhibits the symmetry under the
interchange of the $a$, $b$ labels, we change integration
variables to 
\begin{equation}
 s = {\beta_b m_b \, u \over 1 - u } \, V^2_{ab} \,,
\end{equation}
where
\begin{equation}
V_{ab}^2  =
{ \beta_a m_a + \beta_b m_b \over \beta_a m_a \beta_b m_b }
= {1 \over \beta_a m_a} + {1 \over \beta_b m_b} =
{T_a \over m_a} + {T_b \over m_b} \,,
\label{avev}
\end{equation}
is an average squared thermal velocity. 
This gives
\begin{eqnarray}
{\cal C}^\smC_{ab,\smS} 
 &=& 
-  {\kappa_a^2 \kappa_b^2 \over 4\pi} \,
 { 3 \over \sqrt{2\pi} } \, { (\beta_a m_a \beta_b m_b)^{1/2}
\over \left( \beta_a m_a + \beta_b m_b \right)^{3/2} } \,
 \int_0^\infty ds \, s^{1/2} \,
\left( 1 +  s \right)^{-5/2}
\nonumber\\
  && \qquad\qquad\qquad\qquad\qquad
  \left[   \ln \left( { e_a e_b  \over 4 \pi} \,
K \, { s \over m_{ab} \, V^2_{ab} }\right) 
+ 2 \gamma -2 \right] \,.
\label{wwonderclassicc} 
\end{eqnarray}
The integrals that appear here provide a standard representation
of the beta function,
\begin{equation}
B(x,y) = { \Gamma(x) \Gamma(y) \over \Gamma(x+y) }
       = \int_0^\infty ds \, { s^{x-1} \over (1 + s )^{x+y} } \,.
\end{equation}
This is so because
\begin{equation}
\int_0^\infty ds \, s^{1/2} \, (1 + s)^{-5/2} =
{ \Gamma(3/2) \, \Gamma(1) \over \Gamma(5/2) } = { 2 \over 3} \,,
\end{equation}
and, using
\begin{equation}
\ln s = \lim_{\epsilon \to 0} \, {s^\epsilon - 1 \over \epsilon}
\,,
\end{equation}
we also have
\begin{equation}  
\int_0^\infty ds \, s^{1/2} \, (1 + s)^{-5/2} \, \ln s =
{ \Gamma(3/2) \, \Gamma(1) \over \Gamma(5/2) } \,
\left[  \psi(3/2) - \psi(1) \right] 
 = { 2 \over 3} \, [ 2 - 2 \ln 2 ]  \,,
\end{equation}
and thus the evaluation
\begin{eqnarray}
{\cal C}^\smC_{ab,\smS} 
 &=& 
-  {\kappa_a^2 \kappa_b^2 } \, 
\, \left( {1 \over 2\pi} \right)^{3/2}  \, { 
(\beta_a m_a \beta_b m_b)^{1/2} 
\over \left( \beta_a m_a + \beta_b m_b \right)^{3/2} } \,
  \left[   \ln \left( { e_a e_b  \over 4 \pi} \,
{ K \over 4 \, m_{ab} \, V^2_{ab} }\right) 
+ 2 \gamma  \right] \,.
\label{classicdone}
\end{eqnarray}

The long-distance, plasma screening correction 
$ {\cal A}^\smlt_{b,\smR} $ presented in Eq.~(\ref{nun}) 
gives 
\begin{eqnarray}
{\cal C}^\smlt_{ab,\smR} &=&
  {e_a^2 \over 4 \pi } \,  \beta_a \,
 \int {d^3{\bf p}_a \over (2\pi\hbar)^3}
\, f_a({\bf p}_a) 
\nonumber\\
&&
 { i \over 2 \pi }
  \int_{-1}^{+1} d\cos\theta \, 
	{\rho_b(v_a\cos\theta) \over 
	\rho_{\rm total}(v_a\cos\theta) }
 F(v_a \cos\theta) \ln \left( { F(v_a
  \cos\theta) \over K^2 }\right) \, v_a \,  \cos\theta \,,
\nonumber\\
&&
\label{nnun} 
\end{eqnarray}
where, we recall,  the function $F(u)$  may be expressed in the 
dispersion form
\begin{equation}
  F(u)  =  - \int_{-\infty}^{+\infty} dv \,
{ \rho_{\rm total}(v) \over u
  - v + i \eta } \,,
\end{equation}
with the limit $\eta \to 0^+$ understood.
The spectral weight is defined by
\begin{equation}
  \rho_{\rm total}(v)  = {\sum}_c \,
        \rho_c(v) \,,
\end{equation}
where
\begin{equation}
\rho_c(v) =
\kappa^2_c \,v\, \sqrt{ \beta_c m_c\over 2\pi }
  \exp\left\{
  -{1 \over 2} \beta_c m_c v^2 \right\} \,.
\end{equation}
We insert
\begin{equation}
1 = \int_{-\infty}^{+\infty} dv \, \delta(v - v_a \cos\theta)
\label{insert}
\end{equation}
in the integrand of Eq.~(\ref{nnun}), use the form 
(\ref{partno}) of the particle number density, interchange
integrals, and perform all the integrals save that involved in
the insertion (\ref{insert}). Thus
\begin{eqnarray}
{\cal C}^\smlt_{ab,\smR} &=&
 { \kappa_a^2 \kappa_b^2 \over 2\pi} \,
\left( {\beta_a m_a \over 2\pi} \right)^{1/2} \, 
\left( {\beta_b m_b \over 2\pi} \right)^{1/2} \,
\nonumber\\
&& \quad
 \int_{-\infty}^{+\infty} dv \, v^2 \,
\exp\left\{ - {1\over2} 
\left[ \beta_a m_a + \beta_b m_b \right] \, v^2 \right\}  
 { i \over 2 \pi } \,
{ F(v) \over \rho_{\rm total}(v) }
\, \ln \left( { F(v) \over K^2 }\right) \,.
\label{nunnn}
\end{eqnarray}
Note that this formula exhibits explicitly the symmetry 
$ {\cal C}^\smlt_{ab,\smR} = {\cal C}^\smlt_{ba,\smR} $. 

Since 
\begin{equation}
F(v) - F(-v) = F(v) - F(v)^* = 2\pi i \rho_{\rm total}(v) \,,
\end{equation}
the $\ln K$ dependence of 
$ {\cal C}^\smlt_{ab,\smR} $ 
appears in 
\begin{eqnarray}
{\cal C}^\smlt_{ab,\smR} &=&
 { \kappa_a^2 \kappa_b^2 \over 2\pi} \,
\left( {\beta_a m_a \over 2\pi} \right)^{1/2} \, 
\left( {\beta_b m_b \over 2\pi} \right)^{1/2} \,
 \int_{0}^{+\infty} dv \, v^2 \,
\exp\left\{ - {1\over2} 
\left[ \beta_a m_a + \beta_b m_b \right] \, v^2 \right\}  
\, \ln K^2 + \cdots 
\nonumber\\
&=&
 { \kappa_a^2 \kappa_b^2 } \,
{ \left( \beta_a m_a \beta_b m_b \right)^{1/2} \over
\left( \beta_a m_a + \beta_b m_b \right)^{3/2} }
 \left( { 1 \over 2 \pi } \right)^{3/2} \, \ln K + \cdots \,\,.
\end{eqnarray}
Hence the sum 
(\ref{cclassicall}) 
of Eq's.~(\ref{classicdone}) 
and (\ref{nunnn}) is
independent of the particular value of the arbitrary wave number 
$K$ as it must be.

Although a numerical computation is needed for the general
evaluation of ${\cal C}^\smlt_{ab,\smR} $, it can be found 
when  $\beta_e m_e \ll \beta_i m_i $, or
\begin{equation}
T_i m_e \ll T_e m_i \,,
\end{equation}
where $m_i$ is a typical ion mass and $T_i$ is the common
ion temperature.
This is the case that is usually of interest in applications. 
Since $ m_e / m_i < 10^{-3}$, this constraint
holds unless the ion temperatures are very much larger than the
temperature of the electrons in the plasma.  As a first step, 
we write Eq.~(\ref{nunnn}) as
\begin{eqnarray}
{\cal C}^\smlt_{ei,\smR} &=&
{ \kappa_e^2 \over 2\pi} \,
\left( {\beta_e m_e \over 2\pi} \right)^{1/2} \, 
\nonumber\\
&& \quad
 \int_{-\infty}^{+\infty} dv \, v \,
\exp\left\{ - {1\over2}  \beta_e m_e  \, v^2 \right\}  
{ \rho_i(v) \over \rho_{\rm total}(v) } \, 
 { i \over 2 \pi } \, F(v) 
\, \ln \left( { F(v) \over K^2 }\right) \,.
\end{eqnarray}
This limit under consideration is formally equivalent to the limit
$m_e \to 0$, and so
\begin{eqnarray}
T_i m_e \ll T_e m_i \,: \qquad &&
\nonumber\\
{\sum}_i \, \rho_i(v) &=& \rho_{\rm total}(v) \,.
\end{eqnarray}
Hence, defining
\begin{equation}
{\cal C}^\smlt_{e\smI,\smR} = {\sum}_i \,
 {\cal C}^\smlt_{ei,\smR} \,,
\end{equation}
we have 
\begin{eqnarray}
T_i m_e \ll T_e m_i \,: \quad &&
\nonumber\\
{\cal C}^\smlt_{e\smI,\smR} &=&
{ \kappa_e^2 \over 2\pi} \,
\left( {\beta_e m_e \over 2\pi} \right)^{1/2} \, 
 \int_{-\infty}^{+\infty} dv \, v \,
 { i \over 2 \pi } \, F(v) 
\, \ln \left( { F(v) \over K^2 }\right) \,.
\end{eqnarray}
The integrand that appears here is analytic in the upper-half
complex $v$ plane. Remembering that we are to take the formal
limit $m_e \to 0$ first, for large $v$ we have
\begin{equation}
F(v) \to \kappa_e^2 - {\sum}_i \, { \omega_i^2 \over v^2 } \,,
\end{equation}
where
\begin{equation}
\omega_i^2 = {e_i^2 n_i \over m_i } = {\kappa_i^2 \over \beta_i m_i }
\,,
\end{equation}
are the squared ionic plasma frequencies.  If we take
\begin{equation}
K^2 = \kappa^2_e \,,
\end{equation}
then $\ln \left( F(v) / K^2 \right) $ vanishes for large $v$, and 
\begin{eqnarray}
|v| \to \infty \,: \qquad\qquad\qquad && 
\nonumber\\
 F(v) \, \ln \left( { F(v) \over K^2 }\right) &\to& 
- {\sum}_i \, { \omega_i^2 \over v^2 } \,.
\end{eqnarray}
Thus at large $|v|$ the integral behaves as $ \int dv / v$ . We
add and subtract the corresponding contour integral over a
semi-circle $C$ at infinity in the upper half complex $v$ plane.  The
original integral with the $i\eta$ prescription is
equivalent to one over a straight line just above the real axis.
Adding the integral over the infinite semi-circle gives a closed
contour integral enclosing no singularities which thus vanishes. 
There remains
\begin{equation}
 { i \over 2\pi} \, \int_C {dv \over v} = {i \over 2\pi} \,
\int_0^\pi i d\theta = - {1 \over 2} \,,
\end{equation}
and therefore the evaluation
\begin{eqnarray}
T_i m_e \ll T_e m_i \,: \quad &&
\nonumber\\
{\cal C}^\smlt_{e\smI,\smR} &=&
- \frac{1}{2}\,{ \kappa_e^2 \over 2\pi} \,
\left( {\beta_e m_e \over 2\pi} \right)^{1/2} \, 
{\sum}_i \, \omega_i^2  \,.
\label{smart}
\end{eqnarray}
It should be emphasized that this evaluation is valid only for
the choice $K = \kappa_e$.

\subsection{Quantum Correction}

The complete coefficient is the sum of the previous 
classical result and a quantum correction
\begin{equation}
  {\cal C}_{ab}  = 
  {\cal C}^\smC_{ab} + {\cal C}^{\Delta Q}_{ab}  \,,
\label{fullc}
\end{equation}
corresponding to the separation given in Eq.~(\ref{fulla}).
Using the result (\ref{regb}) for ${\cal A}^{\Delta Q}_b $, 
we have
\begin{eqnarray}
 {\cal C}^{\Delta Q}_{ab} 
  &=&  - {\beta_a \over 2 m_b } 
\int {d^3{\bf p}_a
  \over (2\pi\hbar)^3} f_a({\bf p}_a) \, 
\int {d^3{\bf p}_b
  \over (2\pi\hbar)^3} f_b({\bf p}_b) \, 
{ e_a^2 e_b^2 \over 4 \pi \, v_{ab}^3}
{\bf v}_a \cdot  {\bf v}_{ab} \,
 \left\{ 2 \, {\rm Re} \,
  \psi \left( 1 + i \eta_{ab} \right) - \ln \eta^2_{ab}
  \right\} \,.
\nonumber\\
&&
\label{remainss}
\end{eqnarray}
The relative velocity is defined by
 ${\bf v}_{ab} = {\bf v}_a - {\bf v}_b$, 
and 
\begin{equation}
\eta_{ab} = {e_a e_b \over 4\pi \hbar v_{ab}} \,.
\end{equation}
The function $\psi(z)$ is the logarithmic derivative of the 
Gamma function $\Gamma(z)$.  We use the expression (\ref{partno})
for the particle number densities $f_a({\bf p}_a)$ and $f_b({\bf
p}_b)$.  Making use of the relative velocity,  
we may write the resulting product of exponentials as  
\begin{eqnarray}
\exp\left\{ - {1 \over 2} \, \beta_a m_a v_a^2 \right\} 
&& \exp\left\{ - {1 \over 2} \, \beta_b m_b v_b^2 \right\} 
=
\exp\left\{ - {1 \over 2} \,
 { v_{ab}^2 \over V^2_{ab} } \right\} 
\nonumber\\
&& \qquad
\exp\left\{ - {1 \over 2} \, (\beta_a m_a + \beta_b m_b )
\left[{\bf v}_a - { \beta_b m_b \over 
\beta_a m_a + \beta_b m_b } \, {\bf v}_{ab} \right]^2 \right\} 
\end{eqnarray}
where we have made use of the effective thermal velocity defined
in Eq.~(\ref{avev}). 
We change the integration variables from ${\bf p}_a \,, 
{\bf p}_b$ to ${\bf v}_a \,, {\bf v}_{ab}$ and perform the
resulting Gaussian integral in ${\bf v}_a$ to obtain 
\begin{eqnarray}
 {\cal C}^{\Delta Q}_{ab} 
  &=&  - {1 \over 2 } \, \kappa_a^2 \kappa_b^2 \, 
 { (\beta_a m_a \, \beta_b m_b)^{1/2} \over 
	( \beta_a m_a + \beta_b m_b )^{3/2} } \, 
	\left({1 \over 2\pi}\right)^{3/2}
\nonumber\\
&& \qquad
\int_0^\infty {d v_{ab}^2 \over 2 V_{ab}^2 }
\, \exp\left\{ - { v_{ab}^2 \over 2 V_{ab}^2 } \right\} \,  
\left\{ 2 \, {\rm Re} \,
  \psi \left( 1 + i \eta_{ab} \right) - \ln \eta^2_{ab}
  \right\} \,. 
\label{qcorr}
\end{eqnarray}
We have written Eq.~(\ref{qcorr}) in a form that makes
its dimensions obvious: 
$[\kappa^4 \, v] = {\rm cm}^{-3} \, {\rm sec}^{-1} $.   

The limiting behaviors of the result (\ref{qcorr}) are exhibited
if we make the variable change $ v_{ab}^2 = V^2_{ab} \, \zeta$,
which expresses 
\begin{eqnarray}
 {\cal C}^{\Delta Q}_{ab} 
  &=&  - {1 \over 2 } \, \kappa_a^2 \kappa_b^2 \, 
 { (\beta_a m_a \, \beta_b m_b)^{1/2} \over 
	(\beta_a m_a + \beta_b m_b)^{3/2} } \, 
	\left({1 \over 2\pi}\right)^{3/2} 
\nonumber\\
&& \qquad\qquad
{1\over2} \, \int_0^\infty d \zeta
\, \exp\left\{ - {1 \over 2 } \, \zeta \right\} \,  
 \left\{ 2 \, {\rm Re} \,
  \psi \left( 1 + i \bar\eta_{ab}\, \zeta^{-1/2}  \right) - 
\ln \bar\eta^2_{ab} \, \zeta^{-1} 
  \right\} \,, 
\label{qqcorr}
\end{eqnarray}
in which 
\begin{equation}
\bar\eta_{ab} = { e_a e_b \over 4\pi \hbar V_{ab} } \,.
\end{equation}
Here the exponential damps large $\zeta$ values, and so as far as
evaluating limits are concerned, we can consider $\zeta$ to be of
order unity in the curly braces in the integrand in
Eq.~(\ref{qqcorr}).  

The low temperature limit corresponds to the low velocity limit 
$ V_{ab} \to 0$.  This corresponds to the limit 
$ \bar\eta_{ab} \to \infty $, which is the formal limit 
$ \hbar \to 0$.  This is the formal classical limit.  Indeed,
since
\begin{eqnarray}
x \to \infty \,: &&
\nonumber\\
&&
 \left\{ 2 \, {\rm Re} \, \psi \left( 1 + i x  \right) - 
\ln x^2  \right\} \to 0 \,,
\end{eqnarray} 
the quantum correction 
$  {\cal C}^{\Delta Q}_{ab} $ 
vanishes in the classical limit as it should.  The major case of
interest is the rate at which ions and electrons in a plasma come
into thermal equilibrium.  Because of the very large ion/electron
mass ratio, 
$m_i/ m_e \gg 1$, to a very good approximation 
$V^2_{ei} = T_e / m_e$.
The condition that $ \bar\eta_{ei} \gg 1$ is thus equivalent to
$ T_e \ll (e_i e_e)^2 \, m_e / (4\pi\hbar)^2 $, or that the
temperature is much less than the binding energy of a 
hydrogen-like 
atom.  At such low temperatures, our assumption that we are
dealing with a fully ionized plasma is generally invalid.  Hence
this classical limit is only of limited physical interest.

The more relevant high temperature limit corresponds to the high
velocity limit in which $\bar\eta_{ab}$ becomes small.  Since
$ \psi(1) = - \gamma $, we have
\begin{eqnarray}
\bar\eta_{ab} \ll 1 \,: \qquad &&
\nonumber\\
 {\cal C}^{\Delta Q}_{ab} 
  &=&  - {1 \over 2 } \, \kappa_a^2 \kappa_b^2 \, 
 { (\beta_a m_a \, \beta_b m_b)^{1/2} \over 
	( \beta_a m_a + \beta_b m_b)^{3/2} } \, 
	\left({1 \over 2\pi}\right)^{3/2} 
\nonumber\\
&& \qquad\qquad 
 {1 \over 2} \, \int_0^\infty d \zeta 
\, \exp\left\{ - {\zeta \over 2 }  \right\} \,  
 \left\{ - 2 \gamma - 
\ln \left({ \bar\eta^2_{ab} \over \zeta } \right) \right\}  
\nonumber\\
&&
\nonumber\\
&=& 
    \kappa_a^2 \kappa_b^2 \, 
 { \beta_a m_a \, \beta_b m_b)^{1/2} \over 
	(\beta_a m_a + \beta_b m_b)^{3/2} } \, 
	\left({1 \over 2\pi}\right)^{3/2}  
 \, {1 \over 2} \,
\left\{  3 \gamma + \ln \left({ \bar\eta^2_{ab} \over 2} \right)
	  \right\} \,.
\label{qqlim} 
\end{eqnarray}

The first correction to this result is of relative order
$\eta^2_{ab}$.  The case of interest is the electron-ion energy
exchange rate where, using the notation $e_i = Z_i e$, 
\begin{equation}
\eta^2_{ei} = Z_i^2 \, { e^4 m_e \over (4\pi\hbar)^2 T_e}
\simeq Z_i^2 \, { 27\, {\rm eV} \over T_e }
\end{equation}
is very small.  Hence for the case of interest, the limit
(\ref{qqlim}) suffices.  This limit combines with the previous
calculation (\ref{classicdone}) of ${\cal C}^\smC_{ab,\smS} $ to
give, with sufficient accuracy, 
\begin{eqnarray}
{\cal C}^\smC_{ab,\smS} + {\cal C}^{\Delta Q}_{ab} 
 &=& 
 {\kappa_a^2 \kappa_b^2 } \, 
\, \left( {1 \over 2\pi} \right)^{3/2}  \, { 
(\beta_a m_a \beta_b m_b)^{1/2}
\over \left( \beta_a m_a + \beta_b m_b \right)^{3/2} } \,
  \left[   \ln \left( { 2^{3/2} \, m_{ab} \, V_{ab} \over 
\hbar K }\right) -  {\gamma \over 2}   \right] \,.
\label{C+C}
\end{eqnarray}
The complete coefficient is given by 
\begin{equation}
{\cal C}_{ab} = {\cal C}^\smlt_{ab, \smR} + 
\left[ {\cal C}^\smC_{ab,\smS} + {\cal C}^{\Delta Q}_{ab} \right]
\,.
\end{equation}

For the case of ion-electron relaxation, since 
$m_e / m_i \ll 1$, the electron mass can be neglected relative to
that of the ion.  Assuming that the ion temperature is not more
than an order of magnitude larger than the electron temperature, 
$T_i m_e \ll T_e m_i $.  With this restriction, the expression  
(\ref{C+C}) simplifies, and the result may be expressed as
\begin{eqnarray}
{\cal C}^\smC_{ei,\smS} + {\cal C}^{\Delta Q}_{ei} 
 &=& 
 {\kappa_e^2 \over 2\pi } \, \omega_i^2 \, 
\sqrt{ \beta_e m_e \over 2\pi} \, {1\over2} \,
  \left[   \ln \left( { 8 \, m_e T_e \over 
\hbar^2 K^2 }\right) -  \gamma   \right] \,.
\label{qei}
\end{eqnarray}

\bigskip\bigskip

\acknowledgements

J.~C.~Solem asked a question that sparked this 
work, and R.~F.~Sawyer contributed to a preliminary 
version. We would like to thank Charles Snell for
his work in checking some of our numerical
evaluations.

\newpage
\appendix

\section{Simple Example Illustrating the Method}
\label{hank}

Since the method used in this work is a novel 
one, we include in this Appendix a pedagogical, 
simple mathematical example that illustrates
the basic idea. This is the computation of
the behavior of the modified Hankel function 
$K_\nu(z)$ in the small argument $z$ limit 
with the index $\nu$ also small. The argument 
$z$ will play the role of the small parameter 
in our work; the index $\nu$ will play the 
role of the dimensionality except that in this
simple Bessel function example we shall examine 
the region where $\nu$ is near zero, not three.  
This example already appears in the preliminary
account\cite{brown1} of the new use of dimensional 
continuation, but it worth repeating here so as 
to have a clear, self-contained presentation.

The Hankel function has the integral representation
\begin{equation}
  K_\nu(z) = { 1 \over 2} \, \int_0^\infty {dk \over
  k} \, k^\nu \exp\left\{ - {z \over 2} \, \left( k +
  {1 \over k} \right) \right\}\,.
\label{int}
\end{equation}
Although $k$ is simply a dummy integration 
variable, it is convenient to think of it as 
a wave number or momentum variable.  When $z$ 
is small, $ \exp\left\{ - {z \over 2} \, \left( 
k + { 1 \over k} \right)\right\} $ may be replaced 
by $1$ except when one or the other of the factors 
$\exp\{ - z \, k /2 \}$ or $\exp\{ - z /(2 \, k)\}$ 
is needed to make the $k$ integration converge in 
the neighborhood of one of its end points. When 
$\nu$ is slightly less than zero, the integral
(\ref{int}) is dominated by the small $k$, 
``infrared or long-distance'', region.  In this 
case, only the $\exp\{- z /(2 \, k)\}$ factor 
is needed to provide convergence, and we have
\begin{eqnarray}
  \nu < 0 \, : \qquad\qquad\qquad &&
\nonumber\\
  K_\nu(z) &\simeq& { 1 \over 2} \, \int_0^\infty
  {dk \over k} \, k^\nu\exp\left\{ - {z \over 2
  \,k } \right\} \,.
\end{eqnarray}
The variable change $ k = z / ( 2 t) $ places 
this integral in the form of the standard 
representation of the gamma function, and we
thus find that the leading term for small $z$ 
in the region $\nu<0$ is given by
\begin{eqnarray}
  \nu < 0 \, : \qquad\qquad\qquad &&
\nonumber\\
  K_\nu(z) &\simeq& { 1 \over 2} \, \left(
  { z \over 2} \right)^\nu\Gamma(- \nu)
\nonumber\\
  &\simeq&  -{ 1 \over 2 \nu } \, \left(
  {z \over 2} \right)^{\nu} ( 1 + \nu
  \gamma )  \,,
\label{smaller}
\end{eqnarray}
where $\gamma = 0.5772 \cdots$ is Euler's 
constant.  Note that the second line describes 
the behavior for $\nu < 0$ near $\nu= 0$
including the correct finite constant as well 
as the singular pole term.

When $\nu$ is slightly greater than zero, the 
integral (\ref{int}) is dominated by the large 
$k$, ``ultraviolet or short-distance'' regions.
In this case, only the $\exp\{ - z \, k /2 \}$ 
factor is needed to provide convergence, and we 
have
\begin{eqnarray}
  \nu > 0 \, : \qquad\qquad\qquad &&
\nonumber\\
  K_\nu(z) &\simeq& { 1 \over 2} \, \int_0^\infty
  {dk \over k} \, k^\nu\exp\left\{ - {z \, k \over
  2}  \right\} \,.
\end{eqnarray}
The integral again defines a gamma function, 
and so
\begin{eqnarray}
  \nu > 0 \, : \qquad\qquad\qquad &&
\nonumber\\
  K_\nu(z) &\simeq& { 1 \over 2 \nu } \,
  \left({z \over 2} \right)^{-\nu} ( 1 - \nu
  \gamma )  \,,
\label{biger}
\end{eqnarray}
with again the result containing the correct 
finite constant as well as the singular pole 
term.

The result (\ref{smaller}) for $\nu < 0 $ can 
be analytically continued into the region $\nu>0$. 
In this region it involves a higher power of $z$  
than that which appears in the other evaluation
(\ref{biger}), and hence this analytic continuation 
of the leading result for $\nu < 0 $ into the region 
$\nu>0$ becomes subleading here.  Similarly, the 
result (\ref{biger}) for $\nu>0$ may be analytically 
continued into the region  $\nu<0$ where it now 
becomes subleading.  An examination of the defining
integral representation (\ref{int}) shows that these 
subleading analytic continuation terms are, in fact, 
the dominant, first-subleading
terms.\footnote{
\baselineskip 15pt
For example, subtracting the leading 
term (\ref{smaller}) for $\nu<0$ from the integral 
representation (\ref{int}) gives
$$
  K_\nu(z) - {1 \over 2} \left( {z \over 2} 
  \right)^\nu\Gamma( -\nu)= {1 \over 2} 
  \int_0^\infty {dk \over k} k^\nu  \left[
  e^{-z k/2 } - 1 \right] \, e^{- z / (2k) } \,.
$$
For $ 0 > \nu > -1$, the integral on the 
right-hand-side of the equation converges when 
the final exponential factor in the integrand 
is replaced by unity, the $z \to 0$ limit of this 
factor. Hence this final factor may be omitted in 
the evaluation of the first subleading term. A
partial integration presents the result as
$$
  {z \over 4 \nu }  \, \int_0^\infty dk \, k^\nu\, 
  e^{-z k/2} \,,
$$
whose evaluation gives precisely the analytic 
continuation of the leading term (\ref{biger}) 
for $\nu > 0$.} For $\nu > 0$ one term is leading 
and the other subleading, while for $\nu < 0$ 
their roles are interchanged. Thus their sum
\begin{equation}
  K_\nu(z) \simeq { 1 \over 2 \nu}
  \left\{ \left( { z \over 2} \right)^{-\nu}
  \left[ 1 - \nu \gamma \right] - \left( { z
  \over 2} \right)^{\nu} \left[ 1 + \nu \gamma
  \right] \right\}
\label{about}
\end{equation}
contains both the leading and the first subleading 
terms for both $\nu > 0$ and $\nu < 0 $. In the 
limit $\nu \to 0$ the (``infrared'' and ``ultraviolet'') 
pole terms in this sum cancel, with the variation 
of the residues of the poles producing a logarithm, 
yielding the familiar small $z$ result
\begin{equation}
  K_0(z) = - \ln ( z/2) - \gamma \,.
\end{equation}
It must be emphasized that the correct constant 
terms [$\ln 2 - \gamma $] are obtained by this method 
in addition to the logarithm $ -\ln z$ which is large 
for small $z$.  The result (\ref{about}) was derived 
from the analytic continuation of results that were
easy to compute in one or another region where 
either ``infrared'' or ``ultraviolet'' terms dominated. 
This is the essence of our method. Of course, the 
general result (\ref{about}) could be obtained by 
a more careful computation of both the leading and
first-subleading terms in either of the separate 
$\nu > 0$ or $ \nu < 0$ regions as was done in the 
previous footnote. In the work of the present paper,
however, such an extraction of the subdominant terms 
would be very difficult indeed, although perhaps
possible in principle.

\section{Convergent Kinetic Equations}
\label{equi}

Convergent Boltzmann transport equations have been
discussed by Frieman and Book\cite{Frie}, 
Weinstock\cite{wein}, and by Gould and DeWitt\cite{gould}. 
These are equations of the usual Boltzmann equation 
form, but with modified collision terms on the
right-hand side that account for both the hard,
short-distance collisions and the soft, infrared, 
long-distance scattering that is modified by the 
background plasma medium.  Just as in our work, these
equations are valid only to leading order in the 
plasma density. In this appendix, we shall describe 
these convergent kinetic equations and then sketch 
how they are equivalent to our method which
uses dimensional continuation. But before passing 
to these details, we should again note that our 
method gives only the correct leading order terms 
with no spurious higher-order terms. This simplicity
of computation is to be contrasted with the kinetic 
equation approach which does yield spurious higher-order 
terms that must be identified and discarded to obtain 
a consistent, leading-order result.

Let us first recall that the collision integral 
in the Boltzmann equation is of the generic form 
(\ref{coll}), which may be written as
\begin{eqnarray}
  C_{ab}( {\bf p}_a)\! &=& \!\int {d^\nu{\bf p}'_b
  \over (2\pi\hbar)^\nu} {d^\nu{\bf p}'_a \over
  (2\pi\hbar)^\nu} {d^\nu{\bf p}_b \over (2\pi\hbar)^\nu}
  \left| T \right|^2 (2\pi\hbar)^\nu \delta^{(\nu)} \!
  \left( {\bf p}_b' + {\bf p}_a' - {\bf p}_b - {\bf p}_a
  \right)
\nonumber\\
  && \quad
  (2\pi\hbar) \delta\left( {{p_b'}^2\over 2 m_b} +
  {{p'}_a^2 \over 2 m_a } -  { p_b^2 \over 2 m_b } -
  { p_a^2 \over 2 m_a } \right)
  \Big[ f_b({\bf p}_b') f_a({\bf p}_a')
  - f_b({\bf p}_b) f_a({\bf p}_a) \Big] \,.
\label{colll}
\end{eqnarray}
The papers cited in the preceding paragraph work 
in $\nu = 3$ spatial dimensions and write the total 
collision term as
\begin{equation}
  C_{ab}^{\rm converge}({\bf p}_a) = C^{\rm
  hard}_{ab}({\bf p}_a) + C_{ab}^{\rm soft}
  ({\bf p}_a) \,,
\label{total}
\end{equation}
where each of the two collision terms on the 
right-hand side have the generic form 
given in Eq.~(\ref{colll}). 

The first part $C^{\rm hard}_{ab}({\bf p}_a)$ 
accounts for Coulomb scattering taken to all 
orders with the first Born approximation
subtracted so as to avoid double counting 
since it is contained in the second $C_{ab}^{\rm 
soft}({\bf p}_a)$ term. As Gould and DeWitt note, 
and as we have spelled out in some detail in the
discussion of the hard scattering corrections to 
the cross section weighted momentum transfer integral 
(\ref{difff}), the treatment of this hard collision
contribution requires some care since, in $\nu = 3$, 
the Born and exact Coulomb scattering cross section 
elements are identical.  Gould and DeWitt regulate 
this contribution by taking it to be the scattering 
for a Debye screened Coulomb potential. They write
\begin{equation}
  \left| T^{\rm hard} \right|^2 =  \left| T_\smD
  \right|^2 -  \left| T_\smD^{(1)} \right|^2 \,,
\label{tharddecomp}
\end{equation}
where $T_\smD$ is the full, all-orders amplitude 
for the scattering on a Debye screened Coulomb 
potential, and $T_\smD^{(1)}$ is the first Born
approximation to this amplitude. The squares of
the corresponding amplitudes are subtracted to 
avoid the double counting mentioned above when 
the second collision term $C_{ab}^{\rm soft}(
{\bf p}_a)$ is included. The amplitude squared
$|T_\smD|^2$ is asymptotic to the exact 
Coulomb scattering amplitude squared at large 
momentum transfer
\begin{equation}
  {\bf q} = {\bf p}_a' - {\bf p}_a =
  {\bf p}_b - {\bf p}_b' \ ,
\end{equation}
but its behavior for small $q^2$ does not include 
the correct soft physics which entails 
frequency-dependent, dynamical screening.
The Debye screening makes the separate contributions
of each of the two terms in $|T^{\rm hard}|^2$ to the 
energy loss finite in the infrared region.  The 
subtraction of $|T_\smD^{(1)}|$, however, makes the
contribution of the difference (\ref{tharddecomp})  
to the energy loss
finite when the Debye screening is removed, when
$\kappa_\smD \to 0$.  But the price paid for this is a
spurious unwanted contribution of $-|T_\smD^{(1)}|^2$ 
in the ultraviolet. This piece will take care of 
itself upon adding the correct infrared physics
provided by $|T^{\rm soft}|^2$, where $T^{\rm soft}$ 
is given by the first Born approximation
to the dynamically screen Coulomb amplitude
\begin{equation}
  T^{\rm soft} = {e_a e_b \hbar \over q^2 \, 
  \epsilon({\bf q}/\hbar , \Delta E /\hbar)} \ ,
\end{equation}
with the energy change of the scattering process
being
\begin{equation}
  \Delta E = {{{\bf p}_a'}^2 \over 2 m_a}  -
  {{\bf p}_a^2 \over 2 m_a}  = {{\bf p}_b^2
  \over 2 m_b}  - {{{\bf p}_b'}^2 \over 2 m_b} \ .
\label{deltae}
\end{equation}
The dielectric function $\epsilon({\bf k},\omega)$ 
is that given by the random phase or one-loop, single 
ring approximation\footnote{
\baselineskip 15pt
The soft contribution
$C_{ab}^{\rm soft}({\bf p}_a) $ is also discussed in 
Section 46 in the {\it Physical Kinetics} volume of 
the Landau-Lifshitz series \cite{Lifs}.} (\ref{epsilon}).
Note that $|T^{\rm soft}|^2-|T_\smD^{(1)}|^2$ vanishes in
the ultraviolet since $T^{\rm soft}$ and $T_\smD^{(1)}$
are asymptotic as ${\bf q} \to \infty$. Therefore,
the addition of $|T^{\rm soft}|^2$ to $|T^{\rm hard}|^2$
cancels the unwanted ultraviolet part of $|T_\smD^{(1)}|^2$
mentioned above, leaving only the correct large $q^2$
behavior of $|T_\smD|^2$. To reiterate,
\begin{eqnarray}
\nonumber
  |T^{\rm converge}|^2 
  &=&
  |T^{\rm hard}|^2 + |T^{\rm soft}|^2 
\\[3pt]
  &=&
  \left| T_\smD \right|^2 -  \left| T_\smD^{(1)} 
  \right|^2 + |T^{\rm soft}|^2
\end{eqnarray}
has both the correct ultraviolet and infrared
behavior. The subtraction of $|T_\smD^{(1)}|^2$ 
does not just avoid double counting.
It also, on the one hand, removes the arbitrary 
$\kappa_\smD$ dependence produced by 
$|T_\smD|^2$, and on the other hand, removes the
large $q^2$ contribution of $|T^{\rm soft}|^2$.

This convergent kinetic theory approach is certainly
valid, but it entails spurious higher-order
corrections in the plasma density that must be 
discarded after calculations have been performed 
as Gould and DeWitt correctly do. 
In contrast to regularization (\ref{tharddecomp}), however, it 
is simpler (but ultimately equivalent) to regulate 
the hard scattering contribution by continuing 
it to a spatial dimensionality $\nu$ that is 
slightly above $\nu=3$ and using the
dimensionally continued exact pure Coulomb 
scattering amplitudes. With this regularization
\begin{equation}
  \left| T^{\rm hard} \right|^2 =  \left|
  T_\smC^{(\nu > 3)} \right|^2 - \left|
  {T_\smC^{(\nu > 3)}}^{(1)} \right|^2 \,,
\label{hardnu}
\end{equation}
and the phase-space collision integrals are 
also extended to $\nu > 3$. This regularization 
automatically entails no additional, spurious,
higher-order terms. 

To establish the connection of the convergent 
kinetic approach with our method using dimensional 
continuation, we add and subtract the first Born 
approximation to the scattering with a Debye screened
Coulomb potential so that the soft collision term 
appears as
\begin{equation}
  C_{ab}^{\rm soft}({\bf p}_a) = \bar
  C_{ab}^{\rm soft}({\bf p}_a)  +
  C_{\smD \, ab}^{(1)}({\bf p}_a)  \,.
\label{barr}
\end{equation}
The first term here is given by the Boltzmann 
collision integral with the squared scattering 
amplitude replaced by
\begin{equation}
  \left| \bar T^{\rm soft} \right|^2 =
  \left( { e_a e_b \over \hbar} \right)^2 \,
  \left[ \,\left|{1\over (q^2/ \hbar^2) \,
  \epsilon({\bf q}/\hbar ,\Delta E /\hbar
  ) }\right|^2 - \left|{1\over (q^2 /
  \hbar^2) + \kappa_\smD^2 }\right|^2 \,
  \right] \,,
\label{softt}
\end{equation}
where the second term is produced 
by
\begin{equation}
  \left| T^{(1)}_{\smD} \right|^2 = {1 \over \hbar^2} \, 
\left|{e_a e_b \over (q^2 / \hbar^2) + \kappa_\smD^2}
  \right|^2 \,.
\end{equation}
Since $\left| \bar T^{\rm soft} \right|^2$ 
vanishes rapidly for large momentum transfer 
$q$, the collision integral 
$\bar C_{ab}^{\rm soft}$ 
may be replaced by the Lenard-Balescu 
form which is employed in our method.  This is explained in
detail in the following Appendix \ref{fok}.

We  now give a quick 
proof that the `convergent kinetic equation' 
is equivalent to our method of dimensional
continuation provided that the kinetic equation 
is solved in a consistent fashion and spurious, 
higher-order terms are discarded. To do this, 
we take the hard Coulomb scattering part in
Eq.~(\ref{total}) to be defined by the dimensional
continuation with $\nu>3$, as indicated in the 
discussion of Eq.~(\ref{hardnu}). We take the soft 
part in Eq.~(\ref{total}) to be divided as in
Eq.~(\ref{barr}) with the first term with the over 
bar written in the Lenard-Balescu form.  We then 
extend this Lenard-Balescu part to $\nu <3$ so 
that its dynamically screened and Debye screened 
pieces may be treated separately.  We thus have
\begin{equation}
  C_{ab}^{\rm converge}({\bf p}_a) = C^{\rm our}_{ab}
  ({\bf p}_a) + C_{ab}^{\rm remain}({\bf p}_a) \,.
\end{equation}
Here
\begin{equation}
  C^{\rm our}_{ab}({\bf p}_a) = C^{(\nu > 3
  )}_{ab}({\bf p}_a) - {\partial \over \partial
  {\bf p}_a} \cdot {\bf J}^{(\nu <3)}_{ab}({\bf p}_a)
\end{equation}
corresponds to the result of our method of 
dimensional regularization: the first term
$ C^{(\nu > 3)}_{ab}({\bf p}_a)$ is the 
Boltzmann collision term for pure Coulomb 
scattering in $\nu > 3$ spatial dimensions, 
with the scattering treated to all orders,
while the second term involves 
${\bf J}^{(\nu <3)}_{ab}({\bf p}_a) $, 
which is the number
current of the Lenard-Balescu form (\ref{colbb}) 
with the dynamically screened, first Born 
approximation scattering amplitude given by
Eq.~(\ref{dynamic}). The remainder reads
\begin{equation}
  C_{ab}^{\rm remain}({\bf p}_a) = 
- C_{ab}^{(\nu>3)(1)}({\bf p}_a) 
+ {\partial \over \partial {\bf p}_a} \cdot 
{\bf J}_{\smD \, ab}^{(\nu<3)(1)}({\bf p}_a)
+ C_{\smD \, ab}^{(\nu=3)(1)}({\bf p}_a) \,.
\label{final}
\end{equation}
Here $C_{ab}^{(\nu>3)(1)}({\bf p}_a) $ is the 
Boltzmann collision term for the pure Coulomb 
scattering in $\nu >3$ spatial dimensions in 
the first Born approximation, 
${\bf J}_{\smD \, ab}^{(\nu<3)(1)}({\bf p}_a) $ 
is the Lenard-Balescu number
current for a Debye screened Coulomb potential 
in $\nu < 3$ dimensions, and  
$ C_{\smD \, ab}^{(\nu=3)(1)}({\bf p}_a) $ 
is the Boltzmann collision 
term for a Debye screened Coulomb potential in
first Born approximation in three spatial 
dimensions.  Our final job is to show that the 
remainder  $C_{ab}^{\rm remain} $ vanishes, as 
we now shall do.

If the difference of the Boltzmann collision 
terms that appear in Eq.~(\ref{final}) is written 
as a single integral, then its integrand vanishes 
rapidly at high momentum transfer. The $\nu=3$
Boltzmann collision term for the 
 Debye screened
potential in first Born approximation may be 
extended to $\nu > 3$. To the appropriate leading 
order, this difference can then be expressed in 
the Lenard-Balescu form,
\begin{equation}
 - C_{ab}^{(\nu>3)(1)}({\bf p}_a) 
+ C_{\smD \, ab}^{(\nu>3)(1)}({\bf p}_a) = 
- {\partial \over \partial {\bf p}_a} \cdot 
{\bf J}_{\smD - \smC \, ab}^{(\nu>3)(1)}({\bf p}_a) \,,
\end{equation}
where
\begin{eqnarray}
{\bf J}_{\smD - \smC \, ab}^{(\nu>3)(1)}({\bf p}_a) 
&=&
  \int {d^\nu{\bf p}_b \over (2\pi\hbar)^\nu}
  {d^\nu{\bf k} \over (2\pi)^\nu} \, {\bf k} \,
  \pi \delta\left( {\bf v}_a \cdot {\bf k}  -
  {\bf v}_b \cdot {\bf k} \right)
\nonumber\\
  && \quad
\left\{ 
\left| { e_a e_b \over k^2 + \kappa^2_\smD }\right|^2
  - \left| { e_a e_b \over k^2 } \right|^2 
\right\}
\nonumber\\
  && \quad
  \left[ {\bf k} \cdot {\partial \over \partial
  {\bf p}_b} - {\bf k} \cdot {\partial \over
  \partial {\bf p}_a} \right] f_a({\bf p}_a )
  f_b({\bf p}_b) \,.
\end{eqnarray}
In the method of dimensional continuation, the 
pure Coulomb piece that appears here vanishes. 
This is because the ${\bf k}$ integral carries
the dimensions of $L^{3-\nu}$ and there is no 
length $L$ present to carry this dimension.  
The remaining term involving the Debye screened
Coulomb potential defines, except for poles at 
integer dimensions, an analytic function for 
arbitrary dimensions that is identical in form 
with the current 
$ {\bf J}_{\smD \,ab}^{(\nu<3)(1)}$ 
in the remainder (\ref{final}).  Hence
\begin{eqnarray}
  C_{ab}^{\rm remain}({\bf p}_a) &=& + 
{\partial \over \partial {\bf p}_a} \cdot 
{\bf J}_{\smD \, ab}^{(\nu<3)(1)}({\bf p}_a) 
- {\partial \over \partial {\bf p}_a} \cdot 
{\bf J}_{\smD \, ab}^{(\nu>3)(1)}({\bf p}_a)
\nonumber\\
  &=& 0 \,.
\end{eqnarray}

\section{Fokker-Planck and Lenard-Balescu \\
Limits From Boltzmann Equation}
\label{fok}

In general, if the squared scattering amplitude 
in the collision integral (\ref{colll}) decreases 
sufficiently rapidly at large momenta, or if the 
dimension $\nu$ is sufficiently small such that the
phase-space volume at high energies becomes small, 
then the collision integral (\ref{colll}) may be 
replaced by an equation of the Lenard-Balescu form.  

The Lenard-Balescu equation is a classical equation.  
The mechanical momentum transfer ${\bf q}$ and classical wave
number ${\bf k}$ have the familiar relation
\begin{equation}
  {\bf q} = \hbar \, {\bf k} \,.
\end{equation}
It is the wave number ${\bf k}$ that is the significant variable 
in the scattering amplitude, while the momentum ${\bf q}$ appears 
in kinematical and phase space factors. The classical limit is 
the limit $\hbar \to 0$ with ${\bf k}$ fixed.  Thus we take 
the limit $ {\bf q} \to {\bf 0}$ in kinematical factors in 
which the  momentum role is emphasized.  This limit of small ${\bf q}$ in 
the kinematical factors in the Boltzmann equation produces 
the Fokker-Planck equation, to whose derivation we now turn. 

First we change
variables for the `spectator particle' $b$ by 
writing
\begin{equation}
  {\bf p}_b = \bar {\bf p}_b + {1\over2} \,
  {\bf q} \,, \qquad {\bf p}'_b = \bar
  {\bf p}_b - {1\over2} \, {\bf q} \,.
\end{equation}
Removing the momentum-conserving delta function 
by the ${\bf p}'_a$ integration now presents the 
collision term (\ref{colll}) as
\begin{eqnarray}
  C_{ab}( {\bf p}_a) &=& \int {d^\nu\bar{\bf p}_b
  \over (2\pi\hbar)^\nu} {d^\nu{\bf q} \over
  (2\pi\hbar)^\nu} \left| T \right|^2 \, (2\pi\hbar)
  \delta\left( {{\bf p}_a \cdot {\bf q} \over m_a} +
  {{\bf q}^2 \over 2 m_a }  -  { \bar{\bf p}_b \cdot
  {\bf q} \over m_b} \right)
\nonumber\\
  && \quad
  \left[ f_b\left(\bar{\bf p}_b - {1\over2} {\bf q}
  \right) f_a({\bf p}_a + {\bf q})- f_b\left(\bar
  {\bf p}_b + {1\over2} {\bf q} \right)  f_a({\bf p}_a)
  \right] \,.
\label{colb}
\end{eqnarray}
We make an expansion in the momentum 
transfer ${\bf q}$ when it appears together with
or in comparison with the momenta ${\bf p}_a$ or 
$\bar {\bf p}_b$. As we shall soon see, the leading terms, the
only terms that we shall retain, are quadratic in ${\bf q}$.

Expanding to second order in ${\bf q}$, and performing some
rearrangement to simplify the result, produces
\begin{eqnarray}
  &&\left[ f_b\left(\bar{\bf p}_b - {1\over2}
  {\bf q} \right) f_a({\bf p}_a + {\bf q}) -
  f_b\left(\bar{\bf p}_b + {1\over2} {\bf q}
  \right)  f_a({\bf p}_a) \right] \,.
\nonumber\\
  &&\qquad
  \simeq \left[ 1 + {1 \over 2} {\bf q} \cdot
  {\partial \over \partial {\bf p}_a } \right]\,
  \left[{\bf q} \cdot {\partial \over \partial
  {\bf p}_a } - {\bf q} \cdot {\partial \over
  \partial \bar{\bf p}_b } \right]\, f_b(\bar
  {\bf p}_b) f_a({\bf p}_a) \,.
\end{eqnarray}
Since this factor in the collision integral (\ref{colb}) starts
out linearly in ${\bf q}$ and since we shall work only to second
order in ${\bf q}$, the remaining terms need only be expanded to
first order in ${\bf q}$. Thus we may write 
\begin{eqnarray}
\delta\left( {{\bf p}_a \cdot
  {\bf q} \over m_a} + {{\bf q}^2 \over 2 m_a }  -
  { \bar{\bf p}_b \cdot {\bf q} \over m_b} \right)
 \simeq
  \left[ 1 + {1 \over 2} {\bf q} \cdot {\partial
  \over \partial {\bf p}_a } \right]
 \delta\left( {{\bf
  p}_a \cdot {\bf q} \over m_a}  - { \bar{\bf p}_b \cdot
  {\bf q} \over m_b} \right)  \,.
\end{eqnarray}
For a Galilean invariant theory, the scattering amplitude
depends only upon the squared momentum transfer ${\bf q}^2$ and
the (relative) energy in the center-of-mass $W$. By virtue of 
the conservation of energy [which is enforced by the delta 
function that remains in Eq.~(\ref{colb})] this energy may be
expressed in terms of either initial or final state variables,
\begin{equation}
W = {1\over2} \, m_{ab} \, 
   \left( {\bf v}_a - {\bf v}_b \right)^2 
  = {1\over2} \, m_{ab} \, 
   \left( {\bf v}_a' - {\bf v}_b' \right)^2 \,,
\end{equation}
where $m_{ab}$ is the reduced mass and the velocities have the
usual, generic, form ${\bf v} = {\bf p} / m$.  
Since $\bar{\bf p}_b$ is the integration variable, we must write
\begin{equation}
W = {1\over2} \, m_{ab} \, 
\left( {\bf v}_a - \bar{\bf v}_b - {1 \over 2 m_b} \,
               {\bf q} \right)^2 \,,
\end{equation}
Again, we need only consider the corresponding correction to
linear order in ${\bf q}$. 
According to the chain rule, this ${\bf q}$-dependence of $W$ 
entails the derivative of $|T|^2$ with respect to $W$ times
\begin{equation}
-{1\over2 m_b}\,{\bf q} \cdot {\partial\over\partial {\bf v}_a} 
\, W = - {m_{ab} \over 2 m_b}\,\left( {\bf q} \cdot {\bf v}_a 
           -  {\bf q} \cdot \bar{\bf v}_b \right) 
           + {\cal O}({\bf q})^2 \,.
\label{zeroo}
\end{equation}
To leading order, this term involving the derivative of $|T|^2$
does not, in fact, contribute since the terms above multiply 
$ \delta \left( {\bf q} \cdot {\bf v}_a - 
             {\bf q} \cdot \bar{\bf v}_b\right) $
and thus give a null result. Thus we may consider the scattering
amplitude to simply be a function of 
\begin{equation}
\bar W = {1\over2} \, m_{ab} \, 
\left( {\bf v}_a - \bar{\bf v}_b  \right)^2 \,.
\end{equation}

As we have remarked above, the squared wave number 
${\bf k}^2$ is the relevant variable for the case that we are now
considering rather than the momentum transfer 
${\bf q}^2 = \hbar^2 \, {\bf k}^2 $.  The usual form of the Boltzmann
equation involves a squared scattering amplitude that is Galilean
invariant and thus only a function of $W$ and ${\bf k}^2$.  However, we
need to generalize this a little to take into account the plasma
screening corrections that are included in the Lenard-Balescu
limit of the Boltzmann equation.  The background plasma specifies
a rest-frame coordinate system, and so non-Galilean invariant
variables may now also appear in the scattering amplitude.  Since
the system remains rotationally invariant, the only remaining
variables involve the kinetic energies of the reacting particles.
As we shall shortly see, the only relevant combination is the
energy difference of the initial and final states of a particle
as measured in the plasma rest frame. Since the total energy is
conserved in the collision integral (\ref{colb}), there is only
one energy difference
\begin{equation}
\Delta E = {1\over2} \, m_a  \left( {{\bf v}^\prime}^2_a
                                  - {\bf v}_a^2 \right) 
=  {1\over2} \, m_b  \left({\bf v}_b^2 -
        {{\bf v}^\prime}^2_b \right) \,.
\end{equation}
Since 
\begin{equation}
\Delta E  = {1\over 2} 
     \left( {\bf v}'_a + {\bf v}_a \right) \cdot {\bf q} \,,
\end{equation}  
this energy difference naturally defines a classical frequency, a
frequency determined only by classical quantities, 
\begin{equation}
\omega = \Delta E / \hbar = 
{1\over 2} \left( {\bf v}'_a + {\bf v}_a \right) \cdot {\bf k} \,,
\end{equation}  
and, to the order that concerns us, the squared scattering
amplitude may be expressed\footnote{As in the discussion about
  Eq.~(\ref{zeroo}), the derivative acting upon $\bar W$ gives no
  contribution to our order.} as
\begin{equation}
\left|T\left(\bar W, {\bf k}^2, ( {\bf v}_a + {\bf q} /2 m_a ) 
           \cdot {\bf k}  \right) \right| \simeq
  \left[ 1 + {1 \over 2} {\bf q} \cdot {\partial
  \over \partial {\bf p}_a } \right]
\left|T\left(\bar W,{\bf k}^2,{\bf v}_a \cdot{\bf k} \right)
           \right| \,.
\end{equation}
Since the expansion terms that are linear in ${\bf q}$ involve a
complete integrand that is odd, they do not contribute, and the 
collision term (\ref{colb}) now reduces to
\begin{equation}
  C_{ab}( {\bf p}_a) = - {\partial \over \partial
  {\bf p}_a} \cdot {\bf J}_{ab}( {\bf p}_a) \,,
\label{fokcol}
\end{equation}
in which
\begin{eqnarray}
  {\bf J}_{ab}( {\bf p}_a) &=& \int {d^\nu{\bf p}_b
  \over (2\pi\hbar)^\nu} {d^\nu{\bf k} \over
  (2\pi)^\nu} \, {\bf k} \,
   \left|\hbar T(W,k^2 , {\bf v}_a \cdot {\bf k} )\right|^2 
  \, \pi \delta\left( {\bf v}_a \cdot {\bf k}  -
  {\bf v}_b \cdot {\bf k} \right)
\nonumber\\
  && \quad
  \left[ {\bf k} \cdot {\partial \over \partial
  {\bf p}_b} - {\bf k} \cdot {\partial \over
  \partial {\bf p}_a} \right]f_a({\bf p}_a )
  f_b({\bf p}_b) \,.
\label{colbb}
\end{eqnarray}
Here we have 
removed the overline from the spectator momentum 
variable, $\bar{\bf p}_b \to {\bf p}_b $, and
correspondingly written $W$ in place of $\bar W$.

A trivial algebraic rearrangement of
Eq's.~(\ref{fokcol}) and (\ref{colbb}) expresses the result as
\begin{equation}
C_{ab}({\bf p}_a) = {\partial \over \partial p^l_a }
\left[ B^{lm}_{ab}({\bf p}_a)  {\partial \over \partial p^m_a }
- A^l_{ab}({\bf p}_a) \right] \, f({\bf p}_a) \,,
\end{equation}
which is the Fokker-Planck form.  

The Lenard-Balescu equation entails only leading-order scattering,
but fully dynamically screened. In this approximation, 
\begin{equation}
  \left| \hbar T \right|^2 = \left|
  { e_a e_b \over k^2 \, \epsilon({\bf k}^2 ,
  {\bf v}_a \cdot {\bf k} ) }\right|^2 \,,
\label{dynamic}
\end{equation}
and the reduced form 
(\ref{colbb}) becomes identical with the previous 
Lenard-Balescu collision integral, Eq.~(\ref{lbj}).
Note that the frequency dependence that appears here in the 
dielectric function, $ \omega = {\bf v}_a \cdot {\bf k} $ is just
the $\Delta E / \hbar$ discussed above or, equivalently, the 
$\Delta E$ defined in Eq.~(\ref{deltae}). 
Thus the scattering amplitude (\ref{dynamic}) is just the 
first part of Eq.~(\ref{softt}).

As we have often remarked, the Lenard-Balescu equation is valid
only for dimensions less than three, $\nu < 3$.  This reduced
dimensionality is necessary for the convergence of the wave
number integration at large $k$.  In three dimensions, the
dynamically screened Coulomb scattering amplitude does not vanish
sufficiently rapidly at large ${\bf q} = \hbar {\bf k}$ so as to
permit the $\hbar$ expansion that we have made.  If, however, the
squared Debye Born amplitude is subtracted as in
Eq.~(\ref{softt}), then the Lenard-Balescu reduction may be made
directly in three dimensions.  Indeed, as we have repeatedly
emphasized, this reduction {\em must} be made to consistently
compute only the leading terms.  Since the only kinematical
variables available are momentum variables, the first correction,
which is of order $\hbar^2$, must appear in the form of a squared
length $\lambda^2 = \hbar^2 / p^2$.  A dimensionless ratio can
only be obtained by multiplication with the Debye wave number. 
Hence the first correction
to the leading classical limit that we have just derived is of
order\footnote{Up to an omnipresent, omnivorous, logarithm.}
$\lambda^2 \, \kappa_\smD^2$. This is formally of order of the
plasma density $n$ relative to the leading order result, a
correction that is beyond the order to which we compute. 
Having said all this, one can now continue the Lenard-Balescu
like equation in three dimensions for the subtracted squared amplitude 
(\ref{softt}) to $\nu < 3$.  The two parts of the equation can
then be separately treated as done in the previous Appendix 
\ref{equi}.  

Lifshitz and Pitaevskii \cite{Lifsss} purport to derive the
Fokker-Planck equation from the Boltzmann equation for a dilute
system of very heavy particles moving in a gas of light
particles.  Since our work might be confused as having some
relationship to theirs, we briefly review it here. For the sake
of completeness, we shall show that their work
requires further approximations to become internally
consistent. This we do by providing an explicit example.  But
before starting out to do this, we should again emphasize that our
reduction of the Boltzmann equation starts from the 
assumption that the scattering is restricted to small momentum
transfers because of the {\it dynamics} of the scattering cross
section. Lifshitz and Pitaevskii, on the other hand, assume that
the momentum transfer is small because of the {\it kinematics}
of a heavy particle moving in a light gas.

First we transcribe the description of Lifshitz and Pitaevskii
into our notation.  They write the collision integral in the form   
\begin{eqnarray}
  C_{ab}( {\bf p}_a ) &=& \int d^\nu{\bf q} \left\{
 w_{ab}({\bf p}_a + {\bf q} , {\bf q} )\,  f_a({\bf p}_a + {\bf q} )
   - w_{ab}({\bf p}_a , {\bf q} ) \, f_a({\bf p}_a ) \right\} \,.
\label{lpp}
\end{eqnarray}                  
The ``projectile'' particle $a$ is assumed to be heavy, and the
``gas'' particle $b$ is assumed to be light.
Comparing this structure with our standard form (\ref{colll})
and performing one of the momentum integrals trivially using
the momentum-conserving delta function gives
\begin{eqnarray}
  w_{ab}( {\bf p}_a) &=& \int {d^\nu{\bf p}_b
  \over (2\pi\hbar)^{2\nu}} \, f_b({\bf p}_b) \,
  \left| T({\bf v}_{ab}^2, {\bf q}^2) \right|^2
\, (2\pi\hbar) \,\delta\left( 
{1 \over 2m_{ab} } \, {\bf q}^2  - {\bf q} \cdot {\bf v}_{ab} 
  \right) \,.
\end{eqnarray}
Here, as before, ${\bf v}_{ab} = {\bf v}_a - {\bf v}_b$ is the
relative velocity with ${\bf v}_a = {\bf p}_a / m_a $, 
${\bf v}_b = {\bf p}_b / m_b $, and $m_{ab}$ is the reduced mass.  
Lifshitz and Pitaevskii expand Eq.~(\ref{lpp}) in powers of 
${\bf q}$ when this momentum transfer appears added to the heavy
particle momentum ${\bf p}_a$ and retain terms up to second
order. This formal expansion gives the approximate collision term
\begin{equation}
  C_{ab}( {\bf p}_a ) = {\partial \over \partial p_a^l }
    \left\{ \tilde A_{ab}^l({\bf p}_a) \, f_a({\bf p}_a) 
   + {\partial \over \partial p_a^m} \,
    B_{ab}^{lm}({\bf p}_a) \, f_a({\bf p}_a) \right\} \,,
\label{approxC}
\end{equation}
where
\begin{equation}
\tilde A_{ab}^l({\bf p}_a) = \int d^\nu {\bf q} \,\,
          q^l \, w_{ab}({\bf p}_a , {\bf q} ) \,,
\label{tildeA}
\end{equation}
and
\begin{equation}
B_{ab}^{lm}({\bf p}_a) = {1\over2} \, \int d^\nu {\bf q} \,\,
          q^l \, q^m \,  w_{ab}({\bf p}_a , {\bf q} ) \,.
\label{Blm}
\end{equation}
Lifshitz and Pitaevskii go on to set
\begin{equation}
 A_{ab}^l({\bf p}_a) = \tilde A_{ab}^l({\bf p}_a)
  + {\partial B_{ab}^{lm}({\bf p}_a) \over \partial p_a^m} \,.
\end{equation}
It is easy to see that the approximate collision term
(\ref{approxC}) vanishes when the heavy particle species $a$ is
in thermal equilibrium with the light gas particles $b$ if
\begin{equation}
 A_{ab}^l({\bf p}_a) =  \beta 
              B_{ab}^{lm}({\bf p}_a) \, v_a^m \,.
\label{equil}
\end{equation}
As we shall see, the satisfaction of this constraint requires
further approximation. In this sense, the work of Lifshitz and
Pitaevskii is misleading.

The easiest and clearest way to demonstrate this is to consider
an explicit example.  We examine the case in which the
scattering amplitude $T$ is a constant and the spatial
dimensionality is taken to be two, $\nu = 2$. These restrictions 
lead to trivial integrals and it is
very easy to explicitly evaluate the expressions (\ref{tildeA})
and (\ref{Blm}) using Eq.~(\ref{lpp}) with the distribution of
the light gas particles $f_b({\bf p}_b)$ of Maxwell-Boltzmann
form at temperature $T = 1 / \beta$. The results are
\begin{equation}
\tilde A_{ab}^l({\bf p}_a) =  v_a^l \, n_b 
                     \, \left( {\cal C} \, m_{ab}^2 \right) \,, 
\end{equation}
and \begin{equation}
    B_{ab}^{lm}({\bf p}_a) = \left\{ 
{m_{ab} \over m_b} \, T \, \delta^{lm} 
+ {1\over2} \, m_{ab} \, \left( v_a^l \, v_a^m + {1\over2} \,
                   v_a^2 \, \delta^{lm} \right) \right\}    
             \, n_b \, \left( {\cal C} \, m_{ab}^2 \right) \,,
\label{ugh}
\end{equation}
where $n_b$ is the particle number density of the light gas
particles, and ${\cal C}$ is a constant involving $|T|^2$ and 
geometrical factors such as $2\pi$. These explicit 
results obviously violate the thermal equilibrium constraint 
(\ref{equil}). A consistent result requires the further
approximation that the ``projectile'' mass $m_a$ is much greater
than the ``gas'' mass $m_b$ so that one can replace 
$ m_{ab} / m_b \to 1$.  With this requirement obeyed, and
assuming that the ``projectile'' speed is not too great in the
sense that the kinetic energy $m_a \, v_a^2 /2 $ is not much
greater than the temperature $T$, the second set of terms in 
Eq.~(\ref{ugh}) may be neglected, and we see that the thermal
equilibrium constraint (\ref{equil}) is now obeyed.

Be all this as it may, we must emphasize again that the
Lifshitz-Pitaevskii treatment is based on a kinematical
restriction which is very different than the dynamical condition
that we impose in order to reduce the Boltzmann equation to a
Fokker-Planck equation and ultimately obtain the Lenard-Balescu
limit. 

\section{The Classical Limit}
\label{csavtcl}

Expression (\ref{wave}) for the cross section 
averaged momentum transfer is simply related to 
the classical limit. To see how this goes, we 
write Eq.~(\ref{wave}) as
\begin{eqnarray}
  \int d\sigma \, q^2 =  2 \pi \hbar^2 \sum 
  (l+1) \, 2 \left[1 - \cos 2 \left(\delta_{(l+1)} 
  - \delta_l \right) \right] \,.
\end{eqnarray}
In the classical limit, large $l$-values dominate. 
Thus, with the identification $J = (l + 1/2) \hbar$, 
the sum may be replaced by an integral, $ \hbar^2 
\sum (l+1) \to \int J dJ$, and the phase shifts 
approximated by their WKB evaluation,
\begin{eqnarray}
  \delta_l(p) = \left( l + {1\over2} \right) 
  { \pi \over 2} + {1 \over \hbar} \left\{ 
  \int_{r_m}^\infty dr \left[ \sqrt{ 2m (E-V) 
  - J^2 / r^2 } - p \right]- p \, r_m \right\} \,,
\end{eqnarray}
where $ E = p^2 /2m$ and $r_m$ is the turning 
point, the radial coordinate where the square root 
vanishes.  The classical scattering angle $\theta$ 
may be found in the usual way: The conservation of 
the angular momentum $J = m r^2 d\theta / dt $ is 
used to replace the time increment $dt$ in the 
energy conservation equation by $dt = m r^2 
d\theta/J$. This gives the usual trajectory 
equation for $dr / d\theta $ which may be integrated 
to evaluate the classical scattering angle as
\begin{eqnarray}
  \theta = \pi - 2 \int_{r_m}^\infty dr \, {J 
  \over r^2} \,{1 \over \sqrt{2m(E-V) - J^2 /
  r^2 } } \,.
\end{eqnarray}
Thus we find that, in the classical limit,
\begin{eqnarray}
  \theta = 2 \hbar { \partial \delta_l \over 
  \partial J } = 2 \left[ \delta_{(l+1)} - 
  \delta_l \right] \,,
\end{eqnarray}
so that in this limit
\begin{eqnarray}
  \int d\sigma q^2 = 2\pi \int_0^\infty J dJ \, 
  2 \, (1 - \cos\theta ) \,.
\end{eqnarray}
The classical impact parameter $b$ is defined 
such that $ J = pb $, while $q^2 = 2 p^2 ( 1 - 
\cos\theta )$.  Hence we have found that the 
classical limit of quantum mechanics yields
\begin{eqnarray}
  \int d\sigma q^2 = 2\pi \int_0^\infty bdb \, 
  q^2 \,,
\end{eqnarray}
which is indeed the classical result.

\end{document}